\documentclass[twocolumn]{aastex631}

\usepackage{amsmath}
\usepackage{amssymb}

\shorttitle{Sub-Neptune Mass Loss and Evolution}

\begin{document}

\title{Assessing Core-Powered Mass Loss in the Context of Early Boil-Off: Minimal Long-Lived Mass Loss for the Sub-Neptune Population}




\author[0000-0003-3980-7808]{Yao Tang}
\affiliation{Department of Astronomy and Astrophysics, University of California, Santa Cruz \\
1156 High Street, Santa Cruz, CA 95064, USA}

\author[0000-0002-9843-4354]{Jonathan J. Fortney}
\affiliation{Department of Astronomy and Astrophysics, University of California, Santa Cruz \\
1156 High Street, Santa Cruz, CA 95064, USA}

\author[0000-0001-5061-0462]{Ruth Murray-Clay}
\affiliation{Department of Astronomy and Astrophysics, University of California, Santa Cruz \\
1156 High Street, Santa Cruz, CA 95064, USA}

\begin{abstract}
We develop a python-based state-of-the-art sub-Neptune evolution model that incorporates both the post-formation boil-off at young ages $\leq$ 1 Myr and long-lived core-powered mass loss ($\sim$ Gyrs) from interior cooling. We investigate the roles of initial H/He entropy, core luminosity, energy advection, radiative atmospheric structure, and the transition to an XUV-driven mass-loss phase, with an eye on relevant timescales for planetary mass loss and thermal evolution. With particular attention to the re-equilibration process of the H/He envelope, including the energy sources that fuel the hydrodynamic wind, and energy transport timescales, we find boil-off and core-powered escape are primarily driven by stellar bolometric radiation. We further find that both boil-off and core-powered escape are decoupled from the thermal evolution. We show that, with a boil-off phase that accounts for the initial H/He mass fraction and initial entropy, post-boil-off core-powered escape has an insignificant influence on the demographics of small planets, as it is only able to remove at most 0.1\% of the H/He mass fraction. Our numerical results are directly compared to previous work on analytical core-powered mass loss modeling for individual evolutionary trajectories and populations of small planets.  We examine a number of assumptions made in previous studies that cause significant differences compared to our findings. We find that boil-off, though able to completely strip the gaseous envelope from a highly irradiated ($F \geq 100 F_\oplus$) planet that has a low-mass core ($M_c \leq 4M_\oplus$), cannot by itself form a pronounced radius gap as is seen in the observed population.
\end{abstract}


\section{Introduction} \label{sec:intro}
Since the initial discovery of planet GJ 1214b \citep{Charb09}, we have seen a great number of exoplanets with sizes smaller than Neptune but larger than Earth, thanks to the \emph{Kepler} and \emph{TESS} missions.  Studies of the occurrence rate of these small planets on $<100$ day orbits show a pronounced gap \citep{Fulton17,Vaneylen18, Fulton18}. This gap splits the population of these close-in planets into two populations: super-Earths, with thin or negligible atmospheres atop rocky cores, and sub-Neptunes, with substantially lower bulk densities, which suggests that they may hold thick H/He envelopes. Due to the degeneracy in determining the relative amount of interior elements, an alternative explanation suggests that some water world sub-Neptunes with a steam-dominated atmosphere could also exist within the larger-radius population \citep{Rogers10,Zeng19,Luque22}. The paucity of planets between $1.5\ R_\oplus$ and $2.0\ R_\oplus$ is likely due to the lack of planets with very thin H/He atmospheres \citep{Lopez12,Owen13,Lopez13,Lee21,Lee22}, as such atmospheres are expected to be strongly impacted by ongoing atmospheric escape, with loss enhanced for planets with low masses and high stellar insolation. On the other hand, since increasing the amount of H/He material can significantly inflate a small planet's radius \citep{Lopez14}, planets that survive atmospheric loss retain a primordial H/He atmosphere, larger radius, and low bulk density.

Due to the lack of such equivalent planets in our own Solar System, and with many aspects of the physical properties of sub-Neptunes far beyond the conditions found on or within Earth, their compositions and interior structures still remain unclear. Theoretically, if these planets form in situ, they possess a rock/iron core that makes up the majority of their masses with only a small amount of H/He gas atop, but which contributes a large fraction to their radii. If a large quantity of water is present, they must be formed beyond the snow line and then migrate to their current positions \citep{Rogers11}. Although modeling work has shown the H/He mass fraction for the planets within the radius range of $2-3\ R_\oplus$ is typically a few percent \citep{Lopez14}, the quantitative details vary significantly from model to model, as a small change in H/He mass sensitively alters planet radii. Moreover, how these planets retain these H/He mass fractions over Gyr ages is still not fully understood. We need more detailed knowledge of their atmospheric escape history, which is often coupled with their thermal evolution. This requires us to develop better evolution and mass loss models to assess how their interior thermal states and compositions change as they evolve.

A hydrodynamic wind that drives H/He mass loss from planets occurs when hydrostatic equilibrium cannot be maintained in their atmospheres. How strongly the H/He gas is bound to a planet depends on the planetary internal structure, which sets a planet's radius and hence the potential well from which atmospheric gas must be lifted. Additionally, a planet must be heated sufficiently to have a suitable pressure gradient to drive and sustain a hydrodynamic wind against gravity.

Three physical mechanisms are typically invoked in recent models of atmospheric loss from highly irradiated sub-Neptunes---boil-off, core-powered escape, and photoevaporation.  All three mechanisms result in hydrodynamic wind outflows, but they differ in these two key respects: the source of the energy driving the wind and the impact of the outflow on a planet's structure. 

Stellar ionizing radiation deposited high in the planet's atmosphere heats the upper atmosphere to a few thousand K for a typical sub-Neptune and to $\sim$10,000K for hot Jupiters, giving rise to a hydrodynamic outflow, known as photoevaporation \citep{Lammer03,Baraffe04}. The high energy flux of X-ray and EUV (XUV) photons constitutes an important fraction of stellar radiation in the first 100 Myr of evolution, usually around $10^{-3}$, significantly impacting low-mass planets' physical sizes and their bulk densities. The modeling details of such hydrodynamic evaporation have been widely discussed in the literature for both hot-Jupiters \citep{Yelle04,Garcia07,RMC09,Owen14,Tripathi2015,McCann2019,Debrecht2019} and sub-Neptunes \citep{Lopez12,Owen12,Lopez13,Owen13,Jin14,Chen16,Kubyshkina18,Jin18,Rogers23}. Coupled XUV-driven mass-loss and evolution models yield very similar demographic features to those of observations \citep[i.e.,][]{Fulton18} for low-mass planets \citep{Owen13,Lopez13,Owen17, Jin18}. In terms of direct observations of mass loss, the detection of very large transit radii in Lyman-$\alpha$ has suggested the existence of such hydrodynamic blow-off for both hot-Jupiters \citep{Vidal03,Vidal04,Lecavelier10,Lecavelier12} and warm-Neptunes \citep{Kulow14,Lavie17}. 
 
Alternatively, it has been proposed that the energy released from the primordial thermal energy reservoir of the core can exceed the gravitational binding energy of the envelope, and therefore can drive a complete loss of H/He envelope from a low-mass planet on longer Gyr timescales, known as core-powered escape \citep{Ginzburg2016}. The energy that lifts material from the envelope is argued to be directly from the intrinsic luminosity that eventually leads to core cooling. The stellar bolometric flux also plays a role in powering the outflow within the radiative atmosphere. Compared to photoevaporation, the thermal evolution and atmospheric loss are strongly coupled as the mass loss rate is set by the cooling process. Demographic studies for planet evolution in the presence of core-powered escape have been done, producing very similar features to those seen both in observations and in photoevaporation work \citep{Ginzburg18,Gupta19,Gupta20}. However, studies have shown that we cannot distinguish between photoevaporation and core-powered mass-loss based on the current samples of the confirmed planets \citep{Loyd20, Rogers21, Ho23}, without having access to a 3D radius valley that informs us how the radius gap varies independently with both orbital period and stellar bolometric flux \citep{Rogers21}. 

Furthermore, a more powerful mechanism that occurs with the disk dispersal, much earlier than the others, known as ``boil-off'' \citep{Owenwu16} or ``spontaneous mass loss'' \citep{Ginzburg2016,Misener21}, has been proposed. Observations indicate that the disk gas clears out rapidly from inside out over a short duration of approximately $10^5$ years \citep{Ercolano10}. This occurs when the gaseous disk is nearly depleted after $\sim$3-10 Myr years of evolution, eventually leaving behind a dusty debris disk. As the confinement pressure from protoplanetary disks sharply declines as a result of the disk dissipation, a hot newly born planet possesses a large physical size that suddenly leads to hydrostatic disequilibrium and therefore a tremendous hydrodynamic outflow. This process has a mass loss rate orders of magnitude greater than both photoevaporation and core-powered mass loss. With a short duration of only a few Myr, it can readily strip away a substantial fraction of H/He material, strongly affecting the H/He mass fraction a sub-Neptune would have available at the beginning of the subsequent evolution. As opposed to photoevaporation, boil-off can potentially affect the thermal evolution because of the vast interior thermal energy that is carried away along with the hydrodynamic outflow \citep{Owenwu16}, which consequently shifts the subsequent thermal evolution in the presence of the long-term mass loss away from the evolution yielded from a commonly used ``hot-start'' model that starts evolution with an arbitrarily large H/He envelope entropy at the youngest ages \citep[e.g.,][]{Marley07,Lopez14}. 

Previous work has discussed the importance of all of these physical mechanisms and their impact on small planet demographics. However, a number of simplifications have been made in previous modeling efforts. This includes an isothermal Parker wind for boil-off assuming the energy available from the internal cooling is always sufficient to sustain the wind. In addition, an energy-limited prescription has been used for core-powered escape.  Improving on previous work requires us to better understand the energy source for both boil-off and core-powered escape. Moreover, at what age photoevaporation or core-powered escape comes into play and what physical conditions a planet has when each of these three atmospheric escape mechanisms dominates, are not fully clear. Previous simplifications have led to significant uncertainties in modeling a planet's evolving H/He mass fractions and radii and consequently in studying the nature of physics causing the radius gap. Lastly, observations \citep{Fulton18} show a large decrease in the frequency of planets greater than $4\ R_\oplus$ known as the ``radius cliff''. Although physical explanations for the scarcity of these sub-Saturn size planets are proposed from the formation perspective \citep{Lee19,Kite19}, studies of subsequent planetary evolution, especially in the presence of mass loss that eventually shapes the radius cliff, are critical \citep{Hallatt22}.

Therefore, in this work, we develop a state-of-the-art python-based sub-Neptune evolution model from the deep interior to the radiative atmosphere that includes a self-consistent calculation of the boil-off phase to better assess initial H/He inventories and entropy for the subsequent long-lived physical processes (e.g. thermal evolution, core-powered escape and photoevaporation). In this context, with our numerical model, we focus on the following aspects:
\begin{itemize}
  \item The significance of core-powered escape is reassessed with the results directly compared to the analytical model developed in \citet{Ginzburg2016}.
  \item The energy source and mass re-equilibration for boil-off and core-powered escape and their impact on the thermal evolution are reevaluated.
  \item We constrain the transition time from boil-off or core-powered escape to the photoevaporation-dominated phase. 
  \item We also emphasize the importance of the often-neglected radiative atmosphere atop the convective envelope.  This part of the planet is dominated by stellar heating, and it is a large fraction of the planetary radius, atop the convective H/He envelope that shrinks as the planet's interior cools. 
  \item After discussion of how relevant physics affects sample planets, we study the impact of boil-off and core-powered escape at a population level for a large sample of planets, both analytically and numerically, and shed light on its relation to the observed small planet demographic features and mass loss processes.
\end{itemize}

\section{A new sub-Neptune evolution model} \label{sec:numerical}
\subsection{Thermal contraction}
\label{subsec:thermal}
Planet formation is an energetic process, leading to a hot initial condition and subsequent cooling of the planetary interior. The cooling of the H/He envelope allows gravitational binding energy and internal energy to be released as thermal energy, and as a result, the planet's interior contracts. Such thermal contraction is critical for modeling atmospheric escape, as the planet radius controls the potential well that H/He material is lifted from. Similar to previous planet evolution codes \citep{Fortney07,Lopez14}, the planet's gaseous H/He is assumed to consist of an envelope that is adiabatic and isentropic due to the short mixing timescale of convection, with a radiative atmosphere on top. Given the entropy and the envelope mass for the envelope, the interior thermal state of a planet is completely defined by the Equation of State (EoS), which we take from \citet{Chabrier19}, and hydrostatic equilibrium.

Between each timestep, to evolve the interior we track the energy fluxes that heat the envelope from below and cool the envelope at the top, which change the interior thermal states. The isentropic envelope links the change of specific entropy $\Delta s$ of the envelope adiabat to the net heat transfer $\Delta Q = L \Delta t = \Delta s \int T dm$ to the envelope within the time interval $\Delta t$, which gives:
\begin{eqnarray}
\label{thermalcontraction}
-L &=& \frac{ds}{dt} \int_{M_{c}}^{M_c+M_{env}}  T\,dm   \nonumber \\
&=& - L_{env} + L_{core} \nonumber  \\
&=& - \max(L_{int}, L_{ad}) + L_{radio} - c_v M_{c} \frac{dT_{c}}{dt}
\end{eqnarray}
where $M_{c}$ and $M_{env}$ are the core mass and the envelope mass respectively, and $T$ is the temperature of each mass shell $dm$. On the right-hand side, we sum up each luminosity component to get the total envelope luminosity $L$, where the envelope cooling $L_{env}$ at the top is set by the greater of either the radiative intrinsic luminosity $L_{int}$ or advective luminosity $L_{ad}$, the energy transport from the bulk flow driven by atmospheric escape.  The envelope heating from below is due to the total core luminosity $L_{core}$ from both radiogenic heating $L_{radio}$ and the core heat capacity.

The intrinsic luminosity $L_{int}$, the net cooling rate, is the total energy that a planet's interior radiates per unit time through its radiative atmosphere to space. To evaluate the intrinsic luminosity, we utilize a grid of one-dimensional atmosphere models for solar metallicity, as described in \citet{Fortney07}, over a range of surface gravities, incident bolometric fluxes and interior temperatures. As a planet loses mass through atmospheric escape, the hydrodynamic wind is capable of transporting internal thermal energy out of the interior. This effect is generally minor for a long-lived mass loss, i.e. photoevaporation and core-powered mass loss. However, when a planet is rapidly blowing off its H/He envelope in the boil-off phase, the mass loss rate is a few orders of magnitude greater than that at a later age, making such an energy advection, $L_{ad}$, potentially important. \citet{Owenwu16} argue that the energy advection from boil-off leads to an interior cooling that is much more significant than the radiative cooling, with the advective cooling rate $L_{ad}$ estimated by:
\begin{equation}
\label{advective}
L_{ad} = \frac{\gamma}{\gamma-1} \dot{M} c_s^2
\end{equation}
where $\gamma$ is the adiabatic index, $\dot{M}$ is mass loss rate and $c_s$ is sound speed. This cooling effect is evaluated in this work.

For simplicity, the core is assumed to be isothermal with the temperature found at the bottom of the envelope, essentially meaning it is an efficient conductor. As a planet cools off, the core temperature correspondingly drops, which simultaneously releases thermal energy via the core's heat capacity. For the specific heat capacity at constant volume of the core, we use $c_v = 0.75 \ \mathrm{J K^{-1} g^{-1}}$. To stabilize our numerical calculation, we assess the time derivative of core temperature $dT_{c}/dt$ based on the change of core temperature over the last 5 timesteps. For the same reason, the planets are not allowed to increase specific entropy over time, such that $L_{core} \leq L_{env}$. The consequence of this simplification is evaluated in Section \ref{subsec:core} and \ref{disc:core}. In terms of $L_{radio}$ from radioactivity, the dominant decaying isotopes are ${}^{235}\mathrm{U}$, ${}^{238}\mathrm{U}$, ${}^{40}\mathrm{K}$ and ${}^{232}\mathrm{Th}$. The abundances and radioactive powers of these isotopes at early ages are derived based on their half-lives and the meteoritic abundances at the current solar age \citep{Anders89,Nettelmann11}.

Based on the $\frac{ds}{dt}$ calculated from Eq. \ref{thermalcontraction} at each timestep, we evolve the envelope specific entropy using a fifth-order ODE solver. The resulting specific entropy at a new timestep automatically defines the amount of the total energy (gravitational energy and internal energy) that is allowed to be released through interior cooling and therefore the rate of thermal contraction. The envelope mass $M_{env}$ also controls both the cooling process and the hydrostatic interior structure, as seen on the left-hand side of the equation. In the presence of mass loss, given the mass loss rate  $\dot{M}$, we evolve the total planetary mass $M_p$ in a similar manner simultaneously, as discussed below.

\subsection{Radiative atmosphere}
\label{subsec:radatm}
The radiative atmosphere is a static layer whose radial structure passively evolves with time on top of the adiabatic interior as a planet contracts.  We focus on planets located close enough to their host stars that the radiative-convective boundary (RCB) separating the adiabatic interior from the radiative atmosphere is set by the incident bolometric flux from the star.  At the RCB, the planet's intrinsic luminosity is equal to the stellar energy deposited per time (i.e., stellar flux reduced by inward diffusion of photons since the optical depth at the RCB is generally large).  Such radiative atmospheres typically exhibit two roughly isothermal regions---an outer region optically thin to both incident optical radiation and outgoing infrared radiation and a deeper region that is optically thick to outgoing infrared (and, in its deepest parts, to incoming optical radiation as well) \citep{Guillot10}. The radiative atmosphere's temperature-pressure ($T-P$) profile is thus primarily set by stellar heating. Stellar heating, especially that from M dwarfs and in XUV wavelengths, can decrease over time at young ages. However, the variability in bolometric flux from a Sun-like star is generally modest at the ages relevant to the boil-off phase, $\sim$3-10 Myr after the formation \citep{Baraffe15}. Since we only focus on the history of mass loss and thermal evolution resulting from stellar bolometric flux, independent of stellar spectral types and XUV-driven escape, the time variability of stellar heating is ignored. This assumption only quantitatively affects the time-integrated mass loss by a small fraction. 

For gas opacity to thermal radiation $\kappa_{th}$, the transition from optically thin to thick occurs where optical depth $\tau \sim \kappa_{th}\rho H \sim 1$, for atmospheric density $\rho$ and scale height $H$.  In a hydrostatic atmosphere, $\rho H \propto Pr^2$, so that the pressure where optical depth transitions occur depends only on radius $r$ which, after the initial blow-off period, typically changes  modestly over the planet's evolution, leading to a roughly constant $T-P$ profile in the radiative atmosphere. To capture this behavior, we do not perform a full radiative transfer calculation but rather generate a $T-P$ profile using the widely used analytical two-stream radiative transfer model of \citet{Guillot10}. The thermal opacity $\kappa_{th}$ is chosen to be a constant 0.02 $cm^2\ g^{-1}$, representative of cool H/He solar-composition atmospheres at P$\sim 1$ bar \citep{Freedman14}. In terms of the opacity to visible photons $\kappa_\nu$, we compare the $T-P$ profile of the analytical two-stream model to that from full radiative-convective equilibrium transfer models \citep{Fortney08a,Fortney13} to find the best fit opacity values (in the range of $0.0001-0.006\ cm^2\ g^{-1}$) as a function of incident bolometric flux to well-approximate the $T-P$ profile from the full calculation. The values are found to be only weakly dependent on the surface gravity and the intrinsic luminosity from the interior. Beginning from the outer edge of the envelope, we integrate using hydrostatic equilibrium, taking into account the variation of gravity with radius.

The thickness of the radiative atmosphere is not negligible for low-mass planets because of their low gravity, which decreases with altitude and leads to an outwardly increasing scale height. Interestingly, the radiative atmosphere can contribute to a major fraction of the optical radius of a sub-Neptune, making it important for population studies. Our planetary radius is defined at 20 mbar, a typical pressure level for optical transits. 

\begin{figure}
\centering
\includegraphics[width=0.5\textwidth]{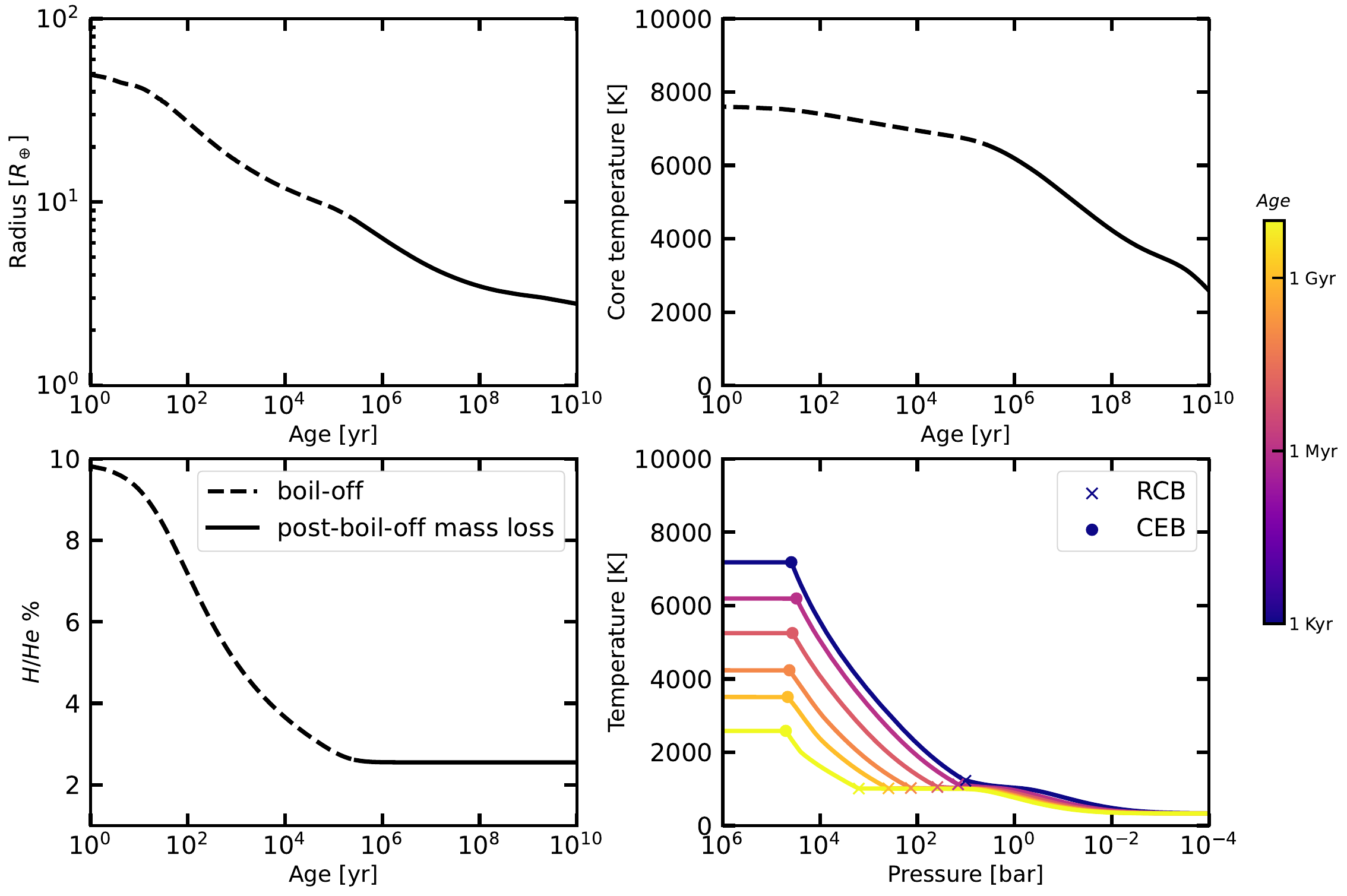} 
\caption{ 
We show an example of the evolution of a $3.6\ M_\oplus$ planet irradiated with $10\ F_\oplus$ flux based on our numerical model. The three black and white panels show the planetary radius, core temperature, and mass fraction of H/He, while the color panel shows the interior and atmosphere $P-T$ profile from 1000 yr to 10 Gyr. The RCB and the core-envelope boundary (CEB) are marked by the cross and circle shapes respectively. Below the RCB is the convective H/He envelope, which is adiabatic, and the rock/iron core, assumed to be isothermal, constitutes the deepest part of the interior below $10^4$ bar. Based on the relative timescales between mass loss and thermal evolution, we define the boil-off phase and long-lived post-boil-off phase plotted in $dashed$ and $solid$, respectively.  The elements of the model are more fully described in Section \ref{sec:results}.
}
\label{evolution}
\end{figure}

The bottom of the radiative atmosphere is separated from what we term the H/He ``envelope'' (Section \ref{subsec:thermal}) at the RCB. Our numerical model estimates the location of the RCB by finding the intersection between the P-T profiles of the adiabatic envelope and the radiative atmosphere. Compared to a RCB pressure estimated from full radiative transfer calculations, our approach only quantitatively shifts the RCB location in radius by a negligible fraction. The RCB pressure for a sub-Neptune varies dramatically with age, from typically at $\sim$ 10 bar (young ages) to 10 kbar (Gyr+ ages). As a sub-Neptune evolves, its interior cools off and the intrinsic luminosity declines. Consequently, the mismatch between the large stellar bolometric flux and the weaker intrinsic flux becomes ever larger, leading to a deeper RCB. Figure \ref{evolution} illustrates this evolution of the RCB location (cross shapes) over time.

Furthermore, we calculate the total atmospheric mass in the static radiative layer at each timestep by integrating the mass shells in the radiative layer. The radiative atmospheric mass does not participate in thermal evolution, and is therefore excluded from the H/He envelope mass to better assess the thermal evolution process. We find that the radiative atmospheric mass fraction is typically at most a few $0.01\%$ of the planetary mass during most of the evolution time contributing a minor difference between the thermal evolution with and without it. However, an exception happens at the early stage of boil-off. The radiative atmospheric mass can constitute up to 20-50\% of the convective envelope mass for a young and inflated planet, thus playing a role in prolonging the thermal contraction timescale. This substantial radiative atmospheric mass results from the slowly-varying density-radius profile ($\rho \sim \rho_{rcb}\exp(-r/H)$ for a roughly isothermal atmosphere), which remains high ($\sim \rho_{rcb}$, the density at the RCB) in the radiative atmosphere due to the large scale height $H$ that is comparable to the planetary radius $r$ above the RCB. Note that the quantitative details remain uncertain here, especially if additional factors such as the opacity contribution from dust in both the atmosphere and the debris disk are considered.

Another exception occurs when the planetary envelope is about to be completely stripped. When a planet loses mass from the deepest part of the radiative atmosphere, right above the RCB, the mass there has to be steadily resupplied from the envelope to sustain an outflow \citep{Ginzburg2016}. By the time that the convective zone is depleted by the mass loss, the radiative atmospheric mass ($\sim 0.01\%$ of the planetary mass) becomes most of the total H/He mass. In this case, thermal evolution of the envelope starts to cease and the planet transitions into an envelope-free super-Earth. We terminate thermal evolution and mass loss once the envelope mass becomes negligible compared to the radiative atmospheric mass, and after that, we treat the planet as a bare core super-Earth assuming its radiative atmosphere is then completely lost in a sufficiently short period.

We evolve the total planetary mass given the mass loss rate by solving the ODE problem $dM_p/dt = \dot{M}$. At each timestep, we estimate the envelope mass $M_{env}$ by subtracting the constant core mass $M_c$ and the time-dependent radiative atmospheric mass $M_{atm}$ from the total mass $M_p$:
\begin{equation}
\label{envmass}
M_{env} = M_p - M_c - M_{atm} \;\;.
\end{equation}
The atmospheric mass, $M_{atm}$, implicitly depends on $M_{env}$ because the envelope mass $M_{env}$ provides the lower boundary condition for the radiative atmospheric mass calculation. Since the atmospheric mass changes slowly compared to the mass loss rate, to improve the computation efficiency we take the value of the atmospheric mass from the last converged solution and treat it as a constant within the next timestep to avoid needing to solve Eq. \ref{envmass} self-consistently using an iterative method at each structure calculation (there are many within one timestep). A full self-consistent calculation of $M_{atm}$ is employed at the first timestep of the evolution.

\subsection{Planet evolution in the absence of XUV-driven escape} 
\subsubsection{Parker wind}
Physically speaking, there is no fundamental difference between the nature of boil-off and core-powered escape. Both processes are directly caused by the large radius H/He envelope and atmosphere in the absence of sufficient ambient confinement pressure, which cannot maintain a hydrostatic equilibrium and therefore leads to a Parker wind outflow. Similar to Bondi accretion in reverse, the Parker wind is a hydrodynamic process that occurs when the protoplanetary disk has dissipated. As long as sufficient energy is available to resupply the H/He material in the atmosphere from below (we evaluate this assumption in Section \ref{subsec:envelope}), the stellar radiation dictates the energy budget for the outflow and therefore the temperature profile in the radiative atmosphere, where the wind advection occurs. For a planet irradiated by a star with an incident flux $F$, its radiative atmosphere is roughly isothermal with the temperature set by the equilibrium temperature $T_{eq}=(\frac{F}{4 \sigma})^{1/4}$ where $\sigma$ is Stefan-Boltzmann constant. For simplicity, we model the hydrodynamic wind using an isothermal Parker wind \citep{Parker1958} beyond the RCB to assess the mass loss rate. This avoids needing an implementation of a numerical hydrodynamic and radiative transfer calculation. The advective-convective boundary is chosen to be the same as the RCB, which is justified later in Section \ref{subsec:envelope}. The wind structure calculation is in parallel with the radiative atmosphere calculation, where the hydrostatic calculation is utilized to assess the optical transit radius (which is not an important variable until boil-off ends), and the steady state Parker wind determines the mass loss rate.

The physical structure of an isothermal Parker wind is given by:
\begin{equation}
\label{parkerwind}
\frac{1}{2} \left( 1- \frac{c_s^2}{v^2}\right)\frac{d}{dr}(v^2) = -\frac{GM_p}{r^2} \left(1 - \frac{2c_s^2r}{GM_p}\right)
\end{equation}
where $c_s=\sqrt{k T_{eq}/\mu}$ is the isothermal sound speed, $k$ is the Boltzmann constant, $\mu = 2.35m_{H}$ is the mean molecular weight of the wind, $m_{H}$ is the mass of atomic hydrogen, $v$ is the wind velocity and $r$ is the radius. At a radius known as the sonic point, a singularity occurs where the wind velocity reaches the sound speed and both sides of the equation vanish:
\begin{equation}
\label{sonicpoint}
R_s = \frac{GM_p}{2c_s^2}
\end{equation}
At the wind base, density and pressure transition continuously between the two layers of the steady-state hydrodynamic wind and the hydrostatic adiabatic envelope. This gives us the mass loss rate:
\begin{equation}
\label{masslossrate}
\dot{M}_{parker} = 4 \pi R_{rcb}^2 \rho_{rcb} v_{rcb} = 4 \pi r^2 \rho v
\end{equation}
from which the density profile $\rho$ for the wind as a function of radius $r$ is obtained, where $R_{rcb}$,  $\rho_{rcb}$ and $v_{rcb}$ are the planet radius, density and wind velocity at the RCB.

\subsubsection{Transition out of boil-off} 
\label{subsec:transitionbf}
Boil-off is a short-lived hydrodynamic process, hence the name, with a mass loss timescale shorter than 1 Myr \citep{Owenwu16} due to the sudden change of the ambient confinement pressure from a protoplanetary disk. The mass loss timescale is given by:
\begin{equation}
\label{tmdot}
t_{\dot{M}} = \frac{M_{env}}{\dot{M}}
\end{equation}
where $\dot{M}$ is the mass loss rate of the isothermal Parker wind, and $M_{env}$ is the envelope mass. If the thermal contraction decreases the planetary size faster than the mass loss can substantially alter the envelope's mass, the wind velocity $v$ at the RCB in Eq. \ref{parkerwind} and \ref{masslossrate} declines quickly over time compared to the rate of mass evolution and consequently, the planet has to stop blowing off. Therefore, we define the end of boil-off at the time when the mass loss timescale becomes comparable to the Kelvin-Helmholtz contraction timescale $t_{cool}$:
\begin{equation}
\label{transitiontime}
t_{\dot{M}} = t_{cool}
\end{equation}
where the Kelvin-Helmholtz contraction timescale is assessed with the total envelope luminosity $L$ in Eq. \ref{thermalcontraction} that incorporates both the envelope cooling $L_{env}$ and the core luminosity $L_{core}$:
\begin{equation}
\label{KHtime_core}
t_{cool} = \frac{GM_c M_{env}}{\alpha R_{rcb} L}
\end{equation}
where $\alpha \leqslant 1$ is a dimensionless factor related to the mass concentration of the envelope. With our initial setup of our model, we find the H/He envelope mass and therefore gravitational binding energy are very slightly inwardly concentrated at the very beginning of boil-off, yielding $\alpha=0.4$. The mass concentration gradually shifts outward to $\alpha=0.8$ for an old planet that has a lower H/He mass and specific entropy. During most of the boil-off phase, the mass and energy exhibit a central to outward distribution with $\alpha \geq 0.5$. In boil-off, the mass loss and thermal contraction are decoupled due to the large timescale difference $t_{\dot{M}} \ll t_{cool}$, leading to negligible thermal evolution and hence a minor decrease in entropy, compared to that in the later evolution. Therefore, the shrinkage in planetary size is due to the mass loss rather than thermal contraction. 

After the transition time defined in Eq. \ref{transitiontime} and \ref{KHtime_core}, the mass loss rate declines and thermal contraction becomes the dominant physical effect in controlling the planetary size. The isothermal Parker wind transitions into the post-boil-off long-lived mass loss phase. The long-lived mass loss history is greatly dependent on the thermal contraction of the planet, in which core luminosity can potentially play an important role. In this case, the mass loss and thermal evolution are coupled. A representative combined evolution of planet radii, temperatures, and H/He mass fractions from our numerical sub-Neptune model is shown in Figure \ref{evolution}, with the boil-off phase in dashed and long-lived mass loss phase in solid.  

\subsubsection{Transition to XUV-driven escape}
\label{subsec:xuv}
As a planet loses mass to the Parker wind mass loss, its physical radius shrinks and the upper atmosphere becomes transparent to XUV photons. Therefore, the high-energy photons are able to penetrate deeply enough into the atmosphere to drive a more efficient wind with a higher temperature. Atmospheric escape starts to transition to XUV-driven when the optical depth to XUV photons, evaluated at the sonic point, is equal to unity \citep{Owen23}:
\begin{equation}
\label{tau}
\tau_{\rm s} = \sigma_{\nu0} H n_0 \sim 1
\end{equation}
where $\sigma_{\nu0}$ is the cross-section for the absorption of XUV photons for hydrogen, $H$ is the scale height of the upper atmosphere and $n_0$ is the neutral hydrogen number density. Based on our numerical model we find such a transition typically happens when boil-off is about to be over, at around 1 Myr. For the purpose of our study, XUV-driven escape is not directly included in the mass loss from our numerical model.

\subsection{Assessing initial conditions} 
\label{subsec:entropy}
The planetary radius dictates the intensity of boil-off. The radiative atmosphere of an inflated planet, with its large scale height, is substantially less bound to the interior making the planet lose mass readily. Since planets with hotter internal thermal states possess larger radii at a given envelope mass fraction, we argue that the strength of boil-off is directly controlled by the envelope entropy. Physically speaking, because the final entropy dictates both the thermal contraction and mass loss timescales in Eq. \ref{transitiontime}, which are equal at the transition time, the final mass fraction is completely determined by the final entropy $s_t$. Given the negligible thermal evolution compared to mass loss during boil-off in Eq. \ref{transitiontime}, the specific entropy remains nearly unchanged during boil-off. Therefore, the initial entropy $s_i \sim s_t$ largely determines the physical states at the end of boil-off and the beginning of the subsequent evolution.
 
However, what entropy a planet is born with is not well determined from previous work. To constrain the initial entropy at the beginning of boil-off, \citet{Owenwu16} suggest that a model planet should not be allowed to thermally contract (due to cooling) faster than the disk dispersal time $\sim 10^5$ yr, the characteristic timescale for the formation of the gas-depleted inner hole, otherwise it is always able to adjust to a new hydrostatic equilibrium during the disk dispersal, leading to no efficient boil-off. 
On the other hand, the planetary initial contraction time should be no faster than the disk lifetime 3-10 Myrs \citep{Mamajek09,gorti2016}, as in the opposite case the planet would shrink rapidly enough while the disk is present to allow for more gas accretion, and consequently, the increased envelope mass would eventually slow down the contraction timescale. 

For these reasons, in this work, the initial entropy and radius are chosen such that the initial Kelvin-Helmholtz thermal contraction timescale in Eq. \ref{KHtime_core} is comparable to the disk lifetime, chosen to be 5 Myr. Note that the core luminosity and advective cooling are initially 0, such that the total envelope luminosity is set by the intrinsic luminosity $L=L_{int}$. As the disk dispersal time of $\sim 10^5$ yr is observed to be substantially shorter than the disk lifetime, our physical choice ensures that the planet's contraction timescale at the onset of boil-off is long compared to the boil-off time and hence the process of boil-off is not sensitive to direct modeling of the disk dispersal. Our initial setup results in an initial RCB radius that is 10-20\% of the sonic radius defined in Eq. \ref{sonicpoint}, and an initial entropy of $9-11 k_b/{\rm atom}$ depending on the bolometric flux and core mass. 



\section{Model results} \label{sec:results}
The outline of this section is as follows. In Section \ref{subsec:boil}, we reassess the assumptions made in the previous core-powered mass loss and boil-off modeling work. In Section \ref{subsec:entropycooling}, we examine the effects of initial entropy and on the role of advective cooling under our self-consistent initial conditions. The results are compared to the behavior reported in \citet{Owenwu16}. The detailed model results for boil-off, including the final mass fractions and transition times, are documented in Section \ref{subsec:finalfraction}. Section \ref{subsec:subsequent} discusses the subsequent planet evolution without the long-lived mass loss and how it is related to the radius cliff. Finally, we evaluate the post-boil-off long-lived mass loss over Gyr timescales in Section \ref{subsec:nobolo}.

\subsection{Boil-off and core-powered escape} 
\label{subsec:boil}
Most of the previous core-powered escape work \citep{Ginzburg2016,Ginzburg18,Gupta19,Misener21} originates from a single analytical model with similar mass loss treatments. According to these authors, although the nature of both boil-off and core-powered mass loss is a Parker wind, core-powered mass loss differs from boil-off in the following three aspects: the energy source for mass re-equilibration, the mass loss timescale (short-lived or long-lived), and whether or not the mass loss rate is enhanced by core luminosity. We focus on assessing each of these three aspects in this section.

\begin{figure}
\centering
\includegraphics[width=0.45\textwidth]{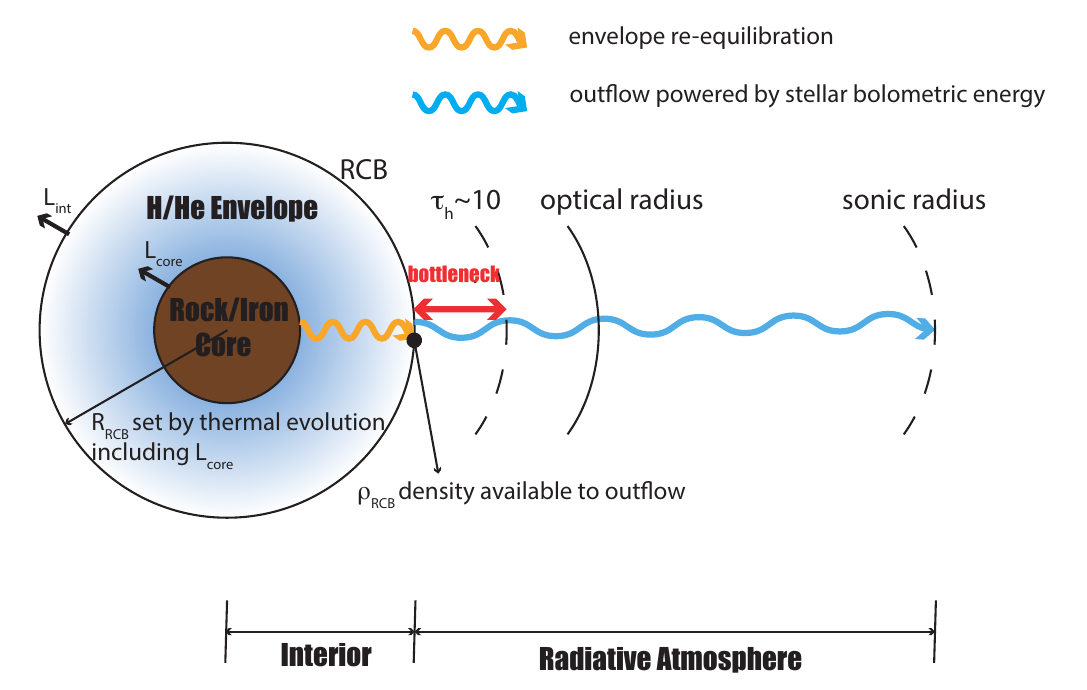} 
\caption{
There are three physical aspects that control the boil-off and core-powered mass loss processes. First, the H/He mass that maintains the RCB density and replenishes the outflow must be lifted from the deep part of the envelope, this is known as the envelope re-equilibration (orange). In GSS16, this was argued to be the bottleneck due to the limited amount of intrinsic cooling energy $L_{int}$ available to overcome the gravitational force, resulting in an energy-limited escape. Second, the outflow in the radiative zone is fueled by the stellar bolometric luminosity and by $L_{int}$(blue). We propose that when $L_{int}$ is insufficient to drive the full outflow, another bottleneck exists in the deep atmosphere below a critical optical depth $\tau_h$, where the absorption of the bolometric energy a planet receives is inefficient, shown in red. Third, an efficient core luminosity keeps the H/He envelope warm and large in radius, which consequently elevates density at the sonic point, encouraging a more substantial mass loss.
}
\label{diagram}
\end{figure}

A primary challenge for modeling these processes regards the energy supply that overcomes the gravitational force, to continuously drive such winds. If the energy input is inadequate, the wind region rapidly cools off when the PdV work drains energy from the internal energy of the outflow leading to a low wind temperature in Eq. \ref{parkerwind}, and therefore the outflow slows down until the energy supply is sufficient to maintain the outflow. This is known as an energy-limited wind. Two layers where energy supply may potentially limit the mass loss rate are invoked: the radiative atmosphere and adiabatic envelope, shown in Figure \ref{diagram}. The energy injection in the radiative zone is primarily from the bolometric luminosity and is assumed in these studies to be sufficient to sustain an outflow of the equilibrium temperature $\sim T_{eq}$. In Section \ref{subsec:bottleneck}, we show that this assumption is reasonable for most though not all planets. However, if mass cannot be replenished fast enough from and within the convective zone, H/He material will eventually be depleted, yielding a low-density layer at the RCB, even if sufficient energy to power the wind is available in the radiative zone. 

Envelope re-equilibration was considered as the bottleneck for boil-off and core-powered mass loss in \citet{Ginzburg2016,Ginzburg18} (GSS16 and GSS18 hereafter). GSS18 assume that most of the H/He mass that replenishes the wind at the RCB starts to expand from as deep as the core-envelope surface, with the energy needed for the re-equilibration of the envelope completely from the interior cooling energy $L_{int}$. This requires a large amount of cooling energy to sustain the outflow, and thus the outflow can be largely energy-limited. The total energy needed per second is estimated by the amount of gravitational energy to overcome:
\begin{equation}
\label{pdv_core}
\dot{E}_{\rm loss,core} = \frac{G M_c \dot{M}}{R_c}
\end{equation}
where $R_c$ is the core radius.  Based on $\dot{E}_{\rm loss,core}$, these authors estimate an energy-limited mass loss rate if the mass re-equilibration is insufficient:
\begin{equation}
\label{mdot_core}
\dot{M}_{e-lim,core} = \frac{L_{int}}{G M_c/R_c}
\end{equation}
Once the intrinsic luminosity becomes more than the energy needed in Eq. \ref{pdv_core}, they treat the wind as non-energy-limited, as in Eq. \ref{masslossrate}, which is the maximum possible mass loss rate given the sufficient energy and mass supply, known as what they call a Bondi-limited regime. They define core-powered mass loss as the later stage of Parker wind mass loss when core cooling $L_{core}$ constitutes a large fraction of the envelope cooling, so $L_{int} \sim L_{core}$. In this case, the energy that liberates H/He mass from the interior is ultimately from the core, rather than the envelope itself. This happens to planets that have a larger core thermal energy reservoir than the gravitational binding energy. On the other hand, they refer boil-off (what they call spontaneous mass loss) to the earlier stage when the envelope energy dominates over the core thermal energy. The above discussion provides a key justification for their mass loss treatment and the necessity of distinguishing between boil-off and core-powered mass loss.

However, an important assumption made in their envelope re-equilibration argument is that the energy released from the envelope internal energy when H/He mass is lifted and cooled from the hot deep interior to the colder RCB, is ignored in Eq. \ref{pdv_core}. We suggest that their argument in Eq. \ref{pdv_core} and \ref{mdot_core} holds true only if the envelope were isothermal (in which case, the contribution from the internal energy can be ignored, corresponding to a steady state isothermal envelope) but not for the adiabatic envelope assumed in their model. We revisit the envelope re-equilibration problem with an adiabatic envelope considered in the energy calculation with particular attention to the internal energy exchange in Section \ref{subsec:envelope}. We identify a different and much physically narrower (thus requiring less energy to overcome) bottleneck region due to the deficiency of $L_{int}$ in Section \ref{subsec:bottleneck} (Figure \ref{diagram}).



\subsubsection{A fast re-equilibration of energy and mass via envelope convection} \label{subsec:re-eq}
\begin{figure}
\centering
\includegraphics[width=0.45\textwidth]{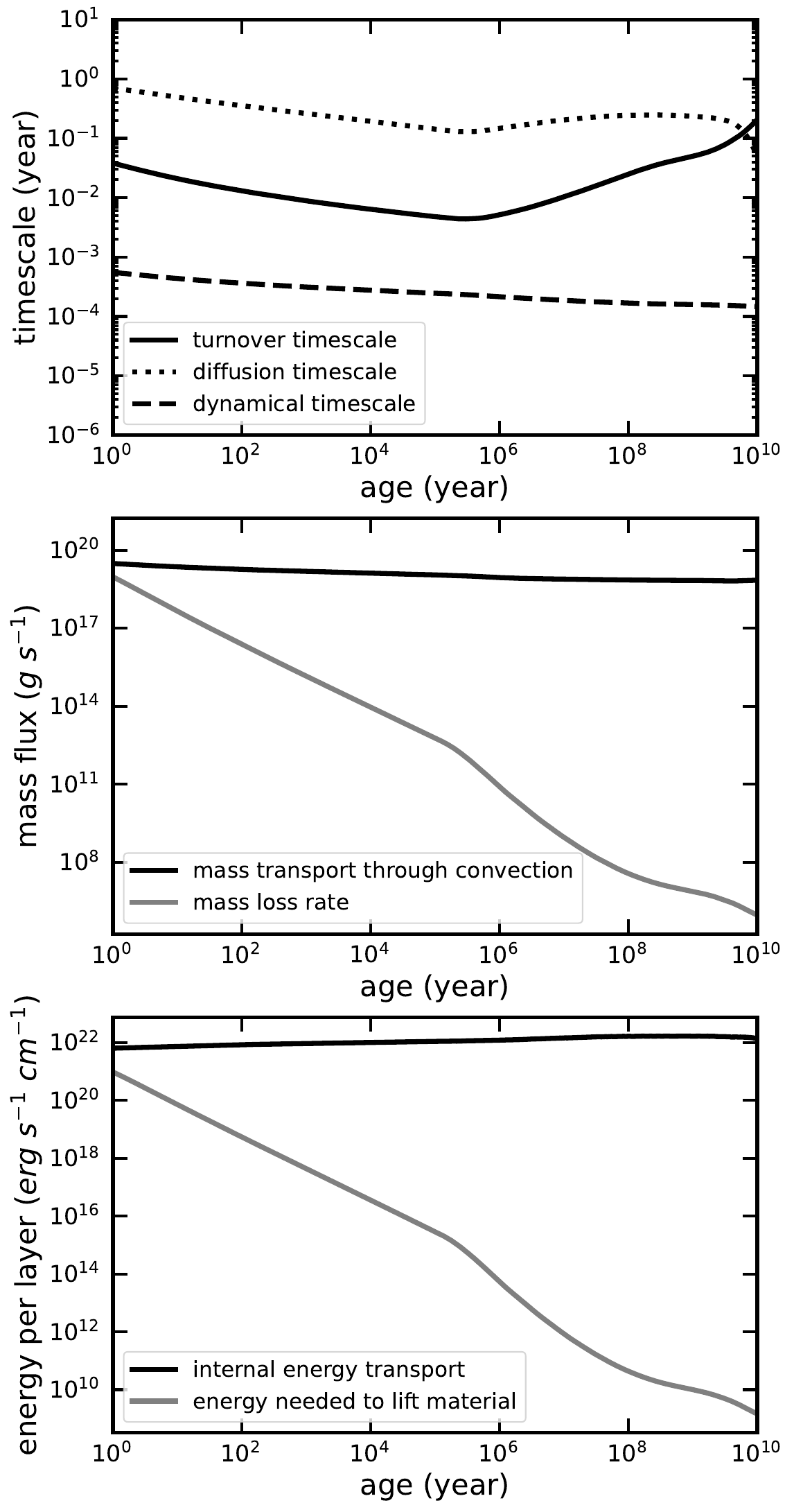} 
\caption{The top panel shows the relevant timescales for energy and radiation re-equilibration in the convective envelope: turnover timescale of the mixing (solid), the eddy diffusion timescale (dotted) and dynamical timescale (dashed). The physical processes that mix and propagate energy and material are faster than the shortest mass loss timescale $\sim$ 1 yr at the beginning of boil-off, validating our isentropic and adiabatic assumption for the envelope (see Section \ref{subsec:re-eq}).  The middle panel shows the convective mass flux versus the mass loss rate of non-energy-limited isothermal Parker wind boil-off, implying that the advective wind cannot penetrate deep into the envelope as the envelope is convection-dominated. In the bottom panel, the internal energy transport through $E_{conv}$ is sufficient to lift H/He material out of the envelope. This is for a $3.6 M_\oplus$ planet irradiated with $100 F_\oplus$. The same model planet is used for Figure \ref{times}-\ref{boiloff_energy}.
}
\label{times}
\end{figure}

We begin by verifying that the envelope remains adiabatic and quasi-hydrostatic during an energetic boil-off mass loss phase.
We calculate the re-equilibration timescales and the mass transport and internal energy transport propagating via convection within the envelope, shown in Figure \ref{times}.  The top panel shows three timescales. These timescales are evaluated at the envelope surface (RCB), as we find that they are either greater than the timescales at the bottom of the envelope or comparable. For any perturbation at the envelope surface, from either removing envelope mass or thermal cooling, the pressure and density of the deep envelope correspondingly reacts and readjusts to a new hydrostatic equilibrium on the dynamical timescale (dashed with $t_{dyn} = R_{rcb}/c_s$), which corresponds to the time that sound waves take to traverse the entire planetary radius (since the atmosphere is close to being in hydrostatic equilibrium). Under our initial physical conditions, the wind crossing timescale $t_{cross}=R_{rcb}/v_{rcb}$, where $v_{rcb}$ is the wind speed at the surface, is always much longer than the dynamical timescale, given the small Mach number $M=v_{rcb}/c_s \ll 1$ at the RCB, so that the density, pressure and therefore radius of the envelope presumably respond instantaneously to the mass loss. During this process the RCB pressure and temperature (and therefore density) physically remain unchanged, as both the outgoing intrinsic luminosity (primarily set by $s$) and the incident bolometric luminosity, which dictates the RCB conditions, are not affected due to their long timescale to change. Consequently, the RCB has to penetrate to a deeper location in radius to react to the pressure decline at the original location. 

However, energy transport is limited by convection. It mixes H/He with different thermal states at different depths, which ultimately determines the temperature (entropy) profile and therefore the mass distribution. In boil-off, as the density and pressure of the envelope decreases, the envelope temperature also has to decrease in order to adjust to the lowered pressure, as shown in the dashed curve from the top right panel of Figure \ref{evolution}. Note that the envelope specific entropy $s$ barely changes due to $t_{\dot{M}} \ll t_{cool}$. This process eventually releases a large amount of the internal energy of the envelope, providing an important amount of energy to the re-equilibration process. This temperature decline is distinct from the interior thermal cooling that drives the thermal contraction and the change of specific entropy $s$, as during this process a negligible amount of net heat transfer happens to the envelope.

Such a temperature change and internal energy transport is via convection on a timescale set by the eddy diffusion timescale (dotted). We estimate this diffusion timescale $t_{conv}$ as the eddy turnover time (solid) $H/v_{conv}$ from mixing length theory times the number of turnovers for the energy and mass to be transported to the envelope surface, which we take to be $\sim {(\Delta R/H)}^2$, where $\Delta R = R_{RCB}-R_{core}$ and $H$ is the pressure scale height of the envelope and $v_{conv}$ is the convective velocity. We calculate $v_{conv}$ using Eq. 3.7 from \citet{InteriorofJupiter}, with the mixing length parameter $\alpha_m=1$ and the specific heat capacity at constant pressure $c_p$ for a diatomic ideal gas. The convective energy flux $F_{conv}$ is chosen to be the intrinsic cooling flux $L_{int}/(4\pi R_{rcb}^2)$ evaluated at the RCB, and the profile parameter $\delta=-(\partial \ln{\rho}/\partial \ln{T})_P$ is estimated using an adiabatic equation of state. The other physical quantities are computed from our numerical model. 

The above discussion therefore gives $t_{dyn} \ll t_{conv} \ll t_{\dot{M}} \ll t_{cool}$ even at a very early age shown in Figure \ref{times} (top). Note that the shortest mass loss timescale is found at the beginning of boil-off $\sim$ 1 yr. Therefore, from above discussion, we see that an adiabatic and hydrostatic envelope that is widely assumed in giant planet evolution models and sub-Neptune models with mild atmospheric escape including photoevaporation and core-powered mass loss still holds true even in the presence of vigorous boil-off, due to the fast mixing effect from envelope convection. This justifies our assumption made in Eq. \ref{thermalcontraction}.

\subsubsection{Lifting mass from convective envelope}
\label{subsec:envelope}
In this section we focus on whether there is enough internal energy transport to lift H/He mass from the deep envelope to supply the outflow at the RCB. Intuitively, since we have shown that the envelope remains adiabatic through the mass loss process, it does not require energy input to sustain gas expansion within the envelope. Instead, the required energy for re-equilibration can be completely taken from the internal energy of the gas. 
The only additional energy required is the energy input to supply the kinetic energy of the wind, which we find is generally negligible in the envelope for an envelope radius $R_{rcb}$ a few times smaller than the sonic radius $R_s$.
Because it has important implications for the rate of core-powered mass loss, we spend the remainder of this section validating this intuition that after mass loss, the convective envelope does not require external energy input to re-equilibrate on a new adiabat, even when that re-equilibration involves expansion of the envelope in the planet's potential well.

Within the envelope, convection is capable of transporting internal energy at a rate even greater than the wind advection is. To demonstrate this, in the bottom panel of Figure \ref{times}, we show that envelope convection advects an amount of internal energy (black, this does not equal the actual net amount of internal energy advected by the wind) much greater than the energy needed to overcome the gravitational force (gray), evaluated per radial distance at the RCB:
\begin{equation}
\dot{M}_{conv} \frac{\partial}{\partial r}\left[\frac{kT}{(\gamma-1)\mu}\right] \gg \frac{GM_p \dot{M}}{r^2} \;\;.
\end{equation}
We find the convective mass flux $\dot{M}_{conv} = 4\pi r^2 \rho v_{conv}$, evaluated at the RCB (black), corresponding to the minimal amount of mass flux that is required to transport the intrinsic radiation out of the envelope, is always a few orders of magnitude greater than the mass loss rate $\dot{M}$ (gray) in the middle panel. It indicates that the redistribution of energy and mass through envelope convection is always sufficient to adjust the envelope adiabatically and hydrostatically against the perturbation generated from the mass loss, without needing an additional energy source, i.e. intrinsic luminosity, as opposed to the suggestion from GSS18. Additionally, this indicates that the wind advection only dominates the radiative zone rather than the envelope, as boil-off requires a mass re-equilibration much less than convection could yield, so we suggest the advective-convective boundary coincides with the RCB.

So far, our re-equilibration arguments are valid for a steady state convective mass/energy flux in the adiabatic envelope, but in reality the H/He mass has to be ultimately taken from each layer of the envelope (i.e. $\partial \rho /\partial t \neq 0$ at any radius in the envelope). In a more general case, we have to assess if enough internal energy can be generated from the re-equilibration compared to the energy needed to lift the H/He material in each layer of the envelope. This involves the Euler energy equation for a fluid:
\begin{equation}
\label{euler_energy}
\frac{\partial E_t}{\partial t} + \rho \boldsymbol{v} \cdot \nabla(h_t+\phi) = h_t\frac{\partial \rho}{\partial t}  \;\;,
\end{equation}
where we have dropped the external heating and cooling terms because they are negligible in the convective envelope.  The combined energy per volume $E_t=\rho e + \rho v^2/2$ is the sum of kinetic energy and internal thermal energy, where $e = c_v T$ is specific internal energy and $c_v$, $\rho$, $\boldsymbol{v}$ and $T$ are the specific heat capacity, density, velocity, and temperature, respectively. We have combined the specific enthalpy of the envelope, $h = e + p/\rho$, with the specific kinetic energy in $h_t \equiv h + v^2/2$,
and $\phi = -GM/r$ is the gravitational potential. Eq. \ref{euler_energy} may be reformulated in terms of the energy per volume $E_{tot} \equiv \rho e + \rho \phi$, so that 
\begin{equation}\label{eqn:bern}
\frac{\partial E_{tot}}{\partial t} = -\rho \boldsymbol{v} \cdot \nabla B + B\frac{\partial \rho}{\partial t} \;\;,
\end{equation}
where we have dropped a term proportional to $\partial \phi/\partial t$ since the gravitational potential is primarily determined by the core mass $M_c$ and does not vary significantly within a mass loss time step.
Eq. \ref{eqn:bern} implies the classic result that the Bernoulli constant $B \equiv h_t + \phi$ is conservative along streamlines (which in our problem are radial) for a steady-state flow ($\partial/\partial t = 0$). While our problem is not in steady-state, the mass-loss timescale is substantially longer than the timescale over which the convective envelope re-equilibrates into a pseudo-steady state (Figure \ref{times}).  We demonstrate in Appendix \ref{app:envelope} that as  a result, the first term on the right-hand side of Eq. \ref{eqn:bern} is small compared to the other two terms in the equation.

Eq. \ref{eqn:bern} may be further simplified by noting that the Mach number of the fluid motion in the convective envelope $\mathcal{M} \ll 1$, so that in the envelope, the kinetic energy $\rho v^2/2 \sim \mathcal{M}^2 \rho kT_{eq}/\mu$ is negligible compared to the fluid's internal thermal energy. 
Consequently, Eq. \ref{eqn:bern} reduces to
\begin{equation}
\label{euler_time}
\frac{\partial E_{tot}}{\partial t} = (h+\phi)\frac{\partial \rho}{\partial t} 
\end{equation}
where $E_{tot} = \rho e + \rho \phi$ 
incorporates the internal thermal energy density and gravitational potential energy density. 
The first and second terms in parentheses on the right-hand side of Eq. \ref{euler_time} correspond to the internal energy advection and change of gravitational energy. We integrate Eq. \ref{euler_time} over the volume of the envelope and obtain the change of the total energy $\Delta E$ over a time interval $\Delta t$ longer than the convection timescale:
\begin{equation}
\label{denergy}
\Delta E = (h+\phi)\Delta M
\end{equation}
where $\Delta E>0$ is the total energy gained by the system in order to expand, $\Delta M < 0$ is the mass loss, and $-(h + \phi)$ is the minimal amount of energy needed per mass to lift material to infinity with temperature at infinity $T_{inf}=0$. Recall that $h+\phi$ is approximately constant with radius so can be evaluated at any point in the convective envelope.

To account for the outflow (Eq. \ref{parkerwind}) in the radiative atmosphere atop the convective zone, we must add several terms to Eq. \ref{denergy}.  The H/He mass taken from the envelope should be at least heated and inflated to $T_{eq}$ and only needs to be lifted to the sonic radius, beyond which it is considered to be completely lost from the planet. The kinetic energy at the sonic radius is also considered here for a more realistic energy budget. This gives the total energy needed per second for the entire system to power the hydrodynamic outflow:
\begin{equation}
\label{pdv}
\begin{split}
\dot{E}_{\rm loss,tot} = (c_p T_{eq} +\frac{1}{2}c_s^2 - \frac{GM_p}{R_s} - h - \phi) \dot{M} \\ =  \left(\frac{G M_p}{R_{\rm rcb}}-\frac{3kT_{eq}}{2\mu}\right)\dot{M}
\end{split}
\end{equation}
where we have evaluated $h=c_p T_{eq} = (7/2) c_s^2$ (with $\gamma = 7/5$ for molecular gas) and $\phi = -GM_p/R_{\rm rcb}$ at the RCB, and plugged in Eq. \ref{sonicpoint}.
This amount of energy equals the energy needed by the non-energy-limited isothermal Parker wind in the radiative atmosphere:
\begin{equation}
\label{pdv_rad}
\begin{split}
\dot{E}_{\rm loss,rad} = \left(\frac{1}{2}c_s^2 + \frac{G M_p}{R_{\rm rcb}} - \frac{GM_p}{R_s}\right) \dot{M} \\ = \dot{E}_{\rm loss,tot} \approx \frac{G M_p \dot{M}}{R_{\rm rcb}}
\end{split}
\end{equation}
For the last equality, as an approximation, we have ignored the kinetic energy $\frac{1}{2}c_s^2$ and the gravitational potential energy at the sonic point $GM_p/R_s$, as they are typically small ($\sim kT_{eq}/\mu$) compared to the gravitational potential energy at the RCB ($G M_p/R_{\rm rcb}$) for planets with our initial RCB radius that is 10-20\% of the sonic radius $R_s$ (Eq. \ref{sonicpoint}). 

The equality of Eqs. \ref{pdv} and \ref{pdv_rad} indicates that the energy for the whole system is due to the outflow above the RCB instead of the envelope, with the change of gravitational energy when mass is transported in the envelope completely from the advection of internal energy. Since the total energy change per mass lost $h + \phi$ during the re-equilibration can be approximated by the specific total energy $e + \phi$ at the RCB, the total energy per envelope mass is nearly conservative during boil-off. Therefore, we argue that the H/He mass can be treated as being taken from the envelope surface between each mass loss timestep. The outflow is not energy-limited by the envelope as opposed to the GSS18 assumption in Eq. \ref{mdot_core}, because the mass redistribution process in the envelope happens on a short convection timescale and releases the exact amount of internal energy required for envelope expansion as the required energy source, leading to a negligible change in the envelope energy budget (specific entropy). Note that the discussion above does not depend on the adiabatic index $\gamma$ and where the envelope mass is concentrated.

To verify this behavior numerically, we set up hydrostatic and adiabatic (with constant adiabatic index) envelopes with different envelope masses but the same outer boundary conditions (as we find the RCB pressure and temperature barely change during the boil-off) assessed without the self-gravity. The exact amounts from the left hand side (the total energy difference) and the right hand side (the analytical estimate we argued) of Eq. \ref{denergy} are found to be equal, as shown in Appendix \ref{app:envelope}. 

\begin{figure}
\centering
\includegraphics[width=0.45\textwidth]{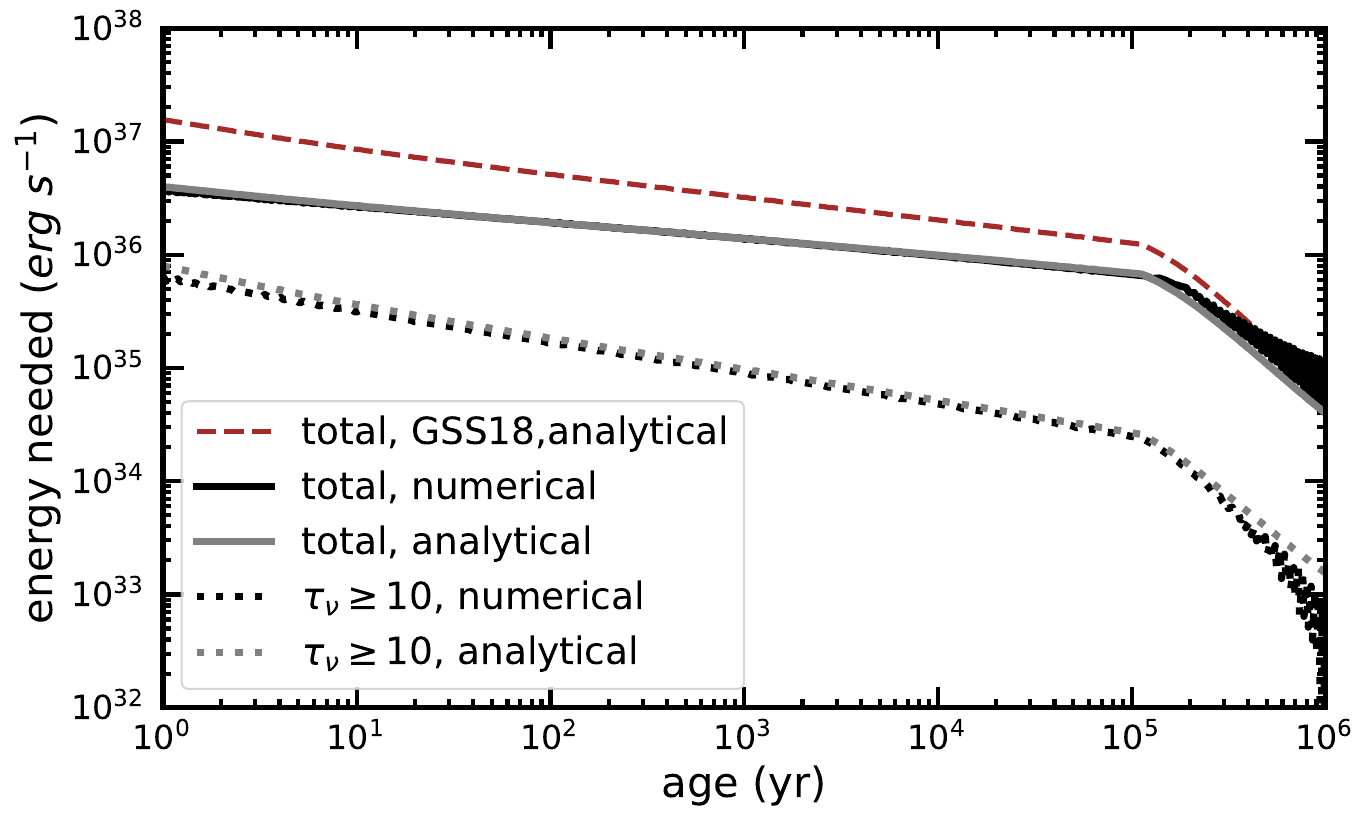} 
\caption{
We show the change of total energy (gravitational binding energy and thermal energy) as a result of mass loss alone (excluding thermal contraction and mass transfer to the radiative atmosphere, see Eq. \ref{dE},\ref{dE_tau}) in between two consecutive static structure models, in solid black.  This is compared (in solid grey) to the analytically derived energy that is needed to power the wind for being lifted from the surface of the convective envelope. Our results demonstrate that the escaping material can be treated as being taken from the surface of the convective envelope (the RCB) rather than the deep interior, as the mass transport within the adiabatic envelope is powered by the internal energy of the envelope (see Section \ref{subsec:envelope}).  As for the deep radiative layer (above $\tau_h$), the energy needed (dashed black) can be approximated by a steady-state isothermal wind (dashed gray) given enough energy supply to power the wind (see Eq. \ref{pdv_rad},\ref{Esteady}). As a comparison, the energy needed from the original GSS16 work is shown in dashed red. Notably, the bottleneck (dashed) suggested in our work is less severe.
}
\label{energy-mdot}
\end{figure}

We display the results of a similar calculation based on our time-evolving numerical model in Figure \ref{energy-mdot} (solid gray and black lines). The analytical expression that estimates the energy needed per unit time $\dot{E}_{loss,rad}$ (as an approximation we used the second equality in Eq. \ref{pdv_rad}, gray solid) for the wind to transport mass in the radiative atmosphere matches the total energy gained in order to lose mass (black solid).

We compute the black curve in Figure \ref{energy-mdot} as follows. Between each numerical time step, we calculate the difference of the total energy $\Delta E$ of the envelope induced by the mass loss alone excluding any consequence from the thermal contraction and the mass transfer into the radiative region:
\begin{equation}
\label{dE}
\Delta E = \Delta U + \Delta E_{th} + L\Delta t + \Delta E_{rcb}
\end{equation}
where $\Delta U$ is the change of the gravitational binding energy, $\Delta E_{th}$ is the change of thermal energy (internal energy), $L\Delta t$ is the energy lost due to thermal cooling represented by the total luminosity $L$ out of the envelope (see Eq. \ref{thermalcontraction}) over a time interval $\Delta t$, and $ \Delta E_{rcb}$ accounts for the total energy lost by the mass transfer into the radiative zone at the RCB. $\Delta E_{rcb}$ is given by:
\begin{equation}
\Delta E_{rcb} = \left(-\frac{GM_p }{R_{rcb}} + \frac{1}{\gamma-1}\frac{kT_{eq}}{\mu}\right) \Delta M_{atm}
\label{eqn:energy_carried}
\end{equation}
where $\Delta M_{atm}$ is the mass transfer into the radiative atmosphere (note that Eq. \ref{eqn:energy_carried} is not the same as Eq. \ref{denergy} but rather is simply the gravitational and internal energy of the material removed from the convective zone and added to the radiative zone). The results are shown in Figure \ref{energy-mdot}. The close overlap between the grey (analytical) and black (numerical) lines shows that the escaping mass can be treated as entirely lost from the RCB without any other energetic consequence for the adiabatic envelope. 

From Eq. \ref{denergy}, we derive the rate of change in total energy of the envelope as $\dot{E} =  \phi\dot{M} + h\dot{M}$. The first term represents the change in gravitational binding energy and the second term corresponds to the internal energy that is advected out of the envelope per unit time through the bulk motion, same as in Eq. \ref{advective}. However, we argue that it does not decrease the specific entropy of the envelope as an envelope cooling, in opposition to the argument in \citet{Owenwu16}. Instead, it is a consequence of mass loss that removes the energy from the internal thermal energy reservoir by decreasing the system mass. The advective energy flux $L_{ad}=h\dot{M}$ then becomes a constant throughout the assumed isothermal atmosphere, leading to no internal energy exchange in the radiative atmosphere, so it is inefficient in powering the outflow under our assumption (see discussion in \ref{disc:adv}).

\subsubsection{Expansion of radiative atmosphere: the bottleneck in the deep atmosphere}
\label{subsec:bottleneck}
\begin{figure}
\centering
\includegraphics[width=0.45\textwidth]{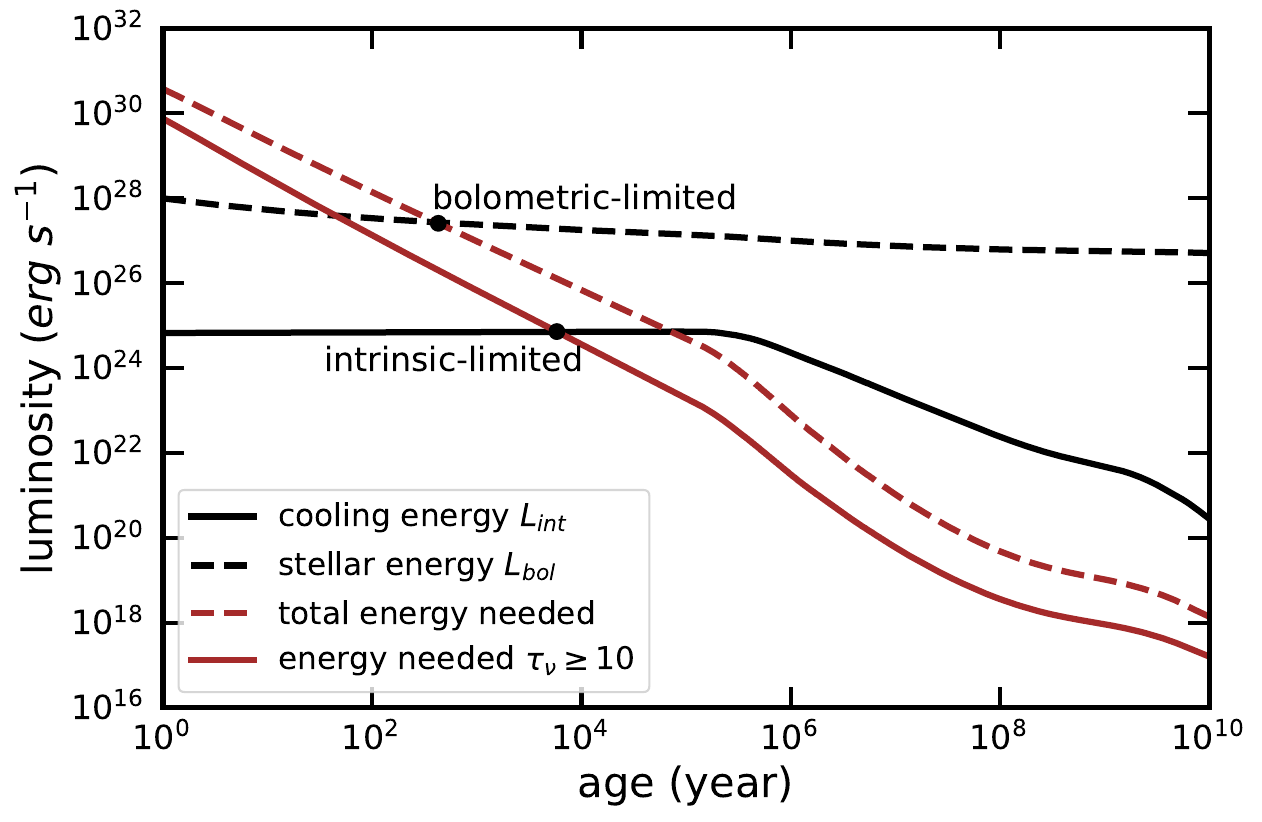} 
\caption{ 
 We show the energy sources (black lines) available to power the hydrodynamic wind: intrinsic cooling energy from the interior radiation (black solid)  and stellar bolometric energy (black dashed). In the red lines, we show the energy required to overcome $PdV$ work in order to lose mass, for both the entire radiative region (red dashed) and the bottleneck layer below the $\tau_h$ surface (red solid). The stellar bolometric radiation limits the efficiency of mass loss before 1 kyr, denoted as bolometric-limited escape. However, the energy barrier that restricts the wind outflow continues until 10 kyr due to the inadequacy of intrinsic luminosity supply in the bottleneck region, which we call intrinsic-limited escape.
}
\label{mdot-luminosity}
\end{figure}

We next focus on the energy supply that sets the wind temperature and powers the outflow in the radiative atmosphere. In Figure \ref{mdot-luminosity}, there are two possible energy sources: the stellar heating $L_{bol}$ from above and the planet's own cooling energy $L_{int}$ from below. The stellar heating is assessed at radius $R_{h}$, where stellar photons entering the atmosphere reach a critical optical depth $\tau_h$, defined to be the depth where external bolometric heating from the star is locally comparable to the energy needed to overcome PdV work. Using this evaluation radius, $L_{bol} = \pi R_{h}^2 F_{bol}$, where $F_{bol}$ is the stellar bolometric flux incident on the planet.  We note that the local flux at this radius is $F_{bol} e^{-\tau_\nu}$, where $\tau_\nu$ is the optical depth to visible photons, but we define $L_{bol}$ at $R_h$ to avoid overestimating the energy available to the wind from stellar photons.

As a comparison, the total energy needed to overcome $PdV$ work (Eq. \ref{pdv_rad}) is shown in the dashed red line. The figure is based on the evolution of a $3.6 M_\oplus$ planet with the non-energy-limited isothermal Parker wind mass loss in Eq. \ref{masslossrate}. The intrinsic luminosity is dwarfed by the stellar bolometric luminosity over the entire evolution, such that the intrinsic luminosity is a secondary energy source in the radiative atmosphere. This finding is independent of the stellar flux we chose. We find that boil-off can be energy-limited due to the deficiency of the stellar bolometric energy deposit in the early stage (the transition is marked by circle shapes). Based on Eq. \ref{pdv_rad}, the mass loss rate is given by:
\begin{equation}
\label{mdot_bol}
\dot{M}_{e-lim,bol} = \frac{L_{bol}}{G M_p/R_{rcb}}
\end{equation}
which we term ``bolometric-limited''. As a planet shrinks as a result of the mass loss, the energy demand declines exponentially and the stellar bolometric energy that is directly deposited in the radiative atmosphere becomes sufficient to power escape of the atmosphere after $\sim$ 1 kyr.

\begin{figure}
\centering
\includegraphics[width=0.45\textwidth]{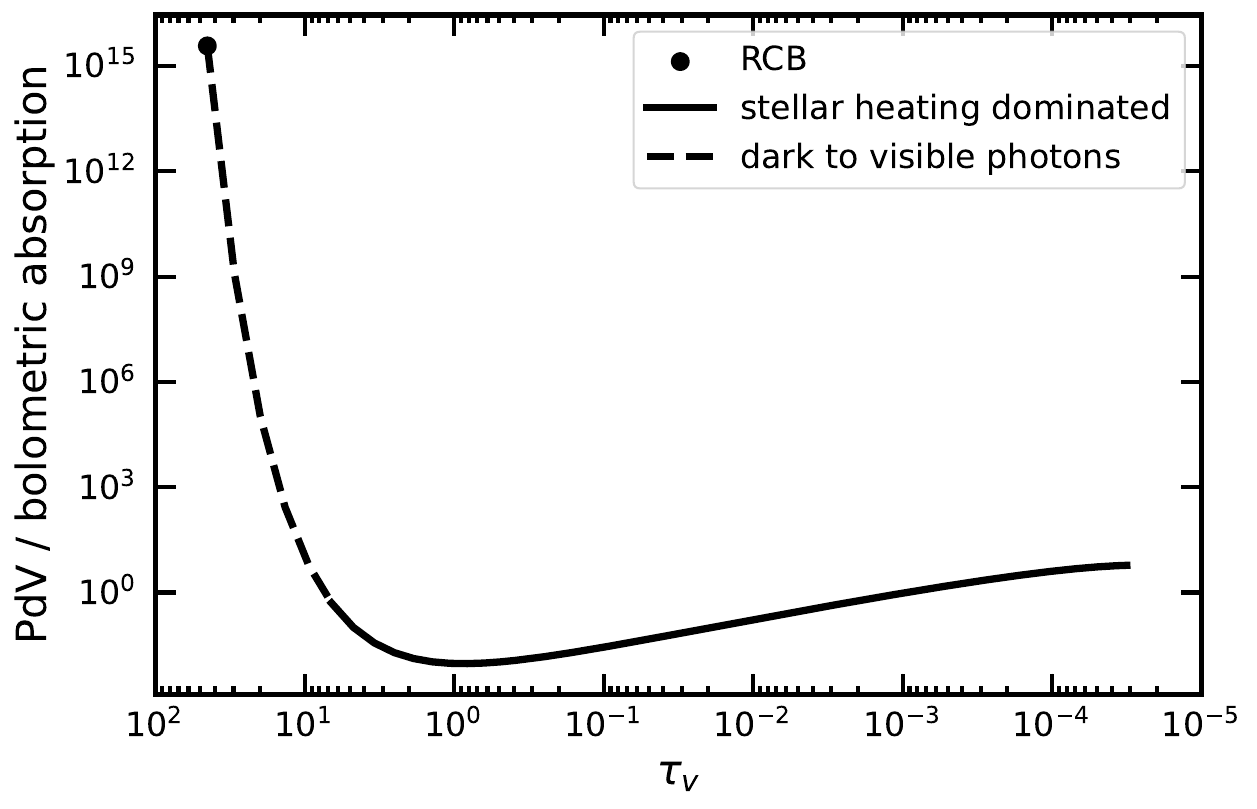} 
\caption{ We show the ratio between the energy needed to overcome the $PdV$ work and the direct stellar bolometric absorption as a function of the optical depth to visible photons $\tau_\nu$, calculated based on Eq. \ref{pdv_layer}. Given the whole radiative atmosphere can get a sufficient amount of the bolometric energy, the bottleneck happens below $\tau_h \approx 10$ (dashed), as the bolometric flux becomes too tenuous to efficiently sustain the energy need to lift the H/He material. In this case, another energy source,  e.g. intrinsic luminosity, is needed in order to further transport material from the RCB (circle) to the radius where $\tau_\nu = \tau_h$.
}
\label{absorption}
\end{figure}

However, a bottleneck region exists in the deeper part of the radiative atmosphere (Figure \ref{diagram}), where the stellar bolometric flux becomes exponentially more diffuse, turning into less useful $PdV$ work. We find that even with sufficient total stellar energy deposited in the atmosphere, the inadequacy of energy injection for the outflow can still happen in a layer that is dark to visible radiation, above the RCB and below a radius that is optically thick to visible photons. In Figure \ref{absorption} we calculate the ratio between the bolometric absorption and energy needed for each layer of the radiative atmosphere at an age after the bolometric-limited phase for the same model planet:
\begin{equation}
\label{pdv_layer}
\frac{PdV}{\rm absorption} = \frac{GM_p\dot{M}dr/r^2}{\pi r^2 \sigma T_{eq}^4 e^{-\tau_\nu}d\tau_\nu} = \frac{4vg e^{\tau_\nu}}{\sigma T_{eq}^4 \kappa_\nu}
\end{equation}
where $g=GM_p/r^2$ is the local gravity, $v$ is the wind velocity and $\kappa_\nu$ is the opacity to visible photons. To compute the second equality in Eq.~\ref{pdv_layer}, we plugged in Eq. \ref{masslossrate}, and $\tau_\nu$ is defined as:
\begin{equation}
\tau_{\nu} = \int \rho \kappa_\nu dr
\end{equation}
Above the critical optical depth to visible photons $\tau_h$, each layer of the radiative atmosphere gets more energy from the direct absorption of the bolometric radiation than the amount needed to expand, whereas the deep atmosphere is inefficient in absorbing stellar energy. We find that the critical optical depth $\tau_h$ is weakly dependent on planetary radius (and therefore age), core mass and stellar bolometric flux. $\tau_h = 10$ typically gives a good estimate (see Figure \ref{absorption} for an example). Note that the re-emission process of the radiative atmosphere at infrared wavelengths is ignored in this calculation, which might assist in the energy absorption in the deep atmosphere and push the bottleneck to a deeper location.

To break through the bottleneck, the only possible energy source is from the radiative interior cooling. In Figure \ref{mdot-luminosity}, we show that the radiative intrinsic luminosity $L_{int}$ (solid black) starts to be ample enough to expand the bottleneck region after $10^4$ yr, before which the mass loss should be energy-limited owing to the bottleneck effect. For ease of comparison with other regimes, we name this process ``intrinsic-limited'' escape. As a comparison, the energy needed for the bottleneck region $\dot{E}_{loss,\tau_h}$ below $\tau_h$ is shown in solid red. The intrinsic-limited regime typically lasts longer and is more energy-limited (with a lower mass loss rate in Figure \ref{boiloff_energy}) than the bolometric-limited regime as long as the $\tau_h$ radius occurs above the RCB (circle). This makes it the common limiting effect for boil-off (in the example in Figure \ref{boiloff_energy}, the bolometric-limited regime is not relevant). We only find the bolometric-limited regime to be relevant for very low-mass $\leq 1 M_\oplus$ planets at young ages, whose $\tau_h$ occurs physically below the RCB. 

Analytically, $\dot{E}_{loss,\tau_h}$ can be estimated by assuming a steady state outflow:
\begin{equation}
\label{Esteady}
\dot{E}_{loss,\tau_h} = \frac{GM_p \dot{M}}{R_{rcb}} - \frac{GM_p \dot{M}}{R_{\tau_h}}
\end{equation}
where $R_{\tau_h}$ is the radius of the $\tau_h$ surface. The validity of this assumption can be verified by our numerical model in Figure \ref{energy-mdot} (dotted lines). Similar to the procedure for the convective envelope, the change of total energy between timesteps is:
\begin{equation}
\label{dE_tau}
\Delta E_{\tau_h} = \Delta U^\prime - \Delta E_{rcb}^\prime + \Delta E_{\tau_h}^\prime
\end{equation}
where $\Delta E_{rcb}^\prime$ and $\Delta E_{\tau_h}^\prime$ are the gravitational binding energy fluxes carried by the bulk flux through the RCB and $\tau_h$ surfaces, respectively. (The primes indicate these quantities are for the deep radiative region, while in Eq. \ref{dE} they were for the convective envelope.)  They consist of the mass transfer through the corresponding surfaces due to both the mass redistribution and the mass flux driven by the hydrodynamic wind. The actual energy need (black dotted in Figure \ref{energy-mdot}) is very slightly higher compared to the steady state approximation (Eq. \ref{Esteady} and dotted grey) since the gravitational contraction of the bottleneck layer produces extra thermal energy to fuel the wind. 

Therefore, in the presence of inadequate intrinsic luminosity deposited in the deep radiative atmosphere, the mass loss rate of such an intrinsic-limited boil-off is estimated by:
\begin{equation}
\label{mdot_int}
\dot{M}_{e-lim,\tau_h} = \frac{L_{int}}{GM_p/R_{rcb} - GM_p/R_{\tau_h}}
\end{equation}
according to the energy needed by the bottleneck region in Eq. \ref{Esteady}. We suggest that a more comprehensive model could employ a mass loss rate:
\begin{equation}
\label{mdot_elim}
\dot{M} = \min(\dot{M}_{parker},\dot{M}_{e-lim,bol},\dot{M}_{e-lim,\tau_h})
\end{equation}
that incorporates both mass loss regimes we discussed.

We have demonstrated the energetics of the radiative atmosphere and H/He envelope. We think bolometric-driven escape might be a better name for the long-lived mass loss phase, rather than core-powered mass loss, as the outflow is not powered by the core energy nor is the envelope re-equilibration limited by the interior cooling energy. The thermal energy from the interior is still an important energy source due to the bottleneck effect, but only in the deep part of the radiative atmosphere and at a young age. 
We emphasize that this conclusion does not imply that core luminosity is irrelevant---when $L_{core}$ efficiently heats up the envelope, the planet remains inflated longer, which can contribute to increased mass loss.  We merely conclude that mass loss is not \textit{limited} by core luminosity.

\subsubsection{Consequence of different wind assumption}
\label{subsec:parker}
\begin{figure}
\centering
\includegraphics[width=0.45\textwidth]{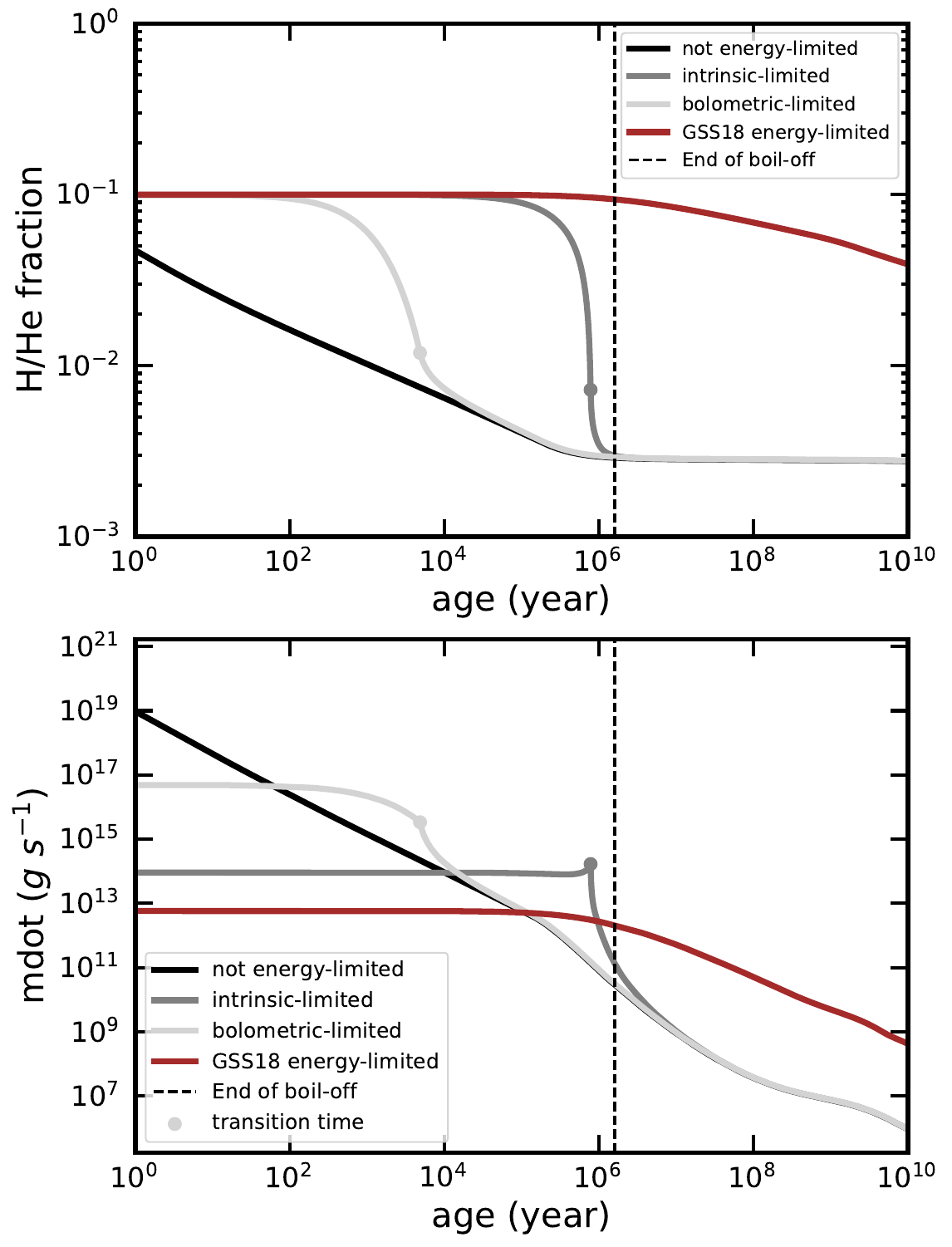} 
\caption{With our numerical model, we show the evolution of the H/He mass fraction (top) and mass loss rate (bottom) under different types of mass loss: non-energy-limited isothermal Parker wind (black), mass loss limited by the total bolometric luminosity available (light gray), and mass loss limited by the ``bottleneck" region below $\tau_h$ (dark gray) where the energy is assumed to be taken from the intrinsic luminosity.  These decoupled Parker winds are compared to the GSS18 energy-limited prescription (red). The first and second types of the decoupled Parker winds, namely ``bolometric-limited'' and ``intrinsic-limited mass'' loss, start to be not energy-limited before the boil-off ends. The transition time is indicated with dots. Consequently, the final mass fractions from all of these decoupled Parker winds converge to a common evolution track by the end of the boil-off at $\sim$ 1 Myr, proving that our default non-energy-limited isothermal Parker wind assumption does not affect the long-term evolution. The vertical dashed line shows the transition time when boil-off ends. The original GSS18 model shows a continued mass loss at old ages.
}
\label{boiloff_energy}
\end{figure}

We next estimate the consequence of using different wind assumptions for the boil-off mass loss rate. Physically speaking, as long as the thermal evolution timescale is much longer than the mass loss timescale $t_{\dot{M}} \ll t_{cool}$, the consequence from different wind assumptions (and initial H/He mass fractions) is eliminated by the end of boil-off, and subsequently the final mass fractions converge to a value dictated by the entropy at the end of boil-off $s_t$ ($\sim$ initial entropy). Following our discussion, the wind can either be limited by the total bolometric luminosity available (light gray, with Eq. \ref{mdot_bol}) or by the total intrinsic luminosity for the bottleneck deep atmosphere (dark gray, with Eq. \ref{mdot_int}), shown in Figure \ref{boiloff_energy}. We demonstrate that the two energy-limited approaches, if used alone instead of Eq. \ref{mdot_elim}, always yield the same final mass fraction by the end of boil-off, converging to the non-energy-limited isothermal Parker wind solution (black), despite the fact that the mass loss is delayed to later times. In this case, we call these three types of wind assumptions in Eq. \ref{parkerwind}, \ref{mdot_bol} and \ref{mdot_int} decoupled Parker winds, as they are insensitive to the thermal evolution and become mutually indistinguishable at a later evolution. Therefore, without specification, we always use a default non-energy-limited Parker wind in the rest of the work.

In Figure \ref{boiloff_energy}, the original energy-limited wind assumption in GSS18 in Eq. \ref{mdot_core} (red), is included as a comparison. Due to the underestimated amount of energy available and the overestimated energy needed, it gives rise to a much lower mass loss rate and therefore prolonged mass loss duration to Gyrs. By this approach, as the thermal evolution is strongly coupled with the mass loss, the evolutionary trajectory never converges to the other solutions and is susceptible to the thermal evolution and therefore available energy supply, i.e. interior cooling. In Figure 8 of \citet{Misener21} a similar behavior of the prolonged boil-off is seen, with the mass loss timescale comparable to the thermal evolution timescale, indicating the strong coupling between their energy-limited mass loss and thermal evolution. We term this type of wind a coupled isothermal Parker wind, hereafter.

\subsubsection{Role of core luminosity}
\label{subsec:core}
Though we have shown that the core luminosity does not limit the mass loss rate by restricting the amount of energy available to supply the H/He mass to the outflow at the RCB, we agree with previous work \citep{Ginzburg2016,Misener21,Rogers24} demonstrating that 
core luminosity plays a role as an energy reservoir to delay the thermal evolution and enhance the mass loss rate by slowing the thermal contraction of the planet (Figure \ref{diagram}). 
This effect especially becomes prominent in the later long-lived mass loss stage when thermal evolution is coupled with the mass loss process. 
We note that this effect is not unique to this particular mass loss mechanism. Any other hydrodynamic wind, like XUV-driven escape, can also be enhanced by the large radius due to the existence of the core \citep{Lopez12}, given the core makes up a large fraction of a planet's energy reservoir. As such, together with our previous discussion on the envelope re-equilibration in Section \ref{subsec:re-eq} and \ref{subsec:envelope}, we prefer to not call the long-lived mass loss core-powered mass loss, but instead we use the term ``core-enhanced effect'' to evaluate the role of core luminosity in amplifying mass loss. 

\begin{figure}
\centering
\includegraphics[width=0.45\textwidth]{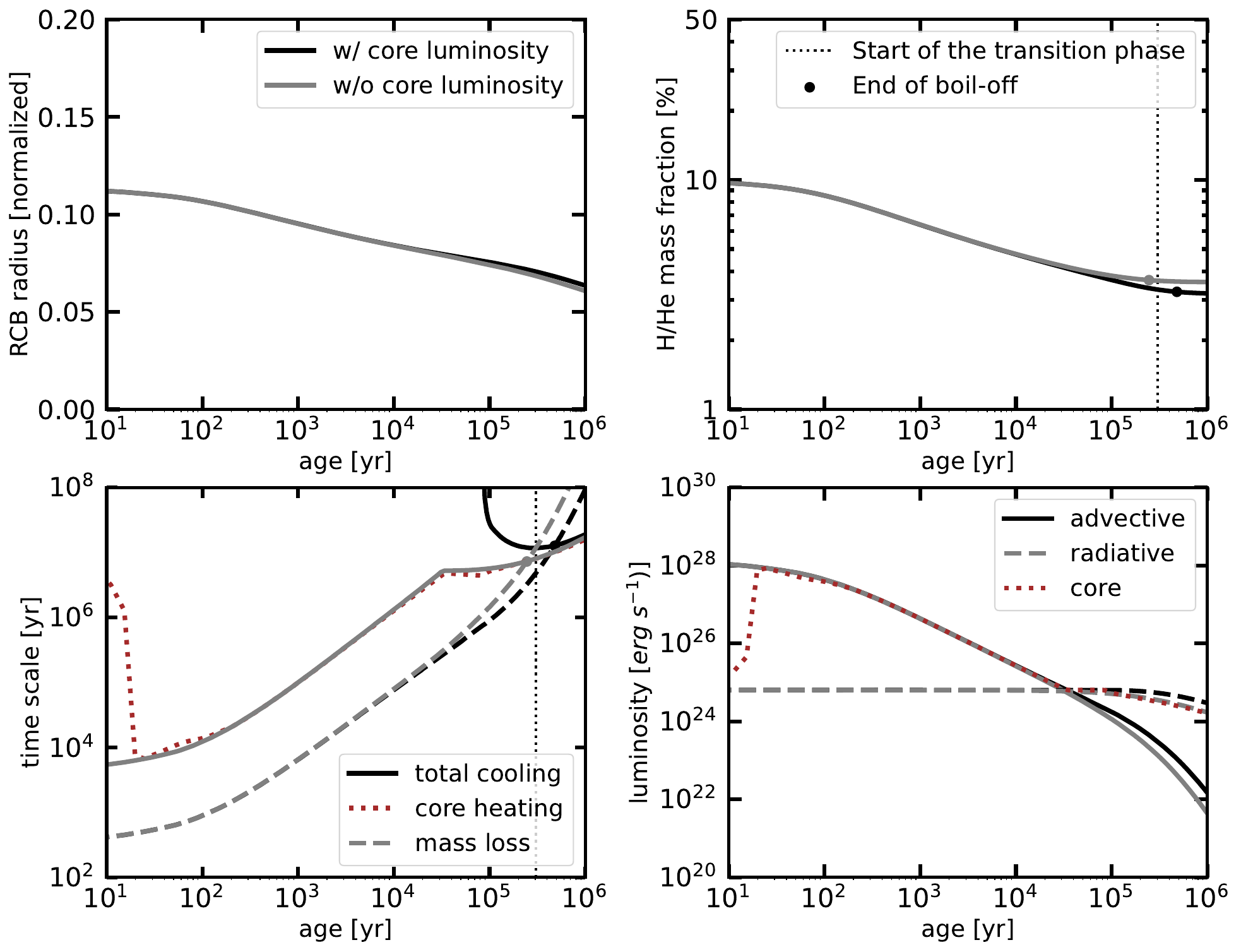} 
\caption{
We compare two models (both with 3.6 $M_\oplus$ and $10 F_\oplus$) with (black) and without core luminosity (gray), both incorporating advective cooling and starting at the same initial entropy and initial mass fractions (top right). The core heat capacity keeps the planet warm and slightly more inflated (top left) compared to the model without core luminosity, leading to a slightly enhanced mass loss. The core-enhanced mass loss only happens after the transition phase (black dotted) starts (Eq. \ref{transitionphase}), when the mass loss (black dashed in bottom left) becomes coupled with the core-luminosity-dominated thermal evolution (dotted red). Note that, before the transition phase, the core luminosity (dotted red) also constitutes the majority of the interior cooling (dashed and solid black in bottom right)  but the mass loss is not core-enhanced due to the decoupling effect.
}
\label{1v3core}
\end{figure}

We now reevaluate the role of core luminosity for our boil-off, which we argue is the mass loss phase that is decoupled from thermal evolution, as discussed in Section \ref{subsec:transitionbf}. \citet{Misener21} argue that the gravitational binding energy of the envelope accounts for the majority of the envelope cooling of boil-off with negligible contribution from the core. As a comparison, in the bottom right panel of Figure \ref{1v3core} we find it is the core luminosity (dotted red) that constitutes the main component of the envelope cooling (dashed and solid) during a vigorous boil-off phase. This is because the increased temperature contrast between the core and the base of the envelope enhances the flux of energy out of the core, when the temperature at the base of the envelope declines (note that this is not a result of thermal cooling but rather due to mass loss, which results in re-equilibration to an adiabat with a lower base temperature as described in Section \ref{subsec:envelope}). Indeed, the core luminosity can be comparable to the advective envelope cooling (solid) before $10^5$ yr, inhibiting any envelope thermal contraction (solid black in the bottom left panel) until the radiative cooling (dashed) starts to dominate the envelope energy budget close to 1 Myr, after which boil-off ceases. This makes the core luminosity a major interior energy source in boil-off (bottom right).

This core luminosity, however, only has a limited role of enhancing the boil-off mass loss due to the decoupling in timescale (bottom left) between the mass loss (dashed) and thermal evolution (solid). Boil-off is essentially not core-enhanced before a transition phase (dotted black), which is defined as when the core luminosity potentially warms up the envelope faster than the mass loss timescale $t_{\dot{M}}$ (Eq. \ref{tmdot}):
\begin{equation}
\label{transitionphase}
t_{core} = \frac{GM_c M_{env}}{\alpha R_{rcb} L_{core}} \leq t_{\dot{M}}
\end{equation}
Once a planet evolves into this transition phase that happens close to the end of boil-off at $\sim$ 1 Myr, the core luminosity becomes effective in enhancing the mass loss. As the thermal evolution and mass loss timescales would be comparable if no core luminosity was present, the presence of the core luminosity and hence the transition phase prolongs boil-off by delaying the thermal evolution, resulting in potentially greater mass loss. However, we find this transition phase is typically shorter-lived than the early boil-off, with a substantially lower mass loss rate, leading to a minor time integrated mass loss enhancement, as shown in Figure \ref{1v3core}, in which we display the planetary evolution with (solid black) and without the core luminosity (solid gray). The decrease in the final mass fraction due to the core luminosity is typically less than 30-40\%. 

After the transition, a hydrodynamic mass loss process is coupled with thermal evolution when $t_{\dot{M}} \gtrsim t_{cool}$. However, in this stage, the long-lived Parker mass loss has an exponentially increasing mass loss timescale. Although the core luminosity can still constitute a large fraction of the planet's energy budget and enhances the mass loss rate by a significant fraction, we find it does not impact the absolute time-integrated mass loss of the later evolution due to the orders of magnitude slower mass loss rate compared to the thermal contraction rate. Consequently, parker mass loss decouples (meaning the mass loss rate becomes negligible, rather than being insensitive to core luminosity) from thermal evolution again (with $t_{\dot{M}} \gg t_{cool}$) after a brief coupling phase lasting only a few Myr post-transition. Therefore, from the timescale aspect, we did not find an efficient long-lived core-enhanced wind, as opposed to GSS18.

To improve the numerical stability for our evolution model, as a simplification, we do not allow a planet to thermally inflate over time, which forces the core luminosity $L_{core}$ to be no greater than the envelope cooling luminosity. It is important to note that in this work, ``thermal inflation'' specifically refers to an increase in envelope specific entropy. Thermal inflation does not necessarily result in radius inflation, as mass loss reduces the envelope mass and consequently decreases the planetary radius. Similarly, \citet{Rogers24} did not find the core luminosity thermally inflates their model planets with their parameterized core luminosity calculation. We find that with the simplification relaxed, the core luminosity can only thermally inflate planets for a brief period of time, lowering the final mass fraction (leading to an overestimation of the final mass fraction in our model) by at most 50\% rather than changing it exponentially. Therefore, we argue our simplification does not impact the general behavior we investigate here. 
The validity of our core treatment and future improvements are discussed in Section \ref{disc:core}. 

\subsection{Role of initial entropy and advective cooling} 
\label{subsec:entropycooling}
A large initial H/He mass means a large radius planet, leading to a vigorous mass loss and more advective cooling \citep{Owenwu16}. With our initial entropy defined in \ref{subsec:entropy}, although we find that the advective cooling rate is orders of magnitude higher than the radiative cooling rate, it is still negligible compared to the large energy reservoir. Due to the decoupling effect discussed in Section \ref{subsec:parker}, our model is not sensitive to the assumed advective cooling when choosing different initial mass fractions, as these models will converge to the same final mass fraction. In this case, the strength of boil-off is only controlled by the initial entropy. 

However, in contrast to our finding, \citet{Owenwu16} reported a more significant advective cooling effect that leads to a cooler interior at the end of boil-off and therefore more substantial final mass fraction when choosing a higher initial mass fraction. This behavior must correspond to a thermal contraction timescale comparable to the mass loss timescale. To explain the different results, we attribute the difference to initial conditions. Overall, the physical parameter that controls the effectiveness of the advective cooling is the initial RCB radius, which largely enhances the advective cooling rate and hence the post-boil-off mass fractions if the initial RCB radius is a large fraction of the sonic point radius. This corresponds to models with a hot interior state (with entropy $> 13\ k_b/atom$ for a 10\% H/He envelope). However, it is unlikely to happen for a realistic planet, as it would correspond to a contraction rate, due to either the mass loss or advective cooling, much shorter than the disk dispersal timescale.

For comparison, starting with what we believe are more appropriate initial conditions, a typical low-mass planet has an initial RCB radius of about only $10-20\%$ of its sonic point radius, and the influence of the advective cooling is then limited. We show that the ratio between the timescales of the advective cooling and mass loss $t_{cool,ad}/t_{\dot{M}}$ represents the relative importance of the advective cooling, or how strongly the advective cooling is coupled with the mass loss, which scales with a planet's initial RCB radius $R_{rcb,0}$ over the sonic point radius $R_s$:
\begin{equation}
\label{coolingvsmdot}
\frac{t_{cool,ad}}{t_{\dot{M}}} = \frac{G M_p M_{env}}{\alpha R_{rcb,0} L_{adv}} \frac{\dot{M}}{M_{env}} = \frac{2(\gamma-1)}{\alpha \gamma} \frac{R_s}{R_{rcb,0}}
\end{equation}
For the planets with our initial radius, the timescale ratio is typically around 10, indicating a decoupling, whereas \citet{Owenwu16} initial conditions with $R_{rcb,0} = R_s$ correspond to a strongly coupled boil-off mass loss. Additionally, we suggest the ignored core-luminosity is important in the  \citet{Owenwu16} setup due to the coupling effect (see Section \ref{subsec:core}). Including the core luminosity would greatly increase the interior cooling timescale (Eq. \ref{KHtime_core}) and weaken the coupling between the thermal evolution and mass loss, leading to a less significant advective cooling effect. The quantitative details and a direct comparison to \citet{Owenwu16} are included in Appendix \ref{app:entropyadv}.

\begin{figure}
\centering
\includegraphics[width=0.45\textwidth]{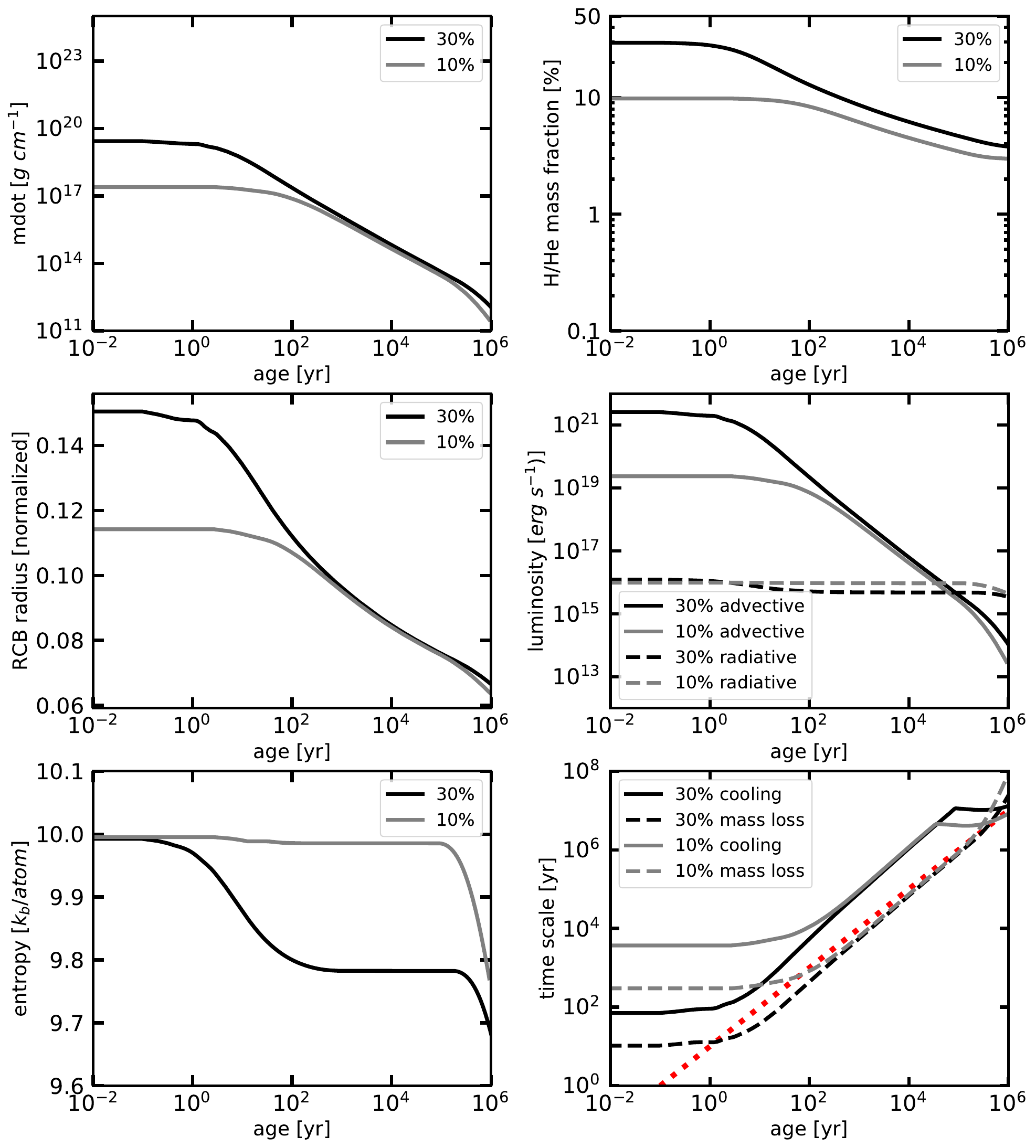} 
\caption{Evolution during boil-off for two $3.6\ M_\oplus$ models both starting at the same entropy, which gives an initial Kelvin-Helmholtz timescale $\sim$ Myr, but with different initial H/He envelope mass fraction ($30\%$ in $black$ and $10\%$ in $gray$). In $10^2$ yr, the $30\%$ model experiences a significant entropy drop due to its fast advective cooling rate while the entropy for the $10\%$ model remains unchanged (bottom left).  As a result, the $30\%$ model has a cooler interior by $10^6$ yr ending boil-off with a slightly larger final mass fraction (top right). After $\sim$$10^2$ yr, the mass loss rate (top left), RCB radius (middle left), intrinsic (radiative) luminosity (middle right, dashed), and advective luminosity (middle right, solid) for the two models converge, essentially meaning the advective luminosity has a negligible impact on the later evolution.  For both mass fractions, after $\sim$$10^6$ yr,  the radiative cooling timescale (bottom right, solid) becomes shorter than the mass loss timescale (bottom right, dashed), and the outflow transitions into the long-lived mass loss phase.
}
\label{1v3entropy}
\end{figure}

Due to this sensitive behavior of boil-off, we always initialize boil-off using our self-consistent initial entropy in the later discussion unless otherwise specified. An example of the evolution track is shown in Figure \ref{1v3entropy}, where the greater role of the advective cooling $L_{ad}$ for the 30\% planet is seen at $10^2$ years, which rapidly cools the planet, slowing mass loss, leading to a larger post-boil-off mass fraction compared to the 10\% model. These enhancements are considered to be small. Note that the initial specific entropy of the 30\% model is set to be the same as the self-consistent entropy of 10\% model to eliminate the initial entropy effect. Moreover, this effect is largely reduced at a lower initial mass fraction $\leq 20\%$ that is predicted by accretion models. In the presence of core heating, the net cooling effect is further diminished. These yield a relative final mass fraction difference smaller than $10\%$ between the models with and without the advective cooling. Therefore, we can safely ignore the advective cooling when choosing a different initial mass fraction $\leq 20\%$ with our determined initial entropy, as the final H/He mass fraction is insensitive to the initial mass fraction.

Our finding above with the assumption that those planets start with the same initial entropy (as shown in Figure \ref{1v3entropy}), still holds true with a more sophisticated initial condition. The specific entropy required to start with a Kelvin-Helmholz cooling timescale of 5 Myr is slightly higher for envelopes with greater initial mass fractions. For planets with reasonable initial mass fractions in the range of $10\%-20\%$, we find that the variation in the self-consistent initial entropy results in only a marginal change of only 10\% in the post-boil-off envelope mass compared to the results using the fixed initial entropy. Therefore, this effect can be neglected.

Moreover, we find that the effectiveness of advective cooling depends on not only the timescale ratio $t_{cool,ad}/t_{\dot{M}}$ but also the ratio between the cooling timescale $t_{cool}$ and the current planetary age. If the cooling timescale is longer than the current age, the entropy is unable to efficiently decline through cooling until the planet becomes older. For instance, in the bottom right panel of Figure \ref{1v3entropy}, we show the evolution of timescales as a function of time. Even though the advective cooling is always 10 times slower than mass loss for the $30\%$ model (solid black) before 1 Myr, the entropy can still significantly drop when the cooling timescale becomes comparable to a certain fraction of the current age between 1-100 yr. We find a 1/10 ratio between the timescales and age (dotted red) gives a good match for the behavior. On the contrary, for the 10\% model, the advective cooling timescale is always much longer than the age to drive significant cooling, which is the reason that the $\leq 20\%$ models are not susceptible to advective cooling. Similarly, around 1 Myr when the cooling timescale becomes shorter than the age, radiative cooling kicks in and thermal contraction becomes considerable again, after which boil-off shuts off. This behavior also applies to mass loss. In the top right panel, the H/He mass fraction for the 30\% model (black) barely declines with time until the mass loss timescale (dashed black in the bottom right panel) becomes comparable to the current age after 1 yr.


To summarize, our treatment of the advective cooling effect is based on the treatment (Eq. \ref{advective}) from \citet{Owenwu16}.  We find that it is negligible for boil-off with our self-consistent initial conditions. Additionally, under our model assumptions and following our theoretical discussion (see Section \ref{subsec:envelope}), we do not find that the energy advection $L_{ad}=h\dot{M}$ has a cooling effect, as the energy advection is a consequence of mass loss from the envelope, which does not affect the thermal evolution. Therefore, our results in the later sections are independent of the treatment of advective cooling.   

\subsection{Final mass fraction from boil-off}
\label{subsec:finalfraction}
Here, we set up our numerical model with self-consistent initial conditions to assess the conditions a planet will have after boil-off. As we showed that the post-boil-off mass fraction is independent of an initial mass fractions $\leq 20\%$, as a simplification, we start our boil-off models with a constant initial H/He mass fraction of $10\%$ to constrain the final mass fraction after boil-off, unless specified. With our initial entropy choice (see Section \ref{subsec:entropy}), the final mass fraction after boil-off depends solely on the scale height of the radiative atmosphere, which is a function of core mass and incident stellar flux.  

We shut off boil-off either at the time $t_{end}$ that sets the transition into the long-lived mass loss in Eq. \ref{transitiontime} or at the time $t_{XUV}$ when the XUV heating starts to dominate the upper radiative atmosphere (section \ref{subsec:xuv}). Practically, we find $t_{XUV}$ happens slightly earlier than $t_{end}$, both around 1 Myr under self-consistent initial conditions, with the time-integrated mass loss negligible between $t_{XUV}$ and $t_{end}$ as the mass loss rate has greatly diminished with time toward the end of boil-off. Therefore, we suggest that photoevaporation should start right after boil-off around 1 Myr when the isothermal Parker wind starts to transition into the long-lived mass loss phase. We find that the post-boil-off Parker wind is briefly as efficient as photoevaporation right after boil-off for $\sim Myr$ and then exponentially decays making it a secondary mass loss mechanism over long timescales. The role of the long-lived, what we call bolometric-driven, escape in the absence of the photoevaporation is discussed in Section \ref{subsec:nobolo}

With the initial conditions we chose, we find the duration of the boil-off phase is typically in the range of 0.5-1 Myr. Right after boil-off, the planet has a contraction timescale of 10-20 Myr due to the radiative cooling that happens around the end of boil-off, compared to the initial contraction timescale of 5 Myr shown in Table \ref{table:1}.  For comparison, \citet{Owenwu16} find a longer 50 Myr timescale starting with a substantially larger radius. We attribute this difference to a different criterion for terminating boil-off.  If we let our boil-off model evolve for another 2 Myr as they do, we find results consistent with \citet{Owenwu16}. Therefore, the cooler appearance of a planet after boil-off is independent of the initial conditions and the types of cooling mechanisms. In fact, we find in our model that it is the radiative cooling that prolongs the contraction timescale in our model, compared to the advective cooling as argued in \citet{Owenwu16}. We note that in our model, most cooling happens near the end of boil-off, when the radiative thermal contraction starts to be efficient, shown in Figure \ref{1v3entropy}. 

Table \ref{table:2} shows a grid of the final mass fraction that a planet can have after boil-off and before photoevaporation or the post-boil-off bolometric-driven escape, as a function of core mass and bolometric flux. Low-mass planets and highly irradiated planets tend to lose more mass due to their large scale height of the radiative atmosphere. Note that most of the models start with an envelope mass fraction of $10\%$ except for heavy cores and cold planets. Since they are less susceptible to boil-off, which allows them to hold a final mass fraction greater than 10\% if born with sufficient amount of H/He mass, we assess them using a higher initial mass fraction. The final mass fraction also represents a critical mass fraction, in that we find that planets with higher initial fractions than the critical mass fractions shown in the table always converge, ``boiling off'' until such final mass fractions are reached. On the other hand, planets formed with initial mass fractions lower than the critical mass fractions will not undergo boil-off. Heavy cores and less irradiated planets are also invulnerable to boil-off, potentially holding any substantial final mass fraction. In this case, photoevaporation and/or post-boil-off bolometric-driven escape starts immediately after the disk dispersal. 

Remarkably, we find that boil-off is so powerful that it is capable of removing the entire convective envelope leaving behind either a bare rock/iron core or a super-Earth with only a layer of radiative atmosphere. This behavior typically happens to planets with core mass $< 4 M_\oplus$ and stellar insolation $>100 F_\oplus$, due to the nature of the large scale height resulting in a less bound and unstable atmosphere. It indicates that boil-off can possibly contribute significantly to the transformation of sub-Neptunes into super-Earths and can potentially have an impact on the formation of the radius valley. On the other hand,  beyond $1000\ F_\oplus$ and $20\ M_\oplus$, our model is consistent with the existence of warm-Neptunes, for instance GJ 436b, which are planets that survived the boil-off phase with heavy envelopes and are rapidly losing mass through photoevaporation. 

Our caveat for our treatment and Table \ref{table:2} is that our boil-off model implicitly assumes the collisional regime for hydrodynamic winds. When the atmosphere is tenuous enough, the gas starts to become collisionless, known as Jeans escape.  The lowest mass fractions (bottom left corner) are so small that they would be subject to a Jeans escape, which would greatly slow down the mass loss rate.  Therefore, the quantitative values for them may not be accurate. Despite that, low-mass and highly irradiated planets can never hold a convective envelope based on our model and will be eventually turned into super-Earths in a short time in the presence of photoevaporation and/or post-boil-off bolometric-driven escape, once the envelope is thin enough. 

\begin{deluxetable*}{c | c c c c c c c}
\tabletypesize{\scriptsize}
\tablewidth{0pt} 
\tablecaption{Kelvin-Helmholtz thermal contraction timescale (Myr) after the boil-off \label{table:1}}
\tablehead{
\colhead{Flux ($F_\oplus$)} & \colhead{1.5 $M_\oplus$} & \colhead{2.4 $M_\oplus$} & \colhead{3.6 $M_\oplus$} & \colhead{5.5 $M_\oplus$} & \colhead{8.5 $M_\oplus$} & \colhead{13 $M_\oplus$} & \colhead{20 $M_\oplus$} 
} 
\startdata 
 1 & 17.6 & 13.4 & 15.5 & - & - & - & - \\              [0.5 ex]
 3 & 15.0 & 12.0 & 12.1 & 16.5 & - & - & - \\        [0.5 ex]
 10 & 8.2 & 12.6 & 10.9 & 17.8 & - & - & - \\        [0.5 ex]
 30 & 8.6 & 11.0 & 10.5 & 12.8 & 21.4 & - & - \\        [0.5 ex]
 100 & 0 & 8.0 & 8.0 & 10.2 & 17.3 & 14.8 & - \\[0.5 ex]
 300 & 0 & 0 & 8.7 & 11.7 & 15.2 & 15.3 & - \\[0.5 ex]
 1000 & 0 & 0 & 0 & 7.6 & 12.5 & 16.8 & 14.3 \\[1.5 ex]
\enddata
\tablecomments{We show the Kelvin-Helmholtz thermal contraction timescale (Eq. \ref{KHtime_core}) after the boil-off process at around $\sim$ 1 Myr is generally 10-20 Myr, longer than the initial contraction time of 5 Myr, due to the radiative cooling effect that comes into play at the very late stage of boil-off.
}
\end{deluxetable*}

\begin{deluxetable*}{c | c c c c c c c}
\tabletypesize{\scriptsize}
\tablewidth{0pt} 
\tablecaption{Maximum final mass fraction allowed after boil-off \label{table:2}}
\tablehead{
\colhead{Flux ($F_\oplus$)} & \colhead{1.5 $M_\oplus$} & \colhead{2.4 $M_\oplus$} & \colhead{3.6 $M_\oplus$} & \colhead{5.5 $M_\oplus$} & \colhead{8.5 $M_\oplus$} & \colhead{13 $M_\oplus$} & \colhead{20 $M_\oplus$} 
} 
\startdata 
 1 & 1.18\% & 3.38\% & 8.62\% & - & - & - & - \\              [0.5 ex]
 3 & 0.49\% & 1.79\% & 4.93\% & 14.64\%** & - & - & - \\        [0.5 ex]
 10 & 0.14\% & 0.77\% & 2.56\% & 8.85\% & - & - & - \\        [0.5 ex]
 30 & 0.02\% & 0.25\% & 1.17\% & 4.44\% & 16.75\%** & - & - \\        [0.5 ex]
 100 & $2\times 10^{-6}$* & 0.03\% & 0.30\% & 1.67\% & 7.21\% & 23.49\%** & - \\[0.5 ex]
 300 & 0 & $6\times 10^{-6}$* & 0.04\% & 0.53\% & 3.14\% & 11.13\%** & - \\[0.5 ex]
 1000 & 0 & 0 & $4\times 10^{-6}$* & 0.04\% & 0.73\% & 4.04\% & 13.50\%** \\[1.5 ex]
\enddata
\tablecomments{A dash symbol indicates that a planet with corresponding core mass and bolometric flux will experience zero or very slight mass loss.\\
* This symbol indicates that the planet has already lost its entire H/He envelope. The final mass fraction is estimated based on a two-layer structure model that only has a layer of radiative atmosphere on top of the core.\\
** For the numbers greater than $10\%$, we start the mass loss with a larger initial mass fraction to properly determine the maximum final mass fraction. All the other models have the same default initial mass fraction of $10\%$.
}
\end{deluxetable*}

\subsection{Subsequent planet evolution}
\label{subsec:subsequent}
To assess the impact on planet population from boil-off, we evolve planets in the absence of subsequent mass loss by continuing thermal evolution for another 10 Gyr with the initial conditions from the end of boil-off (Table \ref{table:2}). This constrains the largest possible radius allowed as a function of stellar bolometric flux and core mass at each given age. Intuitively, low-mass sub-Neptunes tend to possess a large radius and incredibly low bulk density even with a small amount of H/He gas, due to their large scale heights of the radiative atmosphere compared to their physical sizes. For these planets, the boil-off phase results in very thin envelopes or in most cases, a complete loss of envelopes. On the other hand, heavier planets that have experienced a boil-off hold more massive envelopes, but their greater surface gravity allows them to have a high bulk density and remain relatively compact. Therefore, boil-off plays a role in setting an upper boundary in the radius distribution of the sub-Neptune population, which may contribute to carving the radius cliff, the decrease in planets with radii above $\sim 4 R_\oplus$ from observation. 

We show the radius evolution in Table \ref{table:3}. Most of the planets shrink to radii smaller than $5\ R_\oplus$ by 10 Gyr, roughly consistent with the radius cliff but with a larger radius. As a comparison, without a boil-off phase, the radii of low-mass and highly irradiated planets, which dominates the sub-Neptune population, are unbounded $\sim 5-10 R_\oplus$ even at a late age. Note that the massive planets that are invulnerable to boil-off are rare in the observed population and may eventually evolve into the warm-Neptune population. Our result implies that rather than photoevaporation or core-powered escape (the post-boil-off Parker wind), boil-off is likely the main cause to the radius cliff, which significantly removes H/He envelope from the planets with large physical sizes. However, other mass loss mechanisms are needed to further strip planets' envelopes and shrink their radii. The possibility that the post-boil-off bolometric-driven escape sculpts the radius cliff is discussed in the next section. 

\begin{deluxetable*}{c | c | ccccccc}
\tabletypesize{\scriptsize}
\tablewidth{0pt} 
\tablecaption{Maximum radius allowed after the early boil-off at given ages\label{table:3}}
\tablehead{
\colhead{Age (Gyr)} & \colhead{Flux ($F_\oplus$)} & \colhead{1.5 $M_\oplus$} & \colhead{2.4 $M_\oplus$} & \colhead{3.6 $M_\oplus$} & \colhead{5.5 $M_\oplus$} & \colhead{8.5 $M_\oplus$} & \colhead{13 $M_\oplus$} & \colhead{20 $M_\oplus$}
} 
\startdata 
 0 & 1 & 6.03 & 9.84 & 15.80 & - & - & - & - \\              [0.5 ex]
 0 & 3 & 4.12 & 6.65 & 10.34 & 17.75 & - & - & - \\        [0.5 ex]
 0 & 10 & 2.84 & 4.55 & 6.93 & 11.28 & - & - & - \\        [0.5 ex]
 0 & 30 & 2.06 & 3.27 & 4.92 & 7.72 & 13.56 & - & - \\        [0.5 ex]
 0 & 100 & 1.38 & 2.32 & 3.46 & 5.28 & 8.54 & 15.07 & - \\[0.5 ex]
 0 & 300 & 1.14 & 1.71 & 2.54 & 3.86 & 6.04 & 9.92 & - \\[0.5 ex]
 0 & 1000& 1.14 & 1.31 & 1.89 & 2.86 & 4.38 & 6.86 & 11.47 \\[1.5 ex]
 \hline
 0.1 & 1 & 2.91 & 3.77 & 5.26 & - & - & - & - \\              [0.5 ex]
 0.1 & 3 & 2.46 & 3.15 & 4.20 & 6.27 & - & - & - \\        [0.5 ex]
 0.1 & 10 & 2.12 & 2.70 & 3.48 & 5.20 & - & - & - \\        [0.5 ex]
 0.1 & 30 & 1.76* & 2.36 & 2.97 & 4.10 & 6.58 & - & - \\        [0.5 ex]
 0.1 & 100 & 1.38 & 1.98* & 2.52 & 3.29 & 4.77 & 7.37 & - \\[0.5 ex]
 0.1 & 300 & 1.14 & 1.71 & 2.22* & 2.84 & 3.93 & 5.61 & - \\[0.5 ex]
 0.1 & 1000& 1.14 & 1.31 & 1.89 & 2.48* & 3.27 & 4.50 & 6.34 \\[1.5 ex]
 \hline
 1 & 1 & 2.53 & 3.16 & 4.19 & - & - & - & - \\              [0.5 ex]
 1 & 3 & 2.25 & 2.78 & 3.55 & 5.10 & - & - & - \\        [0.5 ex]
 1 & 10 & 2.04 & 2.48 & 3.09 & 4.36 & - & - & - \\        [0.5 ex]
 1 & 30 & 1.76* & 2.25 & 2.74 & 3.63 & 5.48 & - & - \\        [0.5 ex]
 1 & 100 & 1.38 & 1.98* & 2.42 & 3.06 & 4.27 & 6.24 & - \\[0.5 ex]
 1 & 300 & 1.14 & 1.71 & 2.22* & 2.70 & 3.59 & 4.94 & - \\[0.5 ex]
 1 & 1000 & 1.14 & 1.31 & 1.89 & 2.48* & 3.10 & 4.11 & 5.59 \\[1.5 ex]
 \hline
 10 & 1 & 2.22 & 2.75 & 3.59 & - & - & - & - \\              [0.5 ex]
 10 & 3 & 2.06 & 2.48 & 3.13 & 4.43 & - & - & - \\        [0.5 ex]
 10 & 10 & 1.96* & 2.28 & 2.79 & 3.87 & - & - & - \\        [0.5 ex]
 10 & 30 & 1.76* & 2.16 & 2.54 & 3.30 & 4.91 & - & - \\        [0.5 ex]
 10 & 100 & 1.38 & 1.98* & 2.31 & 2.84 & 3.92 & 5.66 & - \\[0.5 ex]
 10 & 300 & 1.14 & 1.71 & 2.22* & 2.54 & 3.29 & 4.52 & - \\[0.5 ex]
 10 & 1000 & 1.14 & 1.31 & 1.89 & 2.48* & 2.89 & 3.73 & 5.08 \\[1.5 ex]
\enddata
\tablecomments{ We show the maximum radius allowed under the subsequent thermal evolution without the post-boil-off mass loss. A dash symbol indicates that a planet with corresponding core mass and bolometric flux will experience a zero or very slight boil-off mass loss, and therefore, an arbitrarily high radius is allowed as long as it is able to acquire enough H/He mass during its formation.\\
* This symbol indicates that the planet's interior has already evolved into its final state with an isothermal interior structure, which typically happens with very negligible envelope mass.}
\end{deluxetable*}

\subsection{Lack of evidence for significant long-lived mass loss}
\label{subsec:nobolo}
After the boil-off ends, planetary thermal evolution becomes more significant than mass loss in controlling the planetary radius. Consequently, the scale height of the radiative atmosphere rapidly diminishes leading to an exponentially decaying mass loss rate \citep{Owenwu16} (see also our Figure \ref{boiloff_energy}, which shows a low mass loss rate after $10^6$ years), which almost evolves into a hydrostatic equilibrium. However, in principle, a planet can still lose a substantial amount of mass on a longer timescale. Here to evaluate the importance of the post-boil-off Parker wind as a long-term mass loss mechanism, we continue to evolve the planets after boil-off but with the Parker wind mass loss turned on for another 3 Gyr. For all of the modeled planets in Table \ref{table:2}, we find the time-integrated mass loss in the presence of the core-luminosity out to a few Gyr due to the post-boil-off Parker wind is a negligible amount ($\leq $0.1\%) compared to that from boil-off ($\sim$ 10\% see also Figure \ref{evolution} for a similar evolutionary track), leading to a nearly constant mass fraction at late ages as demonstrated below in Figure \ref{convergence} and \ref{fraction_comparison}. 

In these figures, we assess the planet's initial mass fraction for the post-boil-off mass loss based on the final mass fraction at the end of boil-off (triangle) to separate the outcomes from the two mass loss mechanisms. This treatment is essential, as different initial mass fractions lead to divergent H/He loss histories after a few Gyrs of evolution owing to the mass loss process being coupled to significant thermal evolution, which greatly depends on the conditions it starts with at the end of boil-off. As a comparison, in GSS18, the post-boil-off (their initial) mass fraction was derived analytically with particular assumptions, such as the envelope radius $R_{rcb}$ is twice the core radius $R_c$, and without directly modeling a boil-off phase. These authors found a scaling function for the H/He mass fraction, $f$, where $f = 0.05(M_c/M_\oplus)^{1/2}$, where $M_c$ is core mass. As a direct comparison to the work developed in GSS18, we initialize planets using their analytic H/He mass fraction scaling function (black curve) to examine the validity of their initial conditions to start the subsequent long-term evolution. This group of planets have the Parker wind turned on. They are compared to the models initialized with our 10\% default initial mass fraction (gray curve), with the Parker wind turned off. The choice of the initial conditions does not affect the evolutionary results from our numerical model, as with self-consistent initial entropies, those models from different initial mass fractions converge to common final mass fraction values at the end of boil-off as seen in Figure \ref{convergence} (see also Section \ref{subsec:entropycooling}). The initial entropies are chosen to be the same for both models.

A single representative evolution curve for the H/He mass fraction between the two models is shown in Figure \ref{convergence}. We find a very tiny difference between the final mass fractions after 3 Gyr (cross shaped, based on the GSS18 initial mass fraction with the bolometric-driven escape turned on) and from the end of boil-off around 1 Myr (triangle shaped, based on our 10\% boil-off models, same data as in Table \ref{table:2}). More examples with a wide range of core masses and incident stellar bolometric fluxes are shown in Figure \ref{fraction_comparison}, showing there is very little Parker wind mass loss during this long time. Therefore, we argue that the post-boil-off Parker wind is negligible on a timescale of Gyrs. This is because the long-lived mass loss does not have enough time to significantly remove envelope mass with $t_{\dot{M}} \gg t$, where $t$ is the planetary age, due to the orders of magnitude difference between the mass loss timescale and thermal contraction timescale (which is comparable to the age $t$) at most times of the post-boil-off evolution.

In rare cases, especially for planets with low core masses and high equilibrium temperatures, our model finds that an envelope can only be totally lost through the post-boil-off Parker wind if the post-boil-off H/He mass fraction is small enough, $\sim$ a few 0.01\% of H/He mass. This happens swiftly within a few Myrs right after boil-off rather than a few Gyrs that is argued for the timescale for core-powered mass loss, if the physical choices are properly made.

Additionally, a self-consistent initial mass fraction should start planetary evolution right out of the boil-off phase, with thermal evolution comparable to or faster than mass loss. However, the GSS18 post-boil-off mass fraction scaling function has a rather flat profile as a function of core mass, in contrast to the steep post-boil-off mass fraction distribution from our model (and \citet{Misener21}) that quickly declines to zero toward the low-mass end (Table \ref{table:2}). This underestimates the post-boil-off mass fraction on the high-mass end and overestimates it on the low-mass end. In Figure \ref{fraction_comparison}, the GSS18 initial mass fraction function is shown in dotted, while the self-consistent initial fractions for the long-term evolution that our model finds correspond to triangle shapes, which indicates that a boil-off would happen, instead of a long-lived core-powered mass loss, for nearly all low-mass planets if they started with the GSS18 initial fractions and were assessed with what we believe are more appropriate decoupled Parker winds (section \ref{subsec:parker}) and self-consistent initial entropies. We let these planets continue to boil-off until these planets enter the ``thin regime", defined as the envelope radius $R_{rcb} = 2 R_c$ becomes twice the core radius in both GSS18 and \citet{Misener21}. We still find that some planets are boiling off with a short mass loss timescale $<$ 1 Myr. This indicates an inconsistency with the GSS18 assumption that the planets have already exited the boil-off phase and entered the long-lived mass loss phase. 

The above initial condition problem is quantitatively identified based on our numerical model. With the GSS18 initial conditions (radius and envelope mass fraction), we calculate the ratio between the mass loss timescale of the decoupled wind and the Kelvin-Helmholtz thermal contraction timescale at the beginning of their thin regime evolution, which we find is many orders of magnitude smaller than 1 for highly irradiated planets and low-mass planets, shown in Table \ref{table:4}. From the previous discussion, a similar behavior is found at their transition time $R_{rcb} = 2 R_c$, if the initial envelope mass fraction is instead self-consistently calculated from boil-off as in \citet{Misener21}, indicating that $R_{rcb} = 2 R_c$ may not be a good criterion for the transition as it can correspond to boil-off physical conditions. 

In summary, in this section, since we show that the long-lived post-boil-off Parker wind is inefficient with a wide range of core mass and incident flux given a boil-off phase is included, we suggest that atmospheric photoevaporation, whose importance of impacting the demographics has been widely studied \citep{Owen13,Lopez13,Owen17}, should be the dominant mass loss mechanism that takes place right after boil-off, which further sculpts the small planet population and in part forms the observational features such as the radius valley and radius cliff. A modeling work shows a good fit of the combined evolution with both boil-off and post-boil-off photoevaporation to the observed radius gap \citep{Affolter23}. Additionally, our numerical model indicates that core-powered escape is not a long-lived atmospheric escape but part of our decoupled Parker wind boil-off under the GSS18 initial conditions if modeled with the decoupled Parker wind, which yields initial planets with too large radii for low-mass planets. A direct analysis of the analytical model will be further discussed in Section \ref{sec:diff}.  

\begin{figure}
\centering
\includegraphics[width=0.45\textwidth]{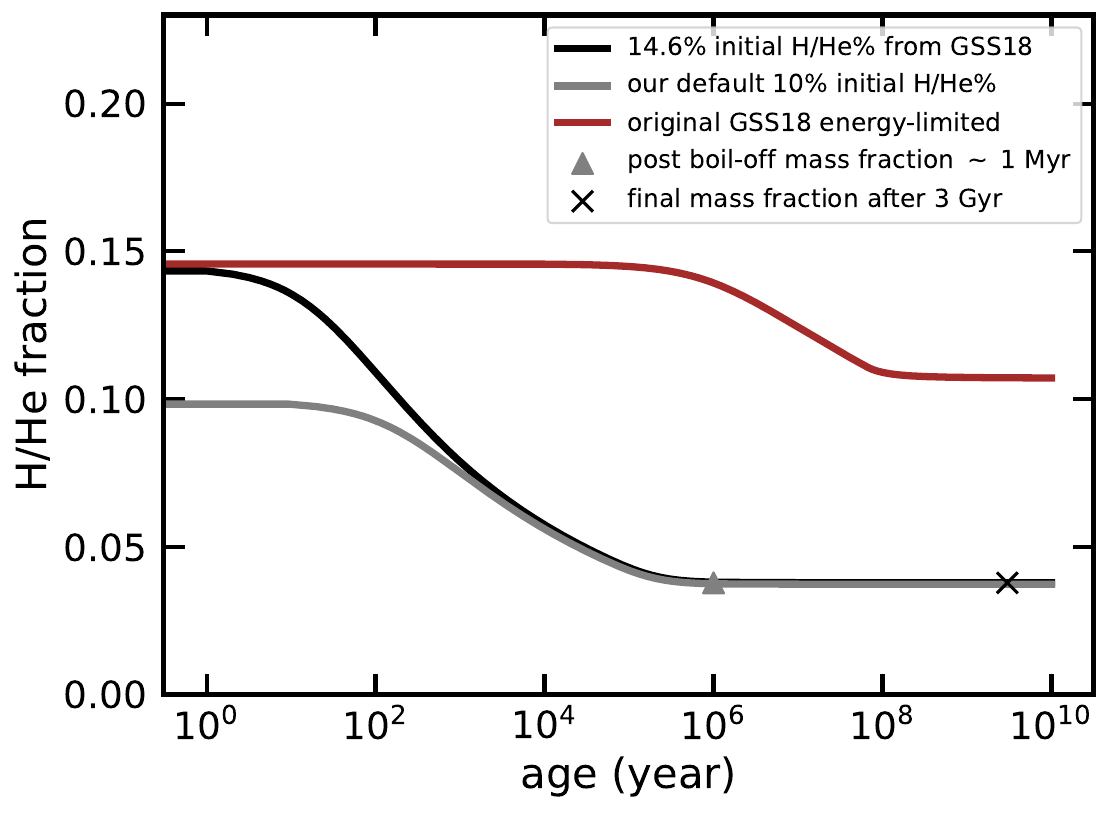} 
\caption{We demonstrate the behavior of the decoupled Parker wind that different initial mass fractions for planets with a given initial entropy, core mass ($8.5 M_\oplus$) and stellar bolometric flux ($300 F_\oplus$) converge to a common final mass fraction value by the end of boil-off. We vary the initial mass fraction from our 10\% default (gray curve, with post-boil-off mass fractions shown in Table \ref{table:2}) to the scaling function developed in GSS18 (black curve).  The mass fraction right after boil-off at 1 Myr is nearly identical to that after 3 Gyr of evolution under the long-lived Parker wind, implying the insignificance of the post-boil-off Parker wind. As a comparison, we include an evolution track for the model under the GSS18 coupled escape, with all other conditions the same as the black model, which never converges in a few Gyrs, due to the coupling effect with the thermal evolution. More examples are shown in Figure \ref{fraction_comparison}.
}
\label{convergence}
\end{figure}

\begin{figure}
\centering
\includegraphics[width=0.45\textwidth]{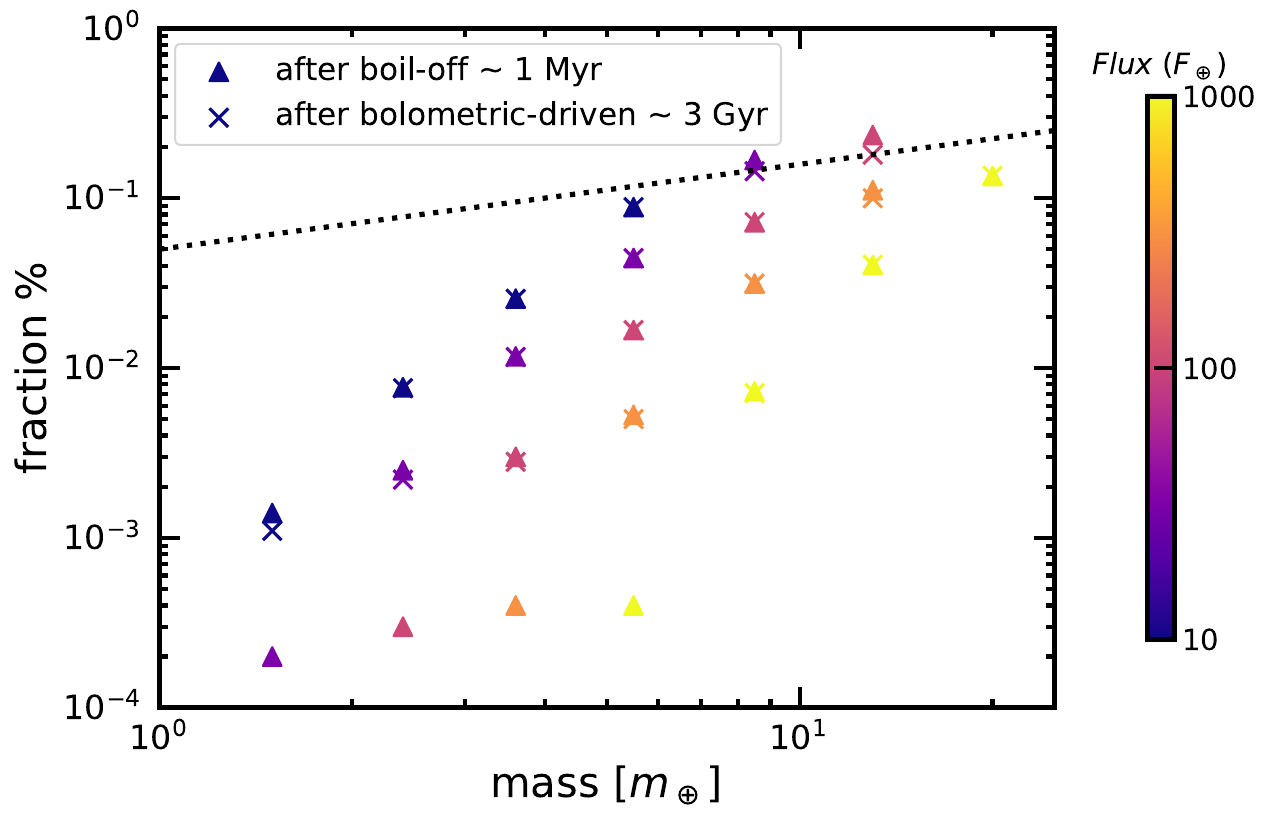} 
\caption{We show the final mass fraction as a function of core mass and stellar bolometric flux (color-coded) after boil-off (triangles) and after 3 Gyr (crosses) evolution of the post-boil-off Parker wind (bolometric-driven), with the model setups similar to those in Figure \ref{convergence}. Note that the post-boil-off mass fractions are calculated using a 10\% default initial mass fraction, while the final mass fractions after 3 Gyr of evolution under the long-lived wind employ the same scaling function $f = 0.05(M_c/M_\oplus)^{1/2}$ developed in GSS18 as initial condition (the initial mass fraction is shown in the dotted line). The long-lived Parker wind is inefficient, only being able to remove the last a few $10^{-4}$ envelope mass after boil-off.
}
\label{fraction_comparison}
\end{figure}

\begin{deluxetable*}{c | c c c c c c c c c}
\tabletypesize{\scriptsize}
\tablewidth{0pt} 
\tablecaption{Timescale ratio between the mass loss and thermal contraction \label{table:4}}
\tablehead{
\colhead{Flux ($F_\oplus$)} & \colhead{1.0 $M_\oplus$} & \colhead{1.5 $M_\oplus$} & \colhead{2.4 $M_\oplus$} & \colhead{3.6 $M_\oplus$} & \colhead{5.5 $M_\oplus$} & \colhead{8.5 $M_\oplus$} & \colhead{13 $M_\oplus$} & \colhead{20 $M_\oplus$}
} 
\startdata 
 10 & $2\times10^{-8}$ & $4\times10^{-6}$ & $5\times10^{-2}$ & $9\times10^{3}$ & $7\times10^{11}$ & $6\times10^{23}$ & $\infty$ & $\infty$ \\        [0.5 ex]
 30 & $5\times10^{-10}$ & $2\times10^{-8}$ & $2\times10^{-5}$ & $1\times10^{-1}$ & $1\times10^{5}$ & $9\times10^{13}$ & $9\times10^{26}$ & $\infty$ \\ [0.5 ex]
 100 & $3\times10^{-11}$ & $3\times10^{-10}$ & $3\times10^{-8}$ & $1\times10^{-5}$ & $2\times10^{-1}$ & $7\times10^{5}$ & $2\times10^{15}$ & $6\times10^{29}$ \\[0.5 ex]
 300 & $9\times10^{-12}$ & $3\times10^{-11}$ & $7\times10^{-10}$ & $5\times10^{-8}$ & $7\times10^{-5}$ & $4\times10^{0}$ & $5\times10^{7}$ & $2\times10^{18}$ \\[0.5 ex]
 1000 & $6\times10^{-12}$ & $6\times10^{-12}$ & $3\times10^{-11}$ & $6\times10^{-10}$ & $7\times10^{-8}$ & $2\times10^{-4}$ & $2\times10^{1}$ & $1\times10^{9}$ \\[1.5 ex]
 3000 & $6\times10^{-12}$ & $1\times10^{-12}$ & $2\times10^{-12}$ & $1\times10^{-11}$ & $3\times10^{-10}$ & $8\times10^{-8}$ & $4\times10^{-4}$ & $2\times10^{2}$ \\[1.5 ex]
\enddata
\tablecomments{ We show the timescale ratio between the decoupled Parker wind mass loss and thermal contraction assessed with the GSS18 initial conditions, such that the RCB radius is twice the core radius $R_{rcb} = 2 R_c$ and the mass fraction is set by $f=0.05 (M_c/M_\oplus)^{1/2}$. The low-mass and highly irradiated planets are still in the boil-off phase with the ratio much smaller than 1, while the massive and cold planets are initially well contracted, indicating that these initial conditions are not the self-consistent conditions that a planet should have at the beginning of the long-lived mass loss phase and the end of boil-off.}
\end{deluxetable*}

\section{Why does the analytical work find that core-powered mass loss is significant?} \label{sec:diff}
Because we find a more limited role for the long-lived mass loss as well as core luminosity compared to that from the seminal work of GSS18, we provide an extended description of the differences between our models. We find that the largest differences come from (1) their energy-limited wind that strongly couples with the thermal evolution, (2) the lack of an energy loss term that removes the envelope energy by the wind, (3) the simplified initial H/He mass fraction and radius values, (4) the criteria used for the transition, (5) the lack of a radiative atmosphere for evaluating planet radii, and (6) different mean molecular weights for the atmosphere.

\subsection{Role of wind assumptions} 
\label{subsec:mdot}

\begin{figure}
\centering
\includegraphics[width=0.47\textwidth]{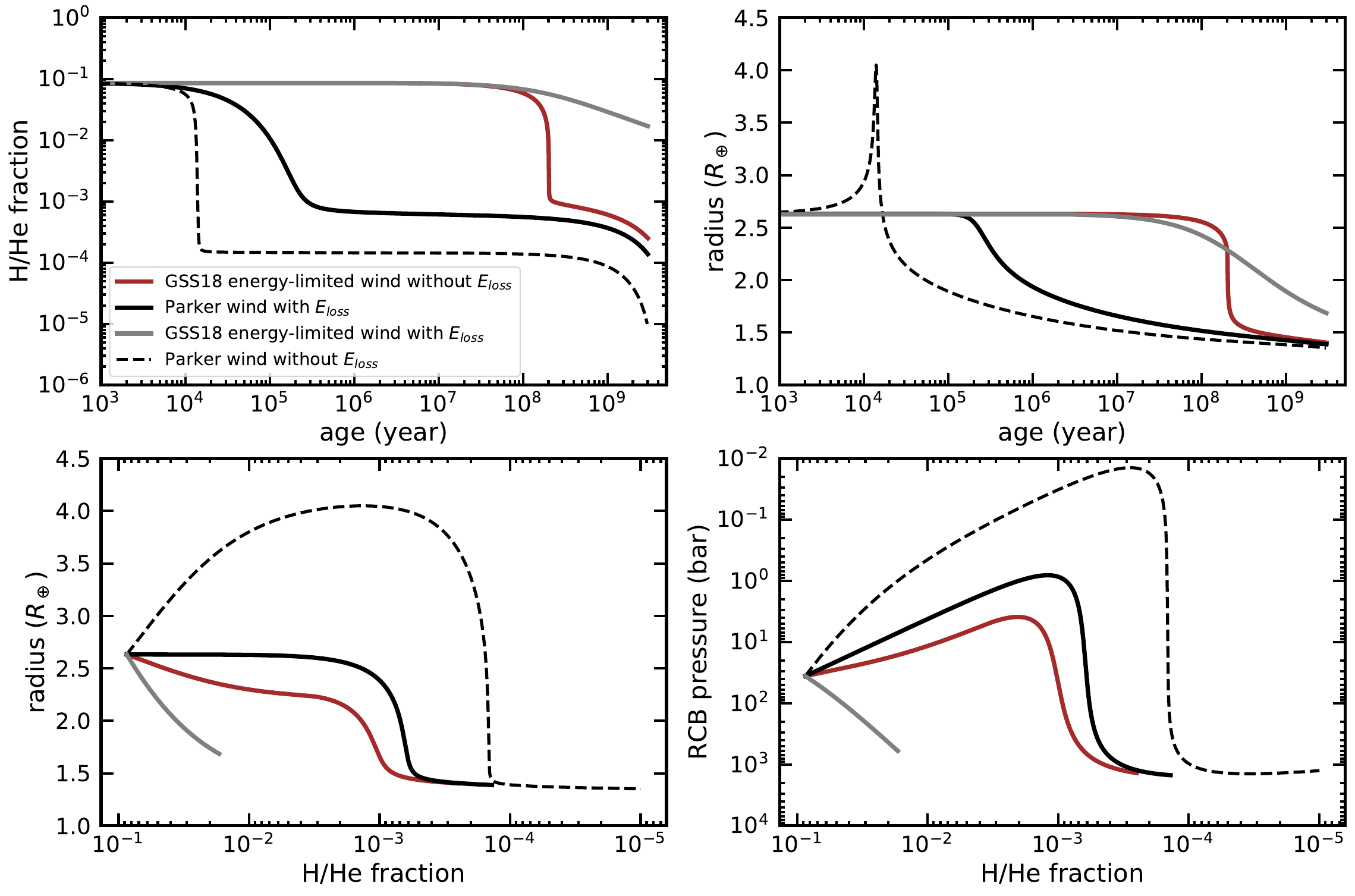} 
\caption{To demonstrate the difference between models with different physical choices, we show the evolution tracks of 3 $M_\oplus$ planets, based on the GSS18 analytical model with their original initial conditions. The model with the original physics is shown in red, with the GSS18 energy-limited wind without the energy loss term $\dot{E}_{adv}$ (see Eq. \ref{energydot}).  A decoupled isothermal Parker wind without $\dot{E}_{adv}$ is in dashed black. 
Including the energy loss term $\dot{E}_{adv}$ is shown in solid black. This black model represents what we believe is more appropriate combination of physics, which is more consistent with the behavior from our numerical model. In gray is an improved GSS18 energy-limited wind, now with $\dot{E}_{adv}$. 
See Section \ref{subsec:mdot} and \ref{subsec:edot} for discussion.
}
\label{mdot}
\end{figure}

We showed in Section \ref{subsec:boil} that the energy needed for adiabatic expansion is only the amount needed to expand the radiative atmosphere from the RCB (see Eq. \ref{pdv_rad}), compared to the GSS18 assumption that energy is required to lift material from the deep envelope right above the core (Eq. \ref{pdv_core}). In Figure \ref{mdot}, we show an experiment based on the implementation for the analytical core-powered mass loss model from GSS18, with different combinations of physical choices. Here we compare the GSS18 wind (red) to our decoupled (by our default non-energy-limited) isothermal Parker wind (dashed black) with the initial conditions the same as the original setup.  A similar delayed and prolonged boil-off is seen with their analytical model, as in Figure \ref{convergence} with our numerical model. We show that by 30-40 Myr the planet with their coupled wind is still quickly boiling off, instead of starting the evolution within the long-lived mass loss phase. This is induced by the overestimated initial H/He mass fractions and/or initial radius, as demonstrated in Section \ref{subsec:nobolo}. Once the extended but less intense boil-off is over when the wind transitions into the Bondi-limited regime and thermal evolution starts to dominate, the planet holds a larger quantity of H/He mass and thermally contracts faster than the mass loss, leading to an insignificant mass loss in the later evolution. As a comparison, the planet under the decoupled isothermal Parker wind holds a substantially less massive envelope at the end of boil-off $<$ 1 Myr with the mass loss nearly completely quenched in the later Gyrs. Planets under their coupled mass loss never converge to the final mass fraction value found by any of the decoupled Parker wind models, due to the coupling with the thermal evolution.

\subsection{Thermal evolution: Energy reservoir is removed only by luminosity, not mass loss} \label{subsec:edot}
As a planet loses H/He envelope mass, the gravitational energy and the internal thermal energy of the escaping matter should be correspondingly lost from the system, leading to a decline of the energy reservoir, which is an energy loss term in addition to the thermal cooling that is not implemented in the original work. From GSS18, the energy reservoir is given by:
\begin{equation}
\label{Ecool}
E_{cool} = g\Delta R \left(\frac{\gamma}{2\gamma - 1} M_{E} +\frac{1}{\gamma}\frac{\gamma-1}{\gamma_c-1}\frac{\mu}{\mu_c} M_c\right)
\end{equation}
where $\Delta R = R_{rcb} - R_{c}$ is the thickness of the envelope, $\mu$ and $\mu_c$ are the mean molecular weights of respectively the envelope and the core, and $\gamma$ and $\gamma_c$ are the adiabatic indexes. The energy loss rate as a result of mass loss can then be estimated by the derivative of the energy reservoir equation with respect to the envelope mass $M_{env}$:
\begin{equation}
\label{energydot}
\dot{E}_{adv} = \dot{E}_{cool} = g\Delta R \frac{\gamma}{2\gamma - 1} \dot{M}_{E}
\end{equation}
$\dot{E}_{adv}$ represents an energy advection, similar to Eq. \ref{advective} and $L_{adv}=h\dot{M}$ in Section \ref{subsec:envelope}. The thermal evolution within a timestep is therefore implemented in our version of the analytical model as:
\begin{equation}
\label{energyevolution}
E_{cool} \to E_{cool} - L\Delta t - \dot{E}_{adv} \Delta t
\end{equation}

In Figure \ref{mdot}, we demonstrate the importance of the energy loss term $\dot{E}_{adv}$ by comparing two models with (solid black) and without it (dashed black), both implemented with decoupled isothermal Parker winds with all other conditions the same as the original GSS18 setup. It is seen in the dashed black line that without $\dot{E}_{adv}$, a planet cannot efficiently contract and can even greatly inflate its radius in a short time interval in the boil-off period, because the total energy $E_{cool}$ stays nearly constant in boil-off phase due to both the inefficient thermal cooling and the lack of energy loss term $\dot{E}_{adv}$, and consequently a decreasing envelope mass $M_{env}$ leads to an expanding envelope $\Delta R$ in Eq. \ref{Ecool}. This unrealistic radius inflation results in an exaggerated time-integrated mass loss. This behavior becomes especially pronounced, which generates a spike around $10^4$ yr as seen in dashed black, as we switch to a decoupled isothermal Parker wind from the GSS18 coupled mass loss (red). With the mass loss-driven energy loss term $\dot{E}_{adv}$ implemented shown in black, we find a higher final mass fraction by 3 Gyr compared to the planet without it (dashed black), with most of the envelope mass lost in 1 Myr, which is consistent with the behavior we find for boil-off based on our numerical model. The unrealistic inflation behavior (dashed black) is also eliminated by $\dot{E}_{adv}$, implying its significance. The black model with both physical improvements is what we argue is a more appropriate setup for modeling boil-off and the subsequent long-lived mass loss.

The inclusion of the $\dot{E}_{adv}$ term leads to an enhanced final mass fraction while a decoupled isothermal Parker wind lowers the final mass fraction. Interestingly, with the physics of both thermal evolution and mass loss improved (black), these two effects largely cancel each other out producing very similar final mass fractions and final radii to those from the original setup without both improvements (red) if GSS18 initial conditions are used. Therefore, the impact from the improved physics is hard to evaluate on a population level (see Section \ref{sec:pop}) at Gyr ages, but from an evolutionary perspective, the net effect with both improvements leads to a mass loss timescale shift from Gyrs to Myrs, indicating a dramatic mass loss timescale difference between our decoupled Parker wind and the GSS18 coupled Parker wind if assessed without the $\dot{E}_{adv}$ improvement. The timescale difference can be tested against the observed population in future studies.

\subsection{Planet radius does not shrink with mass loss} \label{subsec:shrink}
With the improved physics for thermal contraction and mass loss applied to the analytical model, we compared the evolution between different models. Although the analytical model is consistent with our numerical model in terms of mass loss timescale $\sim$ 1 Myr, we find it generally predicts a higher amount of mass loss than our numerical model. As shown with the gray line model in Figure \ref{mdot}, this behavior results from the fact that a planet from the GSS18 formulation of the energy budget in Eq. \ref{Ecool} does not shrink efficiently until its envelope mass fraction falls below 0.1\% as the H/He mass fraction is gradually lost (see radius versus H/He mass fraction on bottom left). As a comparison, all model planets from our numerical model continuously shrink in radius during boil-off, as shown in Figure \ref{evolution}, due to the mass loss (rather than thermal contraction). This holds true even if we account for thermal inflation. 

The constant RCB radius observed in the analytical model must be attributed to a continuous thermal inflation that increases planetary radius more efficiently than the mass loss decreases it. As a comparison, a planet with its mass dominated by the core losing mass at a fixed entropy is expected to undergo a decrease in radius. This inflation behavior corresponds to a tremendous heating from the core to the envelope, which exceeds the intrinsic luminosity with $L_{core} \gg L_{int}$, over the whole mass loss phase. Note that the core luminosity and specific entropy are not explicitly calculated in the analytical model. The increased entropy therefore leads to more mass loss due to the initial entropy effect discussed in Section \ref{subsec:entropy} and \ref{subsec:entropycooling}. This continuous and significant inflation is not found in our model if we relax our assumption and allow the planet to thermally inflate. 

Additionally, the inflation in the analytical model also reduces the RCB pressure by a factor exceeding 1,000 (bottom right panel of Figure \ref{mdot}). This suggests a dramatic increase in specific entropy in their model, since a shallow RCB corresponds to a hotter interior. In comparison, our numerical model suggests that the RCB pressure varies by at most a factor of 2 during boil-off with the thermal inflation included. This indicates that our numerical model predicts a much more modest thermal inflation and a significantly less pronounced role for core luminosity compared to the GSS18 model. 


\subsection{Reassessed coupled Parker wind with improved physics} 
\label{subsec:coupled}
In this section, we reexamine the GSS18 coupled wind and, with improved physics, obtain results that diverge from those of GSS18. In \citet{Misener21} (Figure 2), these authors show that the GSS18 type energy-limited wind exhibits a strong coupling between the mass loss and thermal evolution, with comparable timescales $t_{\dot{M}} \sim t_{cool}$. From Eq. \ref{KHtime_core} and \ref{mdot_core}, we show that the wind treatment leads to a slowly (linearly) changing ratio between the mass loss timescale $t_{\dot{M}}$ and radiative thermal contraction timescale $t_{cool}$ in the boil-off phase: 
\begin{equation}
\label{timescales}
\frac{t_{\dot{M}}}{t_{cool}} = \frac{M_{env}}{L_{int}/gR_c} \frac{L_{int}-L_{core}}{GM_cM_{env}/(\alpha R_{rcb})} = \frac{\alpha (1-\beta) R_{rcb}}{R_c}
\end{equation}
where the luminosity ratio $\beta = L_{core}/L_{int}$ dictates the strength of the core-enhanced effect (this is true only when the thermal evolution is coupled with the mass loss with $t_{\dot{M}}/t_{cool} \sim 1$).

\begin{figure}
\centering
\includegraphics[height=0.8\textwidth,width=0.42\textwidth]{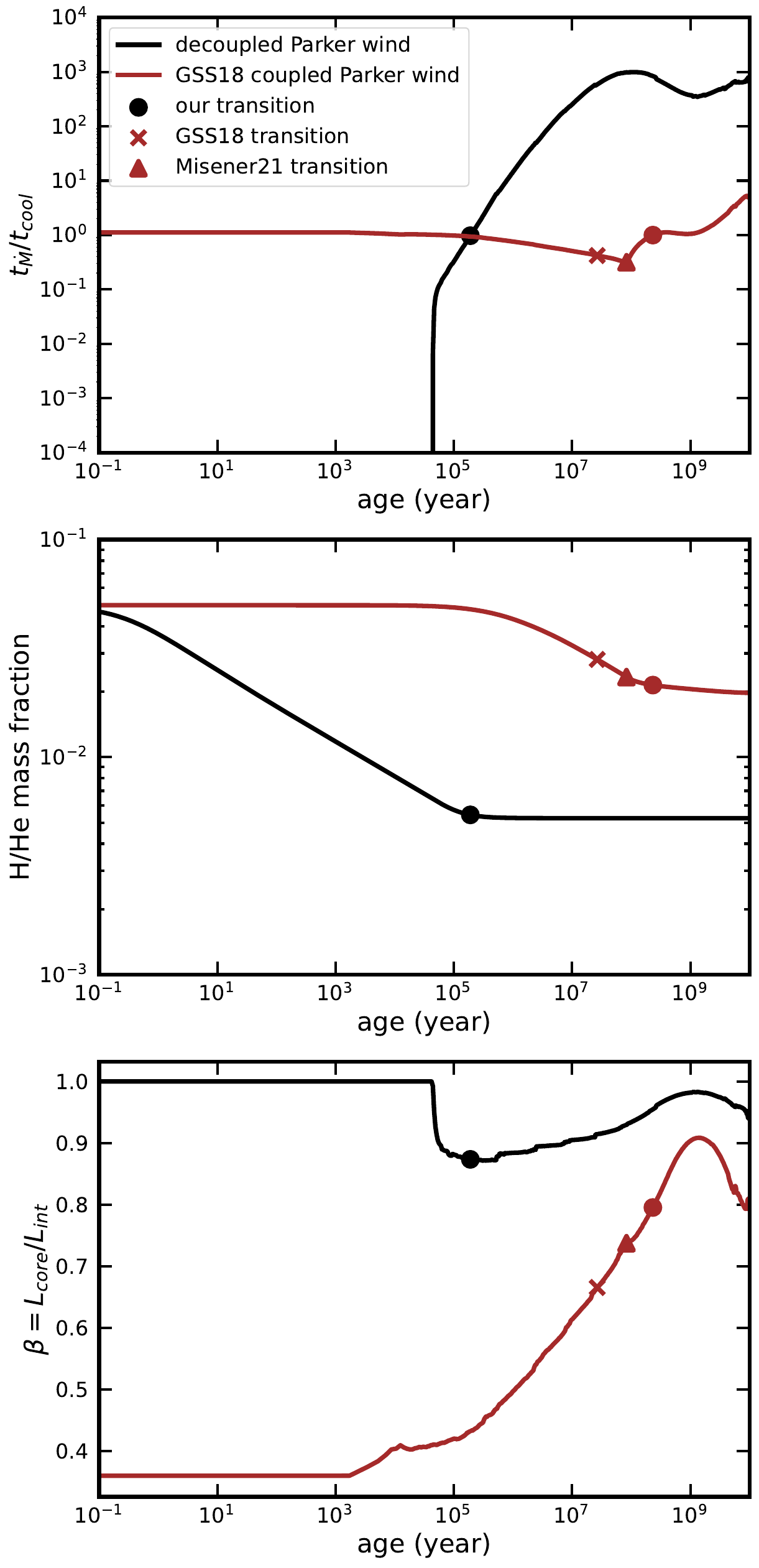} 
\caption{
We show the ratio between the mass loss and thermal evolution timescales (top), envelope mass fraction (middle), and the luminosity ratio between the total core cooling and the intrinsic envelope cooling (bottom) for both the GSS18 coupled Parker wind (red) and our decoupled Parker wind (black). This is computed from our numerical model for a $4 M_\oplus$ planet irradiated with $100 F_\oplus$, initially with 5\% of H/He mass. The top panel shows the linear behavior of the coupled wind and the exponential behavior of the decoupled wind. We compare our transition between boil-off and the long-lived Parker wind (circle) to the GSS18 transition (cross) and \citet{Misener21} transition (triangle). See Section \ref{subsec:coupled} and \ref{subsec:transition} for discussion.
}
\label{transition}
\end{figure}

In the top panel of Figure \ref{transition}, we show the difference in the timescale ratio $t_{\dot{M}}/t_{cool}$ between the linear behavior of the GSS18 coupled wind (red) and the exponential behavior (black) of the decoupled isothermal Parker winds. Note that for the decoupled Parker wind, the ratio only becomes unity (coupled) briefly in the correspondingly core-enhanced transition phase (close to the end of boil-off, see Section \ref{subsec:core}). It is otherwise not core-enhanced despite the greater $\beta$ during most times of the boil-off, as a result of the decoupling between the mass loss and thermal evolution. As a comparison, the GSS18 coupled Parker wind can be much more core-enhanced for a longer period of time, leading to a higher mass loss enhancement. This is because a planet is imposed by the wind assumption to not have a vigorous boil-off phase but instead experience a mild and significantly extended transition phase according to Eq. \ref{tmdot},\ref{transitionphase}, which would be much shorter and  more abrupt if assessed with what we believe is more appropriate decoupled Parker wind.

Since the energy-limited regime happens early in the boil-off phase due to the high energy demand, during which the intrinsic energy $L_{int}$ is argued to be mostly from the gravitational binding energy of the envelope rather than the core thermal energy reservoir \citep{Ginzburg2016,Misener21}, $\beta$ was considered negligible in the previous work in this mass loss phase. In contrast, in our numerical model using the same coupled Parker wind assumption, we find the core luminosity is always a significant fraction of the total cooling. For most planets $\beta$ is typically no greater than 0.9 before transitioning into the non-energy-limited (what they call Bondi-limited) regime, except for those planets nearing complete atmospheric stripping. Consequently, we find that the timescale ratio is typically of order unity during the energy-limited phase. Once the coupled wind transitions into the non-energy-limited phase, the timescale ratio starts to increase exponentially, driving a limited amount of time-integrated mass loss after then, which we find with our numerical model is at most a few $0.1\%$ of H/He mass. 

Compared to GSS18 (red in Figure \ref{mdot}) and \citet{Misener21}, which show a significant number of completely stripped planets, our numerical model (red curve of Figure \ref{convergence}) finds a much lower susceptibility for these planets to the GSS18 coupled wind. A similar pattern emerges when using the GSS18 model with the inclusion of the energy loss term $\dot{E}_{adv}$, as shown in Figure \ref{mdot}, where we find that the GSS18 coupled wind removes the envelope mass significantly slower (gray), compared to the original approach without $\dot{E}_{adv}$ (red). This underscores the importance of the $\dot{E}_{adv}$ term in the analytical model. We find a sub-Neptune is more frequently the final evolutionary product after a few Gyr. The divergent results from the original work is because their lack of $\dot{E}_{adv}$ term magnifies the importance of the core luminosity by thermally inflating planets, leading to an exaggerated demographic feature.

The above behavior is expected from Eq. \ref{timescales} because the mass loss process barely has enough time to entirely remove the envelope $t_{\dot{M}} \sim t_{cool} \sim t$, where $t$ is the planetary age. We find a complete loss of the envelope rarely happens and only happens when the core luminosity is very comparable to or greater than the intrinsic luminosity with $\beta \gtrsim 1$ (the constant radius found with the analytical model as a planet loses mass in Section \ref{subsec:shrink} corresponds to $\beta > 1$), which leads to a significant decrease of the timescale ratio. These stripped planets have very low core masses $<2 M_\oplus$ with low initial mass fractions $<5\%$ and high insolation $> 100 F_\oplus$. This suggests that the GSS18 coupled wind, when combined with improved thermal evolution physics, is not effective in forming a pronounced radius gap.

\begin{figure}
\centering
\includegraphics[width=0.45\textwidth]{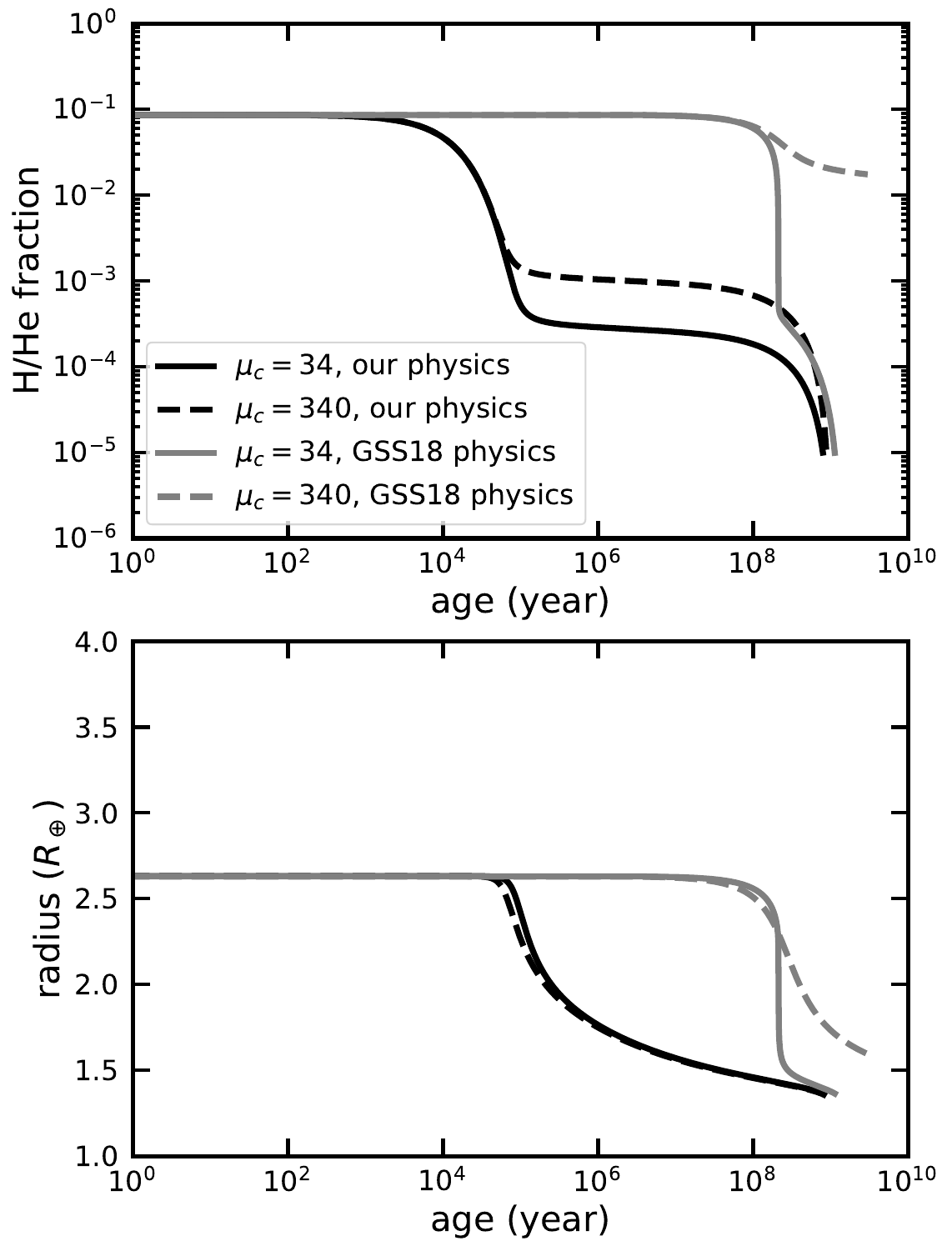} 
\caption{The evolution of radius and envelope mass fraction of 3 $M_\oplus$ models with a low core mean molecular weight $mu_c=34$ (solid, high core heat capacity) and a 10$\times$ times $mu_c$ (dashed,  negligible core heat capacity). The analytical models with the physical improvements (black) shows little impact from the change of the core heat capacity, similar to our numerical model. The models with the original GSS18 physics (gray) show a divergent fate of these planets with a large difference between the final interior H/He compositions.
}
\label{muc}
\end{figure}

In Figure \ref{muc}, we vary the mean molecular weight of the core $\mu_c$ in the analytical model, where $\mu_c$ defines the core heat capacity, from $34\ m_{H}$ (solid, which corresponds to a large core luminosity) to $340\ m_{H}$ (dashed, negligible core luminosity) based on the GSS18 initial conditions with (black) and without (gray) both the physical improvements, i.e. the decoupled Parker wind and the $\dot{E}_{adv}$ term. The planetary evolution with the improvements is only marginally affected by the thermal behavior of the core, which we find is not sensitive to different combinations of initial conditions. Similar to the results from our numerical model in Section \ref{subsec:core}, the core-enhanced mass loss effect only moderately impacts the transition phase around $10^5$ yr rather than the boil-off and the long-lived mass loss phases before and after then respectively (black curves). Consequently, this effect can only change the final mass fraction by a factor of order unity rather than altering the evolutionary fate of a planet. In contrast, without both improvements, the evolution with the GSS18 physics are very sensitive to core heat capacity (gray curves).


\subsection{Criteria for transition} 
\label{subsec:transition}
In GSS16, these authors emphasized that a planet switches to the long-lived core-powered escape from the boil-off phase once it reaches the thin regime with $R_{rcb} = 2R_c$. After that point the thermal contraction is argued to be greatly hindered by the core luminosity, which constitutes the majority of the envelope cooling, leading to a core-enhanced mass loss. Their criterion for the transition however is not self-consistent and is not consistent with our numerical results in Section \ref{subsec:nobolo}. We find the radius ratio at the end of boil-off, that is assumed to be a fixed number by their criterion, can vary greatly with different planetary parameters, i.e. core mass and stellar insolation, if assessed with what we believe is more appropriate decoupled Parker wind, tabulated in Table \ref{table:5}.

\citet{Misener21} used a different criterion, such that boil-off coincides with the energy-limited (coupled) regime whereas the ``core-powered'' mass loss exclusively refers to the Bondi-limited (decoupled) regime. Here we reevaluate these criteria with our numerical model. As their coupled wind assumption (see Section \ref{subsec:coupled}) essentially correspond to a prolonged $\sim$ 1 Gyr and gentle transition phase, both before and after the GSS18 transition $R_{rcb} = 2R_c$ (cross shapes in Figure \ref{transition}) as well as the \citet{Misener21} transition (triangle shapes), we find that the luminosity ratio $\beta$ does not change significantly, shown in the bottom panel, and therefore has a similar impact on the mass loss enhancement at both stages (see section \ref{subsec:core}) shown in the middle panel. This is in contrast with their argument that the core luminosity is only important after their transitions. We suggest that a good criterion for transitioning into a significantly core-enhanced phase is $\beta \sim 1$ (or with $1 - \beta \ll 1$ if a thermal inflation is not considered), which decreases the timescale ratio in Eq. \ref{timescales} by orders of magnitude leading to a significant change of the mass loss.  Since our findings show that $\beta$ hardly exceeds 1 with improved physics, we do not consider a need to distinguish between the core-powered phase and boil-off phase for their coupled wind.

Consequently, both the GSS18 and \citet{Misener21} criteria for the transition give rise to a large uncertainty for the definition of their long-lived core-powered mass loss (shown after the cross/triangle shapes and before the circle shape on the red curve), which we argue both should correspond to our decoupled Parker wind boil-off if the mass loss physics is improved, as shown in Section \ref{subsec:nobolo}. As a comparison, our criterion for the transition in Eq. \ref{transitiontime} is independent of different numerical model setup and planetary parameters, and reflects the characteristic mass loss timescales. According to our transition criterion, the flat sections of the H/He evolution in Figure \ref{mdot} and \ref{transition} account for the actual long-lived Parker wind phase.


\subsection{Need for self-consistent initial conditions}
\label{subsec:conditions}

Due to the coupling effect of the GSS18 wind, any initial conditions would lead to an initial timescale ratio $t_{\dot{M}}/t_{cool}$ very close to 1 (Eq. \ref{timescales}). In this case, the importance of the initial conditions could be easily overlooked. Moreover, as discussed in Section \ref{subsec:transition}, the uncertainty from the criteria for the transition used in the previous work further complicates the initial condition problem for the post-boil-off mass loss. Consequently, with these treatments, the role of boil-off and the long-lived escape are largely entangled. In Section \ref{subsec:radius} and \ref{subsec:fraction}, we show the consequence of the GSS18 initial mass fractions and initial radii. The improvement of the initial conditions is shown in Section \ref{subsec:self-consistent}.

\subsubsection{Initial RCB to core radius ratio is fixed} 
\label{subsec:radius}
With our numerical boil-off model, we determine the self-consistent radius ratio at the end of boil-off, rather than the fixed radius ratio used in GSS16, if assessed with the decoupled Parker wind and show that this ratio varies significantly with core mass and bolometric flux, shown in Table \ref{table:5}. Generally, we find that planets that are cooler and more massive retain more massive envelopes and a higher specific entropy after boil-off, resulting in larger RCB radii. This entropy effect is important as it pushes the RCB outward to a lower pressure, and hence, larger radii, especially combined with a cooler equilibrium temperature, which pushes the RCB outward further. 

Consequently, a fixed initial radius ratio of 2 in GSS18 makes massive $M_c > 5 M_\oplus$ and cold planets $F< 100 F_\oplus$ initially much more compact, making it even harder for those planets to lose mass, while for less massive and hotter planets the constant radius assumption would correspond to a vigorous decoupled boil-off phase. A similar criterion in \citet{Misener21}, would include part of the prolonged boil-off phase as their core-powered escape for low-mass $M_c \leq 5 M_\oplus$ and highly irradiated planets $F \geq 100 F_\oplus$, while heavy and cold planets barely evolve into the thin regime by a few Gyr with negligible mass loss.

Overall, the radius ratio assumption in GSS18 would result in a sharper transformation between sub-Neptunes and super-Earths, consistent with a prominent radius gap, while our model predicts a much more gentle transformation that is caused by boil-off alone as we will see in Section \ref{sec:pop}. 

\begin{deluxetable*}{c | c c c c c c c}
\tabletypesize{\scriptsize}
\tablewidth{0pt} 
\tablecaption{Ratio between the RCB radius and core radius at the end of boil-off\label{table:5}}
\tablehead{
\colhead{Flux ($F_\oplus$)} & \colhead{1.5 $M_\oplus$} & \colhead{2.4 $M_\oplus$} & \colhead{3.6 $M_\oplus$} & \colhead{5.5 $M_\oplus$} & \colhead{8.5 $M_\oplus$} & \colhead{13 $M_\oplus$} & \colhead{20 $M_\oplus$} 
} 
\startdata 
 1 & 2.71 & 3.86 & 5.42 & - & - & - & - \\              [0.5 ex]
 3 & 2.07 & 2.92 & 4.00 & 6.00 & - & - & - \\        [0.5 ex]
 10 & 1.54 & 2.15 & 2.91 & 4.19 & - & - & - \\        [0.5 ex]
 30 & 1.17 & 1.64 & 2.20 & 3.06 & 4.78 & - & - \\        [0.5 ex]
 100 & 1 & 1.23 & 1.63 & 2.22 & 3.21 & 5.30 & - \\[0.5 ex]
 300 & 1 & 1 & 1.25 & 1.68 & 2.35 & 3.50 & - \\[0.5 ex]
 1000 & 1 & 1 & 1 & 1.25 & 1.70 & 2.39 & 3.67 \\[1.5 ex]
\enddata
\tablecomments{ We present the ratio between the RCB radius and core radius at the end of boil-off based on our numerical model assessed with the decoupled Parker wind. The transition is defined as the time when the mass loss timescale equals to the thermal contraction timescale. As a comparison, GSS18 use a constant radius ratio of 2 as the transition condition, which overestimates the time-integrated mass loss of their core-powered escape for low-mass and highly irradiated planets and underestimates it for massive and cold planets.
}
\end{deluxetable*}

\subsubsection{Initial H/He mass fraction scaling function is flat} 
\label{subsec:fraction}
Since the intensity of an isothermal Parker wind is controlled not only by the planet radius but also the RCB density, the initial condition problem is complicated by the initial H/He mass fraction. In GSS16 and GSS18, these authors analytically derived an envelope mass fraction scaling function $f=0.05 M_c^{1/2}$, which they found is roughly $30-50\%$ of the initial mass fractions from planet formation. As a comparison, in Section \ref{subsec:nobolo} (Figure \ref{fraction_comparison}) with our numerical model, we demonstrated a dramatic variation in the final mass fraction at the end of boil-off. This exponential behavior contradicts with their flat power-law envelope mass fraction scaling function. This greatly overestimates the initial mass fraction for low-mass and highly irradiated planets and underestimates the fraction for massive and cold planets in GSS18. 

As a result, the combined effect from both the overestimated initial mass fraction and initial radius would correspond to boil-off initial conditions for low-mass $M_p \leq 4 M_\oplus$ and/or highly irradiated planets with $F \geq 100 F_\oplus$ , leading to an exaggerated amount of mass loss, which is presented as a long-lived core-powered escape in GSS18 (see also Section \ref{subsec:nobolo} for a similar finding based on our numerical model and Section \ref{subsec:transition}). This is evaluated directly based on the analytical model here. In Figure \ref{ic} we show evolution tracks for planets with the physics improved. In the left panels, the model with the original GSS18 initial conditions (red) overestimates both the post-boil-off radius and H/He mass fraction compared to the model with initial conditions that are self-consistently calculated from a boil-off phase (gray) for a $3 M_\oplus$ planet. On the other hand, the mass loss for a planet that is more massive and/or colder tends to be underestimated due to the underestimated initial radius and mass fraction, which however has a less significant consequence as these planets are less susceptible to mass loss. However, this yields different final mass fractions as shown in the right panels.  

\subsubsection{Self-consistent evolution} 
\label{subsec:self-consistent}
To self-consistently initialize the evolution, we investigate two approaches to modify the GSS18 analytical model. One called,  ``fully-analytical'' incorporates both boil-off and the post-boil-off bolometric-driven escape based on the analytical evolution model of GSS18 but with what we find are more appropriate initial conditions. The ``semi-analytical" approach starts the evolution at the transition time after boil-off, which uses the self-consistent post-boil-off conditions from our numerical model, followed by the GSS18 analytical subsequent evolution. 

\begin{figure}
\centering
\includegraphics[width=0.45\textwidth]{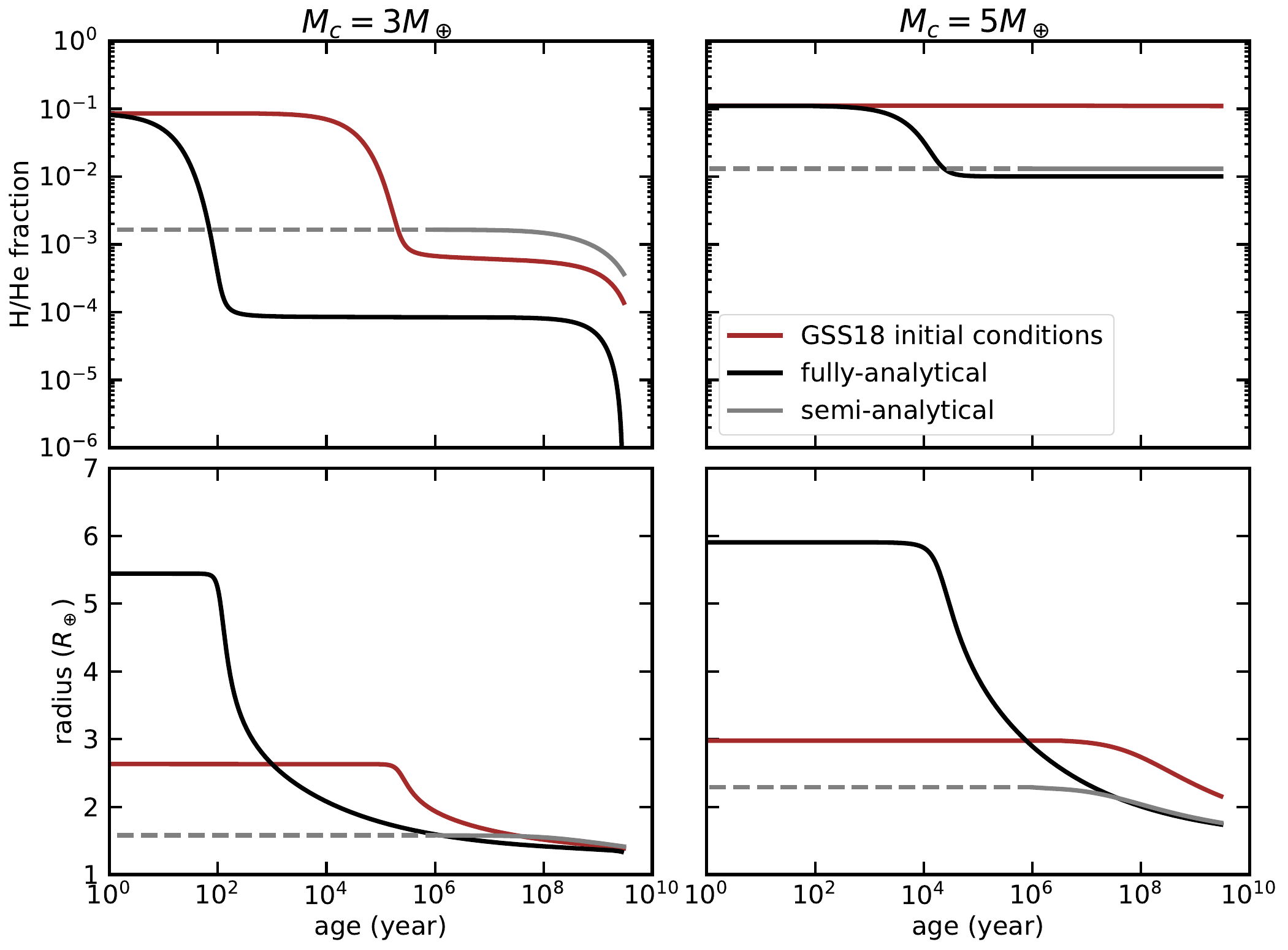} 
\caption{To examine how different initial conditions impact the evolution of low-mass sub-Neptunes, we show the evolution tracks of two planets with low mass (3 $M_\oplus$, left panels) and intermediate mass (5 $M_\oplus$, right panels) receiving $100 F_\oplus$ stellar radiation. We show three tracks with the only difference being how we initialize the evolution. All curves already have the relevant physics improved (i.e. a decoupled isothermal Parker wind with the mass loss driven energy lost term $\dot{E}_{adv}$ included). The black curve shows the evolution from the fully-analytical GSS18 approach, but now including a boil-off phase. The gray curve represents the evolution from the semi-analytical approach with the initial conditions calculated self-consistently from our numerical boil-off model (with the evolution before 1 Myr shown in dashed). 
See Section \ref{subsec:fraction} and \ref{subsec:self-consistent} for discussion.
}
\label{ic}
\end{figure}

To show how different approaches impact the evolution under the long-lived mass loss, in Figure \ref{ic} we compute the evolution of H/He mass fractions (top panels) and RCB radii (bottom panels) with both a low mass core ($3 M_\oplus$, left panels) and intermediate mass core ($5 M_\oplus$, right panels).  These figures show three evolution tracks, where the physics for all three approaches is improved based on the discussion in Section \ref{subsec:mdot} and \ref{subsec:edot}.  The evolution based on the GSS18 initial conditions is shown in red, which leads to a boil-off phase for the low-mass planet when we include the more appropriate decoupled Parker wind.  In solid black, we show the evolution under the fully-analytical approach. These models start right after the disk dispersal with the initial Kelvin-Helmholtz contraction time of 5 Myr. In solid gray, the semi-analytical approach starts from the beginning of the long-lived mass loss phase with an initial thermal contraction time of 50 Myr. Recall that the initial conditions for the semi-analytical model (gray) are after boil-off, hence it starts with a lower mass fraction from our numerical boil-off model (Table \ref{table:2}) and a colder and more contracted interior.

In general, we find that the semi-analytical model is most consistent with the behavior from our numerical model, which we take as what we believe is a more proper analytical approach to study the post-boil-off escape. This improves the ``core-powered escape'' in GSS18, which is a result from simplified physics and initial conditions. The fully-analytical approach (black), however, tends to overestimate the mass loss rate for boil-off especially for the low core mass cases, owing to the behavior of the analytical model that planet radii hardly shrink during boil-off (see Section \ref{subsec:shrink}), which leads to more mass loss. A better match between these two approaches is seen for more massive planets (as in the right panels of Figure \ref{ic}), where the H/He mass fractions and radii nearly converge. These two approaches are assessed on a population-level in Section \ref{sec:pop}. 

\subsection{Importance of radiative atmosphere and mean molecular weights}
\label{subsec:radiative}
\begin{figure}
\centering
\includegraphics[width=0.45\textwidth]{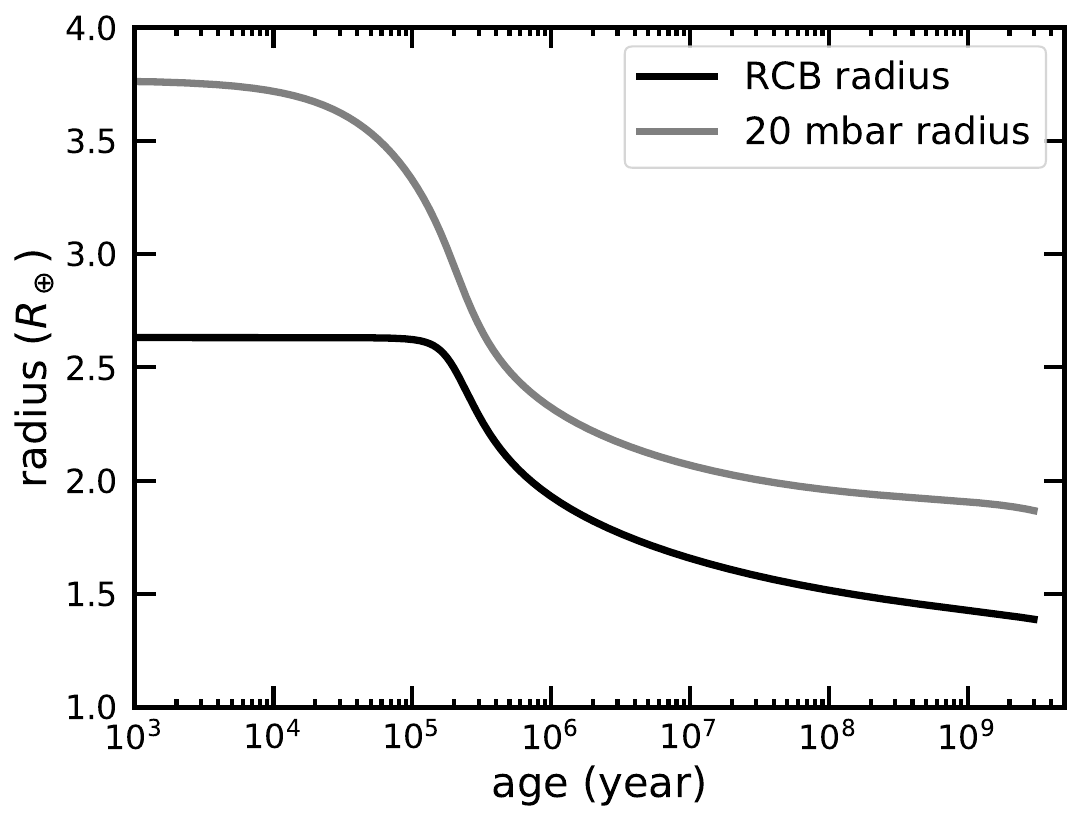} 
\caption{We show the importance of radiative atmosphere for a $3 M_\oplus$ planet that receives 100 $F_\oplus$ stellar bolometric flux initially with a 5\% H/He mass fraction, based on the GSS16 analytical model with the physics improved. The radiative atmosphere is attached on top of the RCB, which enlarges the planetary radius by a considerable fraction.
}
\label{20mbar}
\end{figure}

In the analytical work, the planetary radius is defined at the RCB with the additional radiative atmosphere above ignored in the structure calculation. However, for a low-gravity planet, its scale height is often a large fraction of its total radius, especially with the variable surface gravity effect taken into account, which further increases the scale height at a higher altitude. Moreover, a sub-Neptune can readily possess a deep RCB at old ages requiring a greater number of scale heights to reach its optical radius above the RCB. As a result, we suggest that its radiative atmosphere contributes up to 30-40\% to its total optical transit radius, far different from a giant planet where the thickness of the radiative atmosphere can be typically ignored in modeling work. The importance of the radiative atmosphere becomes more marked for highly irradiated and low-mass planets. In Figure \ref{20mbar}, we directly compare the RCB radius to the radius at 20 mbar, a typical optical transit radius. The difference in radius is significant. Without the radiative atmosphere, the location of the sub-Neptune population on an occurrence diagram shifts to the smaller radius end.

\begin{figure}
\centering
\includegraphics[width=0.45\textwidth]{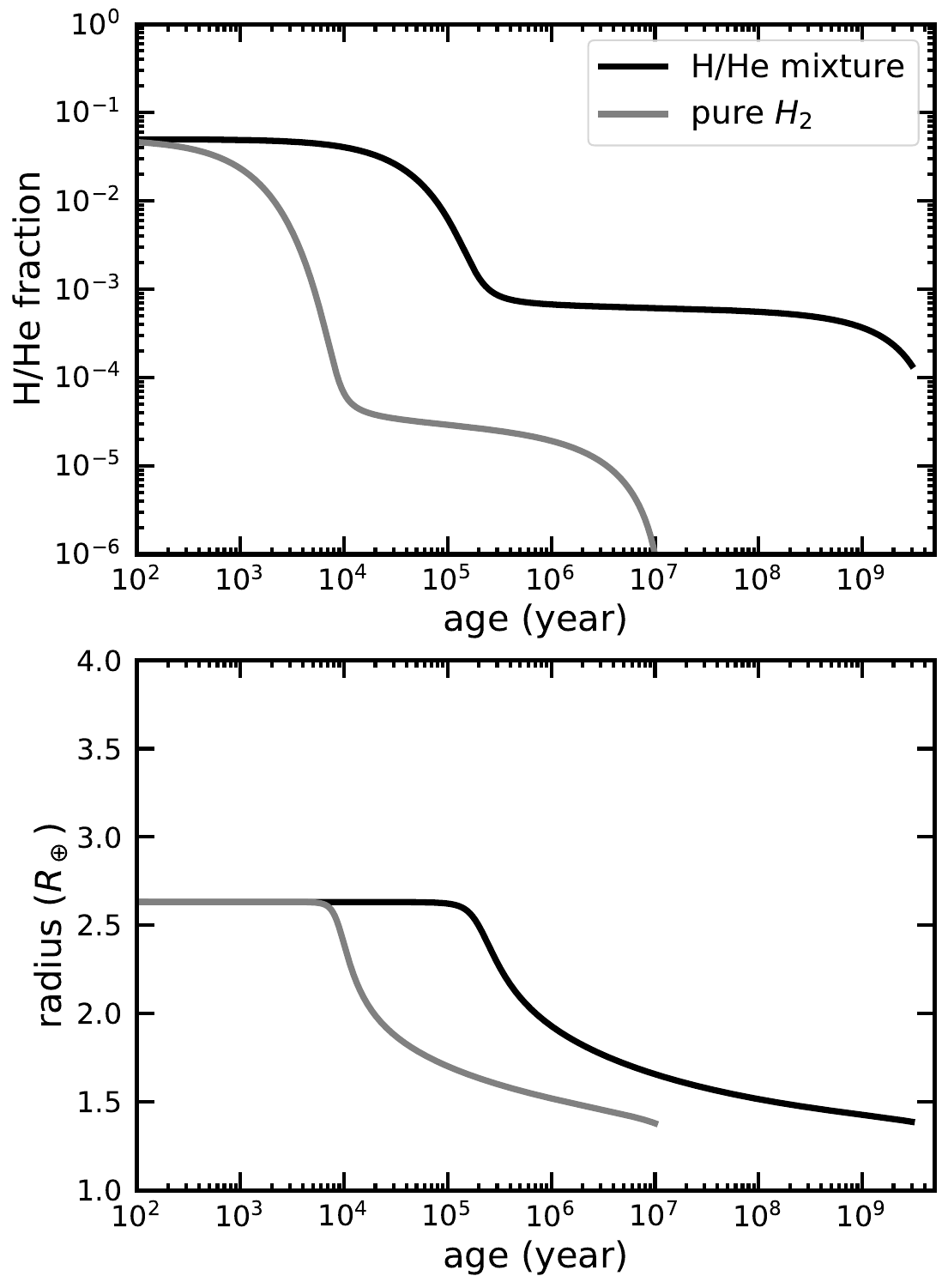} 
\caption{The evolution of radius and envelope mass fraction between a H/He ($black$) and a pure $H_2$ ($gray$) atmosphere, for our nominal planets that have 3 $M_\oplus$ cores with 5\% initial mass fractions and 100 $F_\oplus$ insolations both with the physics improved. A pure $H_2$ atmosphere assumed in GSS18 is much more susceptible to the isothermal Parker wind as discussed in the text. The plot is based on the GSS18 analytical model but a similar effect also affects other models with the isothermal Parker wind including our numerical model.
}
\label{molecular}
\end{figure}

The isothermal Parker winds from both our numerical model and the analytical model are sensitive to the choice of the mean molecular weight of the H/He atmosphere. For a less metal-rich atmosphere, yielding a smaller $\mu$, the sonic point $R_s = GM_p \mu/2k_bT_{eq}$ penetrates deeper into higher density regions of the radiative atmosphere.  With the smaller $R_s$, the density there $\rho_s$ increases exponentially, which actually leads to a more vigorous mass loss $\dot{M} = 4 \pi R_s^2 \rho_s c_s$ for the more metal-poor planet. This effect is especially marked when the planet is still young and inflated with a large scale height. In the GSS18 work, a pure H$_2$ atmosphere was assumed ($\mu=2\ m_{H})$, which we find is much more susceptible to mass loss compared to our default H/He atmosphere with $\mu=2.35\ m_{H}$.  This is shown in Figure \ref{molecular}, in terms of the mass loss timescale. Perhaps surprisingly, the pure $H_2$ atmosphere is completely stripped at least two orders of magnitudes faster, which could result in a more pronounced radius gap. 

\section{Population-level comparison}
\label{sec:pop}
\subsection{Need for a population-level study}
\label{subsec:pop_dist}
Based on the GSS18 analytical model, a range of population-level work \citep{Ginzburg18,Gupta19,Gupta20,Gupta21,Gupta22} has been done with similar physical treatments and initial conditions. With certain choices for core mass and equilibrium temperature distributions, they reproduce a radius valley and radius cliff very similar to that from observations. However, as we discussed above, with a number of physics and initial conditions improved, the final population from the analytical model can be potentially impacted. In this section, the planet occurrence rate distributions from these improved versions of the analytical evolution model are produced and directly compared to our numerical model on a population level, to reassess the importance of the post-boil-off bolometric-driven escape or ``core-powered escape.''

To evaluate how different approaches impact the demographics of small planets, we run a population of 10,000 randomly generated planets following the same equilibrium temperature distribution used in GSS18 that has a constant probability density $dN/d\log{T_{eq}}$ for $T_{eq}$ in the range of 500--1000 K and a power-law of $T_{eq}^{-6}$ within 1000--2000 K. For the core mass distribution, a variety of distribution functions are evaluated. The simplest one is a Rayleigh distribution that peaks at $3M_\oplus$:
\begin{equation}
\label{rayleigh}
\frac{dN}{dM_c} \propto M_c \exp{\left( -\frac{M_c^2}{2\sigma_M^2} \right)}
\end{equation}
with $\sigma_M=3M_\oplus$, which truncates with a negligible probability density beyond $10 M_\oplus$. In GSS18, the authors also find that substituting a flat core mass function below $5M_\oplus$ also yields a similar final distribution. To match observations, these authors find that a high mass tail of core masses that extends out to 100 $M_\oplus$ in addition to the Rayleigh distribution is needed to replicate the ``radius cliff" feature. With this core mass and equilibrium temperature function, we compute the final mass fractions and planetary radii at 3 Gyr by different approaches, namely our numerical model-based approach, the fully-analytical approach and the semi-analytical approach. The consequence of a variety of physics choices will also be evaluated in this section.

\subsection{Analytical core-powered distribution with improved physics and initial conditions}
First of all, we implemented the GSS18 analytical evolution model with the same physics, initial conditions and planet distribution functions. To validate our implementation, a single evolution curve was directly compared to the \citet{Gupta19} version of the analytical model, which has very slight differences between this version and that in GSS18 (Gupta, pers. comm.).  Our implementation yields identical evolution tracks. To further gain confidence, we ran a simple test distribution with 10,000 planet models with the same setup as in GSS18 for $3 M_\oplus$ planets with $T_{eq}$ = 1000 K, producing a final distribution that is a clear match to Figure 1 in GSS18. In this test run, the initial envelope mass fraction follows an even distribution ranging between $10^{-5}$ and 100\%, the same as in GSS18

\begin{figure}
\centering
\includegraphics[width=0.45\textwidth]{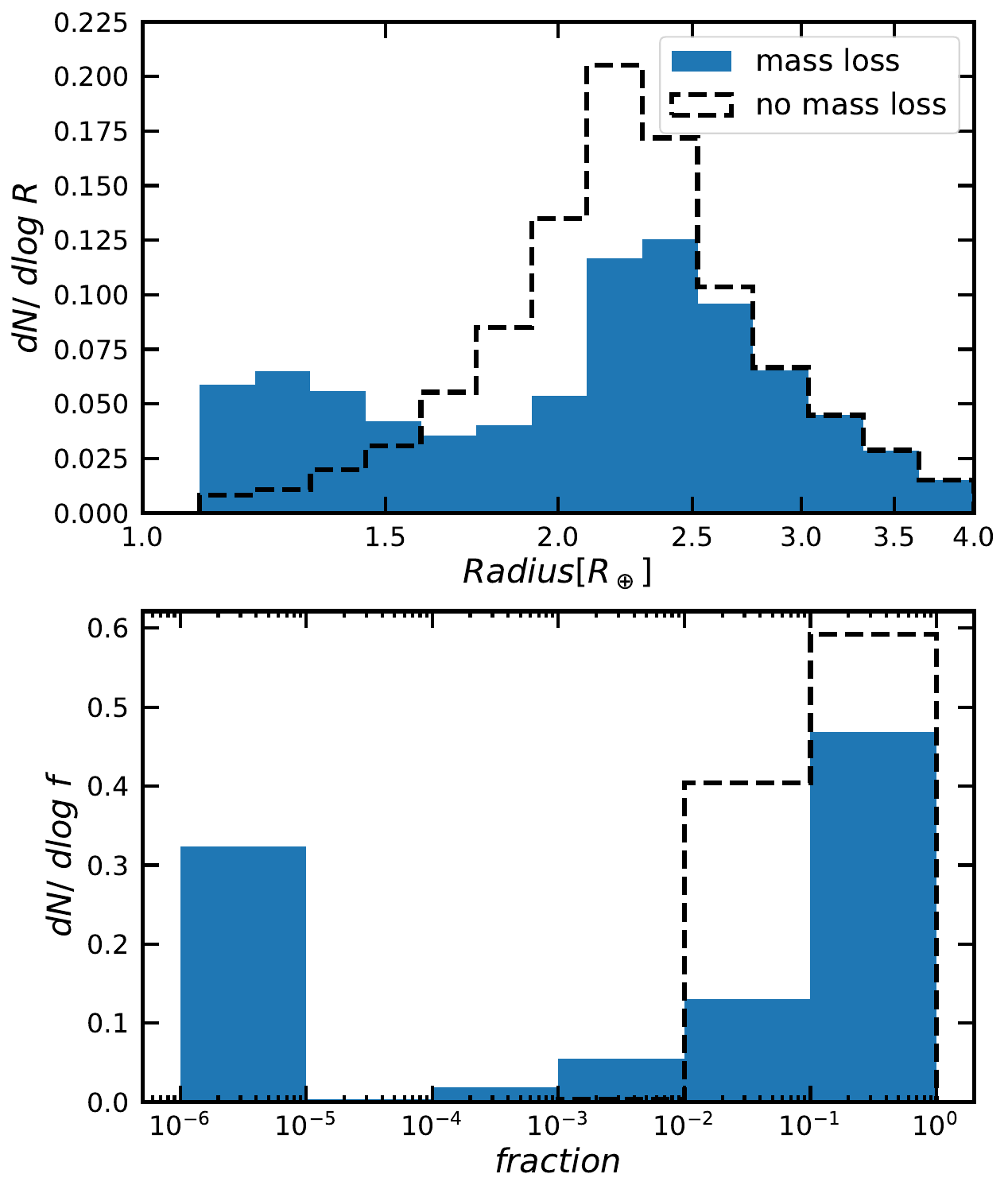} 
\caption{From our implementation of the analytical model, we show the distribution of the final planet radius and envelope mass fraction with ($blue$) and without ($black$) mass loss using the same original GSS18 analytical setup. A similar bimodal distribution is replicated.
}
\label{original}
\end{figure}

For a more comprehensive population-level comparison, using the same physical choices and the planet distribution functions as documented in Section \ref{subsec:pop_dist} and GSS18, our version of the analytical model reproduces a highly similar (but not identical) bimodal distribution to that from GSS18. The planet occurrence rate as a function of planet radii and H/He mass fractions with and without the mass loss is shown in Figure \ref{original}. Compared to their Figure 3 (the grey bars), our distribution shows a slightly higher sub-Neptune population occurrence in the range of $2.0-3.5 M_\oplus$ while the GSS18 distribution exhibits two nearly even peaks. Since the original model is not publicly available, the observed differences may stem from specific details in the realizations of the core mass distribution and equilibrium temperature distribution that we do not have access to.

\begin{figure}
\centering
\includegraphics[width=0.45\textwidth]{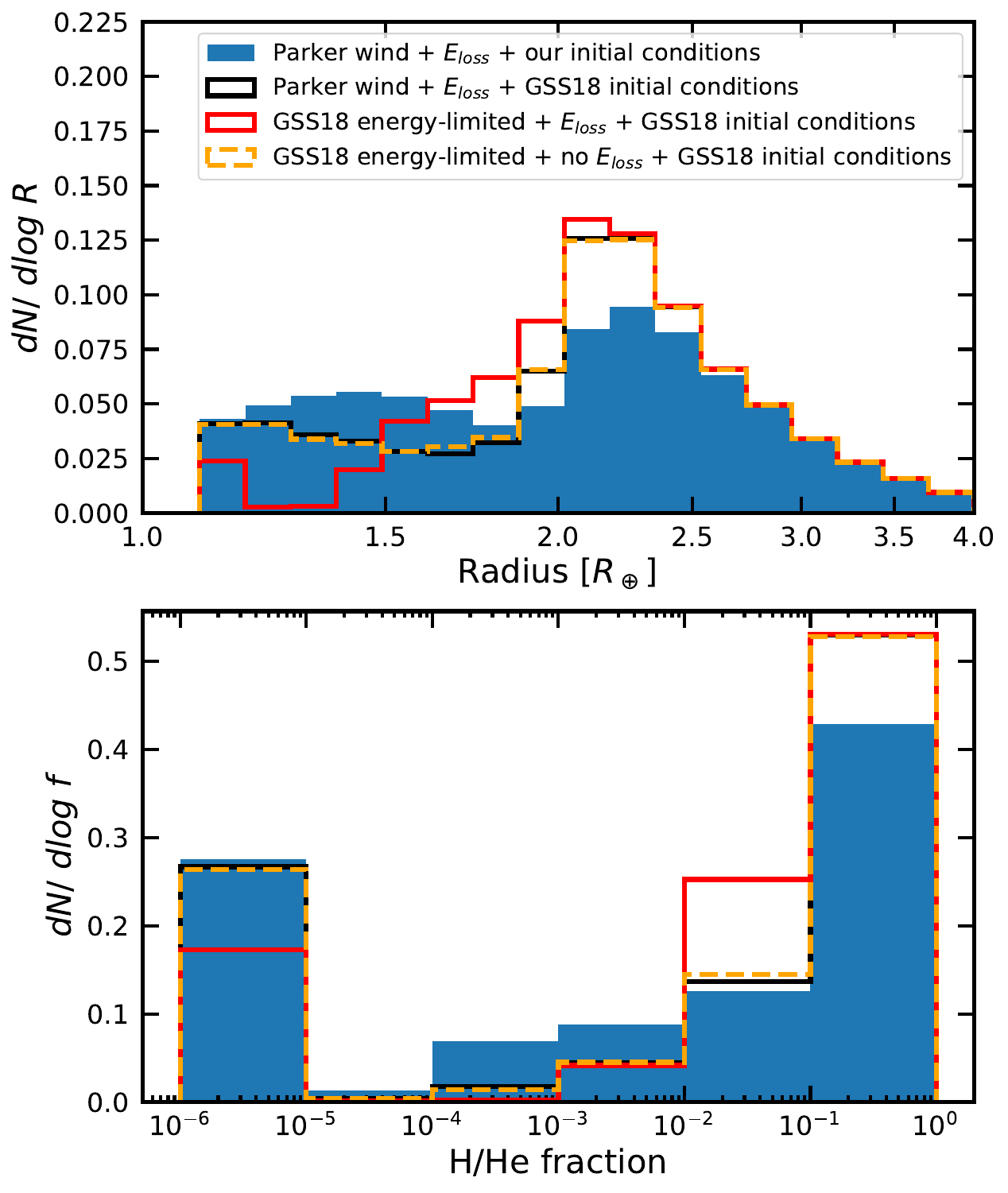} 
\caption{The final distributions of the planet radiii and H/He mass fractions with a range of different physics choices used in the analytical model. By including the $\dot{E}_{adv}$ term that is missing from the GSS18 model (yellow) with all other conditions the same, there is no sign of a radius gap in the range of $1.5-2.0 R_\oplus$ (red). A decoupled isothermal Parker wind with the $\dot{E}_{adv}$ term (black) yields a very similar final distribution to the original one (yellow), but with a dramatic shift in the mass loss timescale (Figure \ref{mdot}). The blue distribution corresponds to the fully-analytical model that additionally has our boil-off initial conditions, showing a bimodal distribution but with a less prominent radius gap.
}
\label{physics}
\end{figure}

After having gained confidence in our implementation, we then step through a number of physical effects that we believe could be improved, as discussed in Section \ref{sec:diff}. With the same planet distribution functions employed as in GSS18, the computed final distributions of planetary radii and H/He mass fractions with different combinations of physics are shown in Figure \ref{physics}. For all of the population-level comparison runs, we always use a mean molecular weight of the atmosphere for a H/He mixture with $\mu=2.35\ m_{H}$ unless specified otherwise. The planetary radius is defined at the RCB, following GSS18. 

Our version of the original GSS18 setup is in yellow. Next, we add in the $\dot{E}_{adv}$ energy loss term that is missing from the GSS18 model, leading to a single peak around $2-2.5 M_\oplus$ without a radius gap feature (shown in red). This demonstrated that the GSS18 energy-limited mass loss (``core-powered escape") cannot form a radius gap with what we believe is more appropriate thermal evolution, which was suggested earlier in Eq. \ref{timescales}, Section \ref{subsec:coupled}, and Figure \ref{mdot}). Then, with the $\dot{E}_{adv}$ term, we switch to a decoupled isothermal Parker wind (the non-energy-limited wind unless otherwise stated) instead of the GSS18 coupled wind, which interestingly yields a distribution (black) nearly identical to that from the original setup in yellow. However, a significant timescale difference is seen on an evolutionary level in Figure \ref{mdot}, with the mass loss timescale shifted to very young ages ($\sim$ 1 Myr), corresponding to the boil-off timescale, with negligible mass-loss thereafter.

Lastly, with the revised physics, we address the initial condition problem for the post-boil-off mass loss phase by self-consistently including a boil-off phase in the analytical evolution model, which starts with a larger initial planetary radius and faster initial K-H contraction timescale of 5 Myr. This blue histogram corresponds to our fully-analytical approach, which we find leads to fewer planets with massive envelopes $>10\%$, due to boil-off. The corresponding mass fractions are shown in the lower panel. Although a pileup of blue planets in the range of $1.2-1.8 R_\oplus$ is caused by boil-off, this radius grouping has more planets than in the original yellow model, due to the larger sample of sub-Neptunes in the $10^{-4}$ to $10^{-2}$ envelope range that never lose their envelopes, rather than making a larger super-Earth population (envelope-free) compared the yellow model.   

\subsection{Comparing different approaches}
\label{subsec:comparison}
As discussed in Section \ref{subsec:shrink}, we find that the GSS18 analytical model yields a RCB radius that barely decreases throughout the whole boil-off phase, overestimating the mass loss rate for low-mass planets resulting in a deeper radius gap. The nearly constant RCB radius, however, is not seen in our numerical model. Moreover, since the total energy available for cooling estimated in Eq. \ref{Ecool} is only valid when the thickness of the envelope is comparable to or smaller than the physical size of the core, extrapolating beyond this regime may lead to an evolutionary uncertainty. Therefore, the fully-analytical model may not appropriately model the population.

To overcome the downside of the fully-analytical approach, we investigated a semi-analytical approach as described in Section \ref{subsec:self-consistent}. We set the initial mass fractions by interpolating in the post-boil-off results from our numerical boil-off model (Table \ref{table:2}). We then start the analytical evolution at the beginning of the long-lived bolometric-driven escape phase with boil-off properly incorporated.  The initial radius is set by imposing an initial Kelvin-Helmholtz thermal contraction timescale of 50 Myr, which is the post-boil-off conditions found in Section \ref{subsec:finalfraction}.  

Additionally, we ran a grid of evolution calculations based on our numerical model with self-consistent physical conditions incorporating both boil-off and the long-lived escape, and record the final mass fractions and 20 mbar optical radii at 3 Gyr as a function of planet core masses and bolometric fluxes. We then interpolate within the grid of evolution models for those 10,000 randomly generated planets to compute the final distributions, which represents our fully numerical approach.

\begin{figure}
\centering
\includegraphics[width=0.45\textwidth]{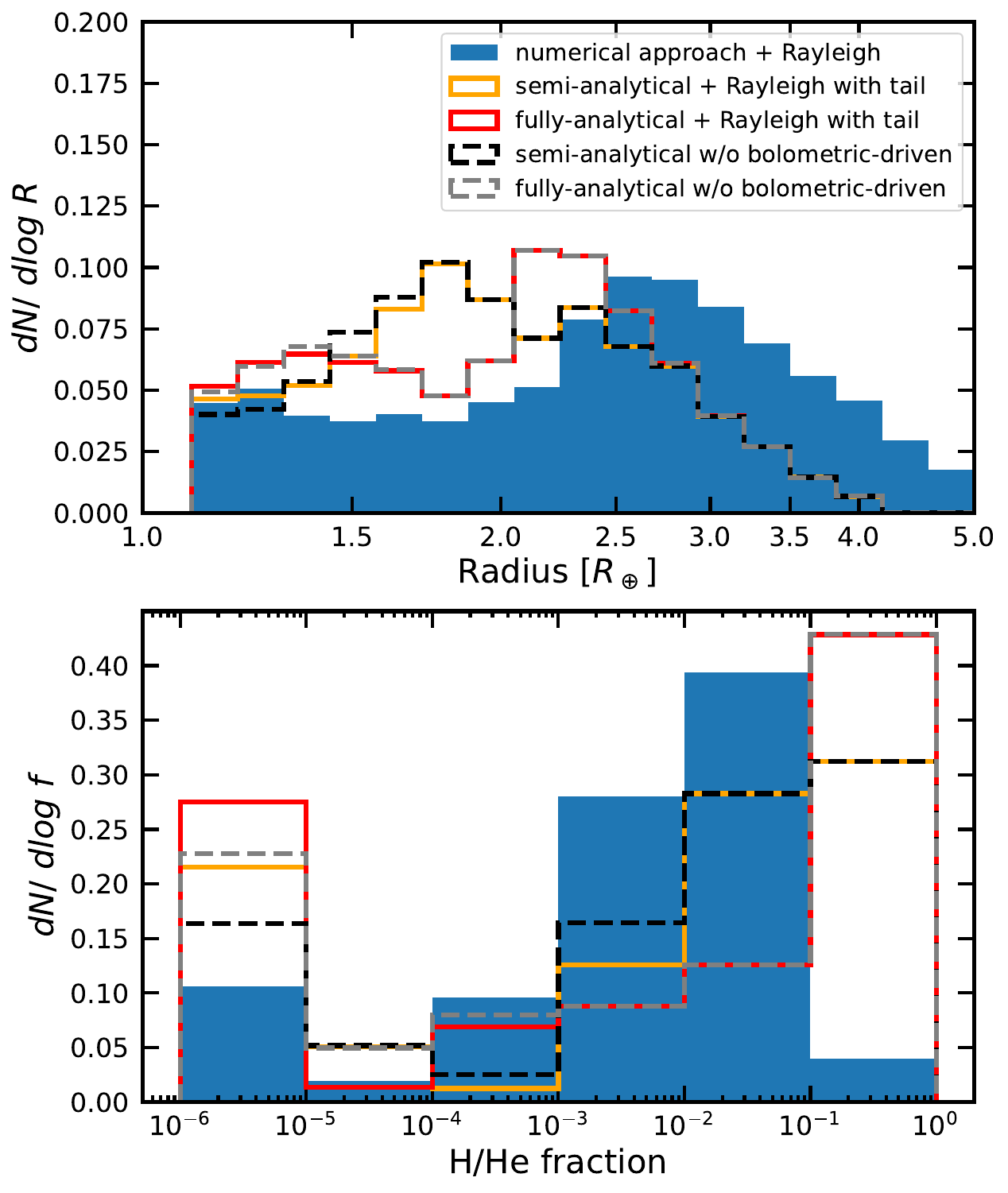} 
\caption{The distributions of the final planetary radius and envelope mass fraction by 3 Gyr from the models discussed in the text. The fully numerical model with a Rayleigh core mass distribution in blue shows a flat distribution in $<2.0 M_\oplus$ with no significant radius gap. The semi-analytical model in yellow uses initial conditions from our numerical boil-off calculation and the GSS18 core mass distribution that has a high mass tail.  It does not exhibit a radius gap but instead a peak at $1.5-2.0 M_{\oplus}$. The fully-analytical model that calculates both the boil-off and long-lived mass loss (bolometric-driven) phases self-consistently is shown in red. As a comparison, similar model setups but with the long-lived escape turned off are shown in dashed black and dashed gray for both analytical and semi-analytical models, showing no significant long-lived mass loss.  All of the analytical models have the radius defined at the RCB, while the numerical model has a radiative atmosphere.
}
\label{numerical}
\end{figure}

In Figure \ref{numerical}, we compare the distributions of the final planet radii and envelope mass fractions at 3 Gyr from our fully numerical model (blue), the fully-analytical approach (red, which was the blue distribution in Figure \ref{physics}) and the semi-analytical approach (black). Note that our numerical model employs a different core mass function, which does not have the high mass tail of GSS18 but a simpler Rayleigh distribution. As a comparison, we also show the final distributions without any long-lived escape for both the semi-analytical approach (yellow) and the fully-analytical approach (gray), finding no significant changes compared to the black and red distribution with the long-lived wind, suggesting its insignificance. 

Importantly, we find no clear radius gap between $1.5-2 R_\oplus$ for the numerical distribution but a flat distribution down to $1 R_\oplus$, which we argue mostly results from boil-off. The planets that are on the left of the peak, in the range of $1.0-2.0 M_\oplus$, however, are a mixture of low-mass planets with very little H/He mass and bare cores (super Earths) with a relatively smaller super-Earth population compared to the analytical models. Furthermore, the analytical approaches have a much greater population with high initial mass fractions $>10\%$ largely contributing to the sub-Neptune population and constituting the radius cliff, which is a result of too many heavy planets $>8 M_\oplus$ concentrated in the high mass tail of the core mass function, as most of them remain intact after boil-off. 

Additionally, the numerical model without the high mass tail for the core mass distribution exhibits a cutoff at $5 R_\oplus$, shifted to a higher radius compared to the radius cliff feature in \citet{Fulton18}. This requires other mass loss mechanisms to shrink planetary radii further, which suggests a role for XUV-driven escape. The radius cutoff we found is not a result of the massive planets $>10 M_\oplus$, as is needed in the analytical models, but a consequence of our numerical boil-off. Lastly, we find most sub-Neptunes have an intermediate H/He mass fraction in the range of $10^{-3}-10^{-1}$ after boil-off and by 3 Gyr. 

Also in Figure \ref{numerical}, we find the semi-analytical approach (black) cannot form a radius gap but a peak around $1.5-2.0 R_\oplus$. It produces a H/He mass fraction distribution more similar to the numerical one, which has a larger population of planets that have intermediate H/He mass fraction $10^{-3}-10^{-1}$ compared to that from the fully-analytical model. This population of planets corresponds to the peak at $1.5-2.0 R_\oplus$. As a comparison, the semi-analytical approach still has the feature from the massive planet $>10 M_\oplus$ tail, which forms another minor peak around $2.5 M_\oplus$. However, we argue that the peak around $1.5-2.0 R_\oplus$ accounts for the sub-Neptune peak around $2.0-2.5 R_\oplus$ from observations \citet{Fulton18} but is shifted to a smaller radius by about 30 \% due to the lack of radiative atmospheres in the analytical model. Lastly, below $1.5 M_\oplus$ the super-Earths do not form a clear peak, but a gentle slope.

In summary, each of our models suggests that there is no significant long-lived bolometric-driven escape (``core-powered escape"). Boil-off is a powerful mass loss process that greatly sculpts the planet population (blue) somewhat into the shape that we see today but it alone is not able to reproduce the exact shapes of the observational features including the deep radius gap and the more contracted radius cliff. Other mass loss mechanisms such as XUV-driven escape are required to further shrink planet radii, deepen the radius gap, and shift the radius cliff to smaller sizes. Since the planets that are right beyond the observed radius cliff in the range $4-5 R_\oplus$ are sub-Saturn planets with more massive core masses $> 5 M_\oplus$, their mass loss is only subject to strong XUV heating \citep{Hallatt22}. We expect the hot (with an orbital period $\sim$ a few 10 days) and less massive $<10 M_\oplus$ planets to eventually shift below the radius cliff, with the cold and more massive objects constituting the observed sub-Saturn population, which can be readily tested in future studies.

\subsection{Core mass distribution effect and radiative atmosphere}
In Section \ref{subsec:radiative} and Section \ref{subsec:comparison}, we show the importance of the radiative atmosphere and the role of the core mass distribution. In this section, we evaluate the impact on the population distribution from these two choices.

In order to replicate the radius cliff, the cutoff in the frequency of planets with  radii above $4.0 R_\oplus$, GSS18 used a power-law high mass tail for the core mass distribution with a relatively large population of core masses $> 5 M_\oplus$, concatenated to the right of a Rayleigh core mass distribution. However, in our numerical model, we find that a $20 M_\oplus$ core by 10 Gyr has a large radius far beyond the ``radius cliff." Moreover, these massive planets that are modeled remain unaffected after boil-off with a massive H/He envelope $>15\%$. These larger-radius planets should be regarded as Neptune-class or larger and therefore excluded from a small planet population study. 

\begin{figure}
\centering
\includegraphics[width=0.45\textwidth]{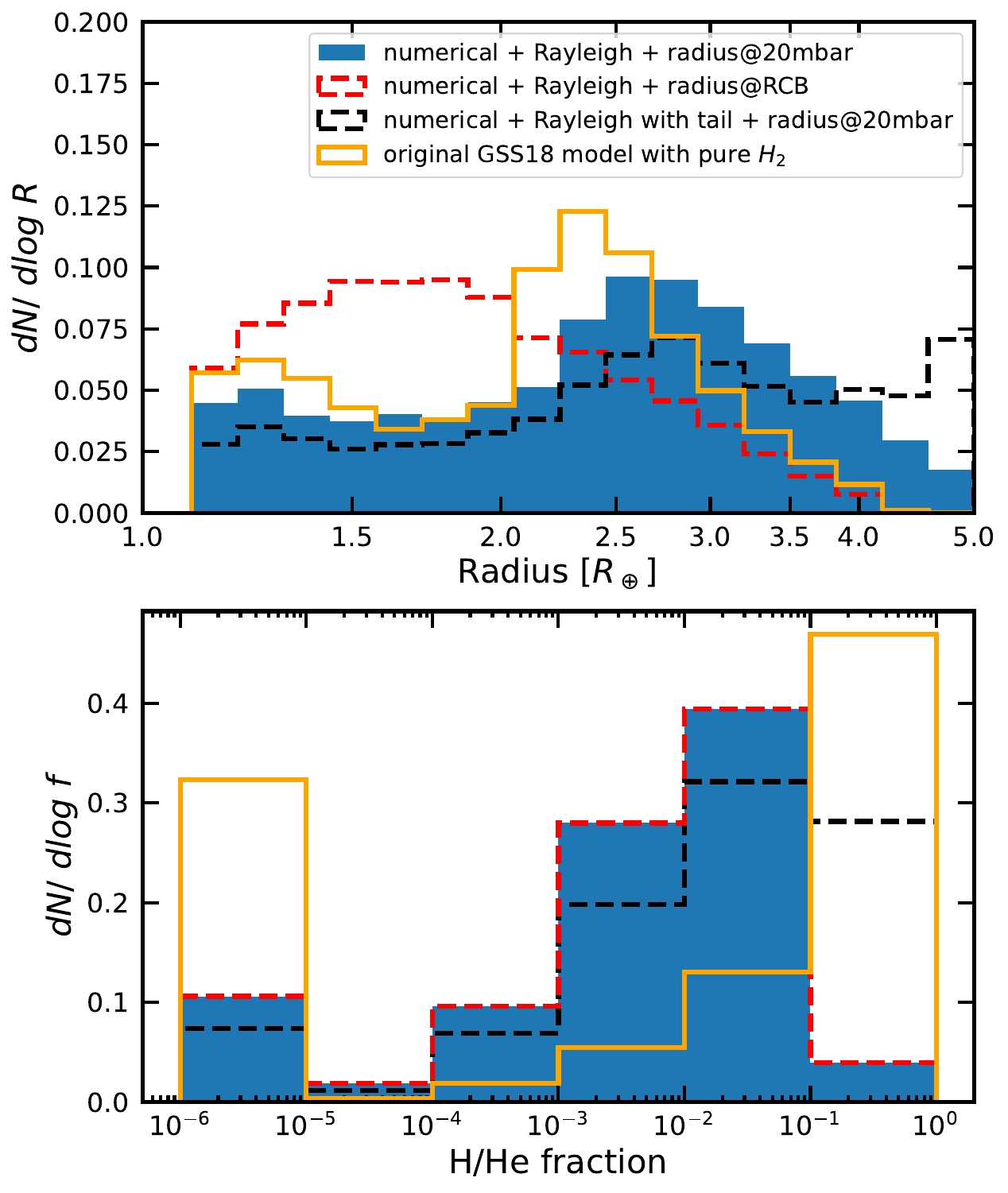} 
\caption{To demonstrate the importance of the radiative atmosphere, we show a comparison between the distribution of 20 mbar radii in blue (the same blue distribution in Figure \ref{numerical}) and RCB radii in red. Without a radiative atmosphere, the red model has a radius peak that is greatly shifted to a much smaller radius at around $1.5-2.0 R_\oplus$, without any radius gap feature, which is reminiscent of the semi-analytical model in Figure \ref{numerical}. The radius cutoff also shrinks by 20\%, coincidentally similar to that from the analytical model that has a different core mass distribution and no radiative atmosphere (orange, with GSS18 setup). We then illustrate the core mass distribution effect by incorporating the high mass tail into our numerical model (black), which leads to an unrealistic flat distribution that extends outside $5 M_\oplus$. 
}
\label{core_radiative}
\end{figure}

In Figure \ref{core_radiative}, we show the distributions from our numerical model in blue and the original GSS18 version of the analytical model in yellow. Our numerical model does not need the high mass tail of core masses to present a radius cliff feature. On the contrary, including the high-mass tail in our numerical setup (shown in black) fails to replicate a radius cliff and unrealistically extends a flat radius distribution beyond $5R_\oplus$, which suggests that the massive planets with H/He mass fraction $>10\%$ are unnecessary in our model while it is important to the analytical models. Additionally, we demonstrate that the radiative atmosphere is a key to properly determining the shape of the planet distribution, due to its significant contribution of the physical size of a low-mass planet (see section \ref{subsec:radiative}). As a comparison, we show a distribution of the RCB radii without the radiative atmosphere calculation based on the blue numerical model, which is shown in red. The radius peak shifts significantly to a smaller radius at $1.5-2.0 M_\oplus$. This peak location is consistent with that from the semi-analytical model that does not a have radiative atmosphere (see black in Figure \ref{numerical}). It exhibits a similar cutoff at $4 R_\oplus$ as also seen in the analytical models (yellow) but without the need for the high mass tail of core masses in the numerical work. 

\begin{figure}
\centering
\includegraphics[width=0.45\textwidth]{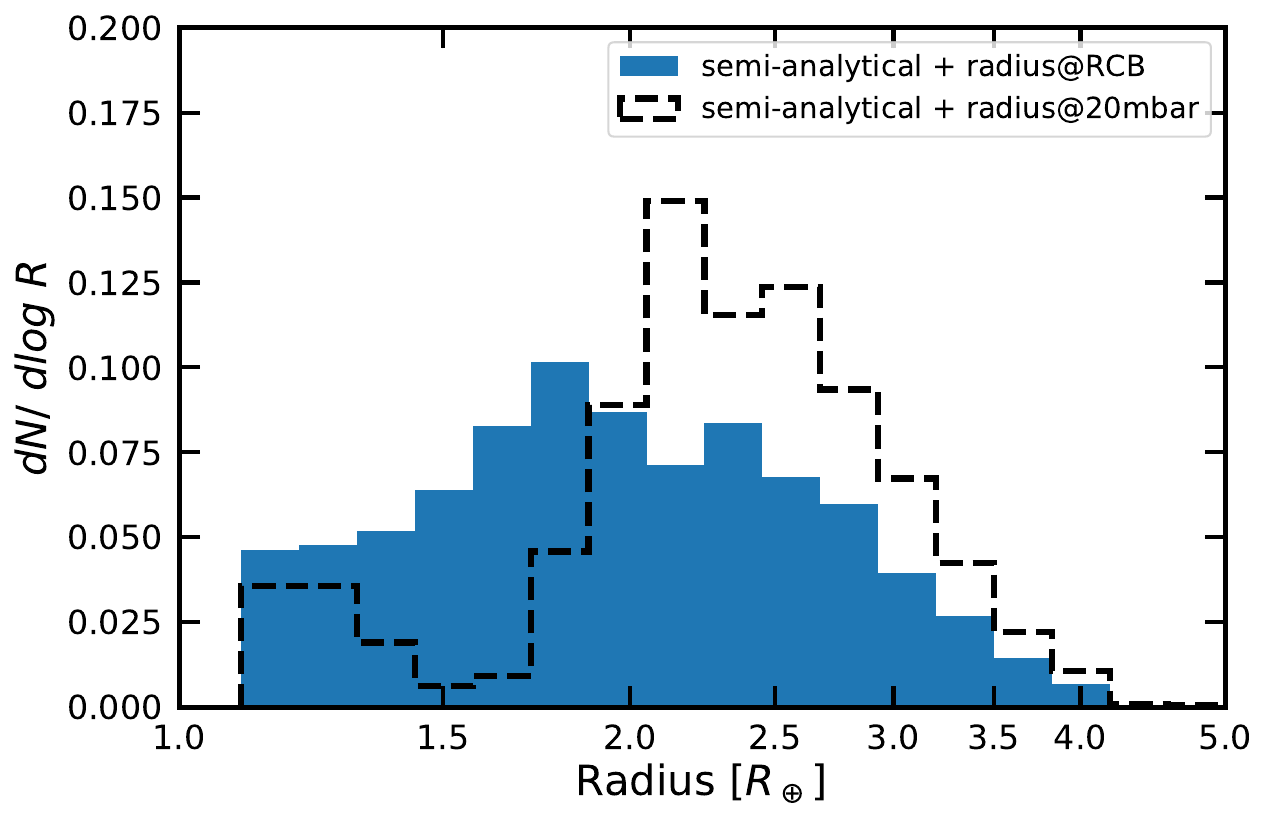} 
\caption{We show the importance of the radiative atmosphere for the GSS18 analytical model by attaching a layer of isothermal radiative atmosphere on top of the RCB and calculating the distribution of 20 mbar radii in black. As a comparison, the radius distribution without it is shown in blue, the same as the black in Figure \ref{numerical}, which exhibits a shape far from the observations. We find that the existence of the radiative atmosphere enlarges the planet radii for low-mass planets by 20-30\%, which shifts the peaks from $1.5-2.0 R_\oplus$ to a more reasonable $2.0-2.5 R_\oplus$. 
}
\label{analytical_radiative}
\end{figure}

Lastly, in Figure \ref{analytical_radiative}, we include the radiative atmosphere in the semi-analytical model (black) and compare the distributions to the one without it (blue, the same as black in Figure \ref{numerical}). We find that with the radiative atmosphere the peak shifts to a reasonable location at a larger radius with a radius gap feature shown in the range of $1.3-1.7 R_\oplus$, although the envelope-free population $<1.3 M_\oplus$ seems small compared to the observation. This is different from our numerical results that show a flat distribution (blue in Figure \ref{numerical}). The discrepancy is likely caused by the fact that the semi-analytical model lacks the population with intermediate envelope mass $10^{-4}-10^{-2}$, which in comparison smooths out our numerical distribution. Therefore, the improved semi-analytical model indicates that boil-off alone can not reproduce the observed radius gap, consistent with our numerical results. Another mass loss mechanism is required to further deepen the gap and shape the distribution into the one from observations. 

\section{Discussion} 
\label{sec:discussion}
\subsection{Need for coupling current model to a state-of-the-art hydrodynamic wind model}
In our model, the boil-off atmospheric escape is assumed to be a non-energy-limited isothermal Parker wind at the planet's equilibrium temperature throughout the entire radiative atmosphere (see Eq. \ref{parkerwind}). As we discuss in Section \ref{subsec:boil}, in the case of an energy deficiency, the internal energy of the wind that is quickly dissipated by the adiabatic expansion would lead to a cooler and lower density deep atmosphere (below $\tau_h$), and would therefore hinder the hydrodynamic outflow. To tackle this problem in our model, the energy-limited prescriptions in Eq. \ref{mdot_bol} and \ref{mdot_int} are employed to assess the mass loss rate, where the ratio of energy being turned into usable work is assumed to be 100\%. Although it is shown to not significantly impact the post-boil-off mass fraction and therefore the later evolution after boil-off, this might introduce uncertainty into the early stage evolution.  This calls for a real hydrodynamic calculation to further study this issue.

Moreover, in reality, the wind is not completely isothermal as we assumed, such that the modestly warm deeper atmosphere has more bolometric energy deposited whereas the optically thin region is a cooler isotherm known as the ``skin temperature" region. Especially at a late age when the mass loss rate has been greatly reduced, the wind base is pushed to a higher altitude, mostly located in the colder ``skin temperature" region, resulting in a less efficient wind. This requires a detailed energy transfer model integrated into a hydrodynamic code to capture the physics discussed. 

\subsection{Need for more sophisticated core thermal evolution}
\label{disc:core}
A wide range of sub-Neptune evolution models have assumed a simplified isothermal core that is set to have the same temperature as the bottom of the adiabatic envelope \citep{Nettelmann11,Lopez12,Ginzburg2016} over the entire evolution, which we term a coupled core evolution, which corresponds to a perfectly conducting core. This simplification, however, can leads to a large uncertainty 
for the core evolution as it may lead to different core cooling timescales \citep{Vazan18a}, which can potentially impact our model results. This is another assumption that needs more attention and we discuss three points below.

First, we discuss the consequence of our coupled core evolution. In reality, the core energy transport is limited by convection in the silicate outer layer (the geophysical ``mantle'') instead of conduction. The efficiency of energy transport depends on the physical conditions at the thermal boundary layer at the top of the mantle, e.g. the temperature contrast between the core temperature $T_c$ and the temperature at the bottom of the envelope $T_b$, as well as the mantle viscosity. Therefore, our current coupled core evolution can potentially overestimate the core luminosity by instantly releasing all of the core thermal energy for cooling available, especially when the temperature at the bottom of the envelope $T_b$ rapidly declines as a results of the vigorous boil-off mass loss. Note that such an overestimation typically does not affect our boil-off evolution, as the core luminosity can hardly impact the mass loss process due to the decoupling between the thermal evolution and the mass loss as demonstrated in Section \ref{subsec:core}. 

Second, in the numerical model, we do not allow the core luminosity $L_{core}$to exceed the intrinsic luminosity $L_{int}$ (see Section \ref{subsec:thermal}, \ref{subsec:core}), which may underestimate its role of thermally inflating a planet, potentially impacting the boil-off mass loss.  This impact would typically occur in the transition phase when the core luminosity can thermally inflate a planet faster than the mass loss decreases planetary radius in Eq. \ref{transitionphase} and with $L_{core} > L_{int}$ (i.e. $\beta > 1$). This would couple the mass loss with the thermal inflation, leading to more mass loss. We find such thermal inflation can potentially affect a low-mass $\leq 4 M_\oplus$ and highly irradiated $\geq 100 F_\oplus$ planet, which are the ones more vulnerable to boil-off. To capture this behavior requires more sophisticated core evolution physics in future work, with a proper choice of the decoupled Parker wind as we suggested in section \ref{subsec:boil} because of the coupling behavior. Additionally, we argue that the thermal inflation can only last for a brief period of time during boil-off, as vigorous mantle convection rapidly brings down the temperature contrast, terminating the thermal inflation. With the mantle convection considered, we expect a factor of order unity change of the final mass fraction after boil-off, rather than meaningfully affecting the evolutionary fate of a planets (e.g. super-Earths or sub-Neptunes) that would otherwise correspond to an exponential change of the final mass fraction. Our treatment in this work therefore does not qualitatively impact our numerical results.
 

Third, if the initial core temperature is higher than that at the bottom of the envelope at the beginning of boil-off ($\sim$ 8,000 K), compared to what we assumed, this leads to a large thermal energy reservoir to cool and potentially more mass loss during the core-enhanced transition phase (see Section \ref{subsec:core}). We assess the consequence of using such a potentially underestimated initial core temperature (dictated by the initial temperature at the base of the envelope), by starting the evolution with a higher core temperature assessed with a decoupled core evolution such that $T_c \geq T_b$. We evaluate the maximum possible mass loss enhancement with the highest core temperature $T_{c,max}$ that is theoretically allowed, such that the thermal energy $kT_{c,max}$ is comparable to the gravitational binding energy of the core $GM_c/R_c$:
\begin{equation}
kT_{c,max} = \frac{GM_c}{R_c}
\end{equation}
which yields $T_{c,max} \sim 18,000 K$ (Such a scenario assumes no core cooling or energy loss during planet formation and corresponds to an initial Bondi radius as small as the core size.) 

\begin{figure}
\centering
\includegraphics[width=0.45\textwidth]{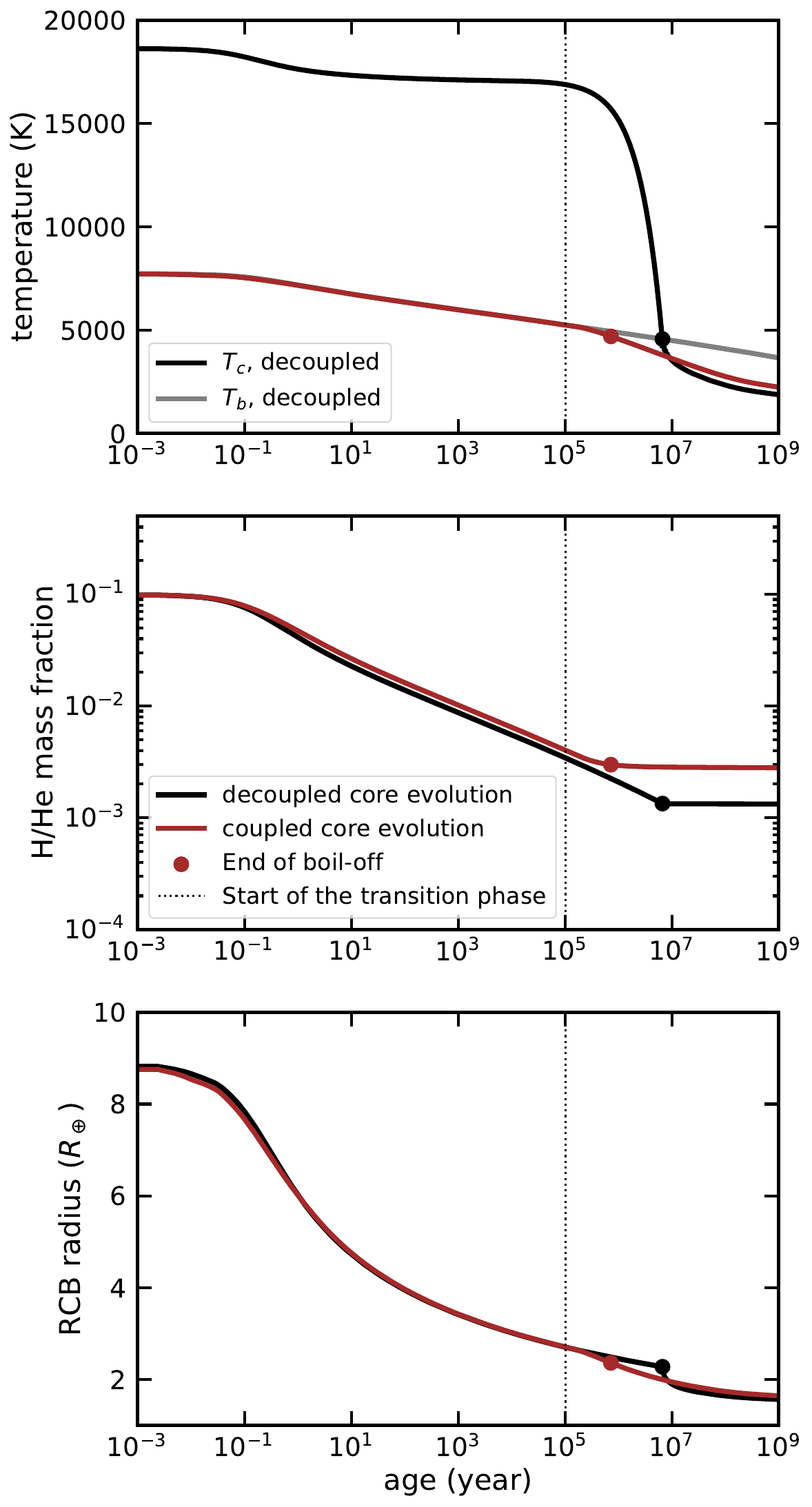} 
\caption{We show the temperature of the isothermal core $T_c$ and at the bottom of the envelope $T_b$ (top panel), H/He mass fraction (middle) and RCB radius (bottom), in the presence of a coupled (red) and decoupled core evolution (black). This is for a planet with a $3.6 M_\oplus$ core initially with 10\% envelope mass receiving $100 F_\oplus$ stellar flux. The decoupled core evolution ceases once the core temperature drops to match the temperature at the bottom of the envelope, leading to a coupled subsequent core evolution. In the top panel, the core temperature (black) under the decoupled evolution declines quickly and connects to the temperature at the bottom of the envelope (gray) within the transition phase (which starts after the dotted line) until the boil-off is ended (with circles) and therefore switches to a coupled core evolution. The prolonged transition phase of 10 Myr, which is regarded as the core-enhanced part of the late boil-off, leads to a slight enhancement of the mass loss, after which the long-lived Parker wind is inefficient due to the efficient thermal cooling and contraction, similar to the behavior of red. 
}
\label{core_temperature}
\end{figure}

In Figure \ref{core_temperature}, we examine a decoupled core evolution (black) with a $T_{c,max}$ start and show the evolution of temperature, H/He mass fraction, and the RCB radius. The results are compared to the coupled core evolution (red) that starts with the lower core temperature assessed at the base of the envelope. To decouple the core evolution without implementing a convective/conduction transport scheme within the core (which is beyond the scope of this work), and to maximize the core-enhanced mass loss effect without thermally inflating a planet and therefore the total mass loss, we assume that the core heating keeps the envelope at constant specific entropy, and what would be a constant planet radius, meaning any radius evolution is due only to boil-off mass loss. Under this assumption, the envelope is unable to thermally contract until the core temperature equals that at the bottom of the envelope (after which the core evolution transitions to being coupled). In this case, the core luminosity simply equals to the envelope cooling luminosity $L_{core} = L_{env}$. 

In the presence of such an extreme initial core temperature, we find the transition phase (which starts after the dotted black, assessed with Eq. \ref{tmdot}, \ref{transitionphase}) between boil-off and the long-lived Parker wind is prolonged to at most 10 Myr. In the top panel, we show that before the transition phase, the core temperature remains nearly constant, essentially meaning negligible core-enhanced effect. The core temperature under the decoupled core evolution quickly declines and converges back to the temperature at the bottom of the envelope (gray) within the prolonged transition phase. This leads to a coupled core evolution thereafter with only a small amount ($\sim 0.1\%$) of mass loss enhancement shown in the middle panel. As a comparison, our fully coupled evolution for the same model planet has a negligible mass loss enhancement within the transition phase (middle) and a faster contraction rate (bottom) within 1-10 Myr as it exits the boil-off phase (the transition times are denoted with circle shapes, see Eq. \ref{tmdot}, \ref{KHtime_core}) due to the thermal contraction.  We note that such a mass loss enhancement under the decoupled evolution and high initial core temperature is even smaller for a more massive planet and for a planet with a lower core mass as boil-off would already remove its entire envelope. A similar behavior applies to a colder or hotter planet, respectively. Therefore, we argue such decoupled core evolution with a much higher core temperature does not significantly affect our numerical model results.

In future work, a more detailed treatment of energy transport and core cooling is needed to better understand the interior and evolution of sub-Neptunes \citep{Vazan18b}. For instance, an inner iron core could be partially liquid or solid, and could crystallize over time.  Similarly, parts of the rocky mantle could be liquid or solid, necessitating convective and or conductive energy transport.  Furthermore, there could be heat production from crystallization as well as decompression as a result of boil-off. These processes can be better modeled in the future. 

\subsection{Relation to other core-powered mass loss work}
In a wide range of work, with mutually similar initial conditions and physical assumptions \citep{Ginzburg18,Gupta19,Gupta20,Gupta21,Gupta22}, a significant role for core-powered mass loss was found. We find that the results are in contrast with our model because their initial conditions would correspond to a boil-off condition, which was taken as core-powered mass loss. With an improvement of initial conditions, including a boil-off phase to self-consistently calculate the initial conditions for core-powered mass loss, and interior physics, \citet{Misener21} still found an efficient long-lived core-powered mass loss phase, as a result of both the wind assumption, which is similar to that from GSS18 (Eq. \ref{mdot_core}), and the lack of the $\dot{E}_{adv}$ term. Those treatments makes their mass loss sensitive to the core luminosity and lead to an overestimation of the mass loss, discussed in Section \ref{subsec:coupled}. 

In \citet{Rogers24}, this key assumption for the core-powered mass loss, however, was not assumed. But they instead used a similar non-energy-limited (decoupled) isothermal Parker wind assumption for both boil-off and core-powered mass loss with similar initial entropies to those from our work. The energy source for the Parker wind was argued, without a detailed validation, to be from the interior cooling, which ultimately releases the gravitational binding energy when a planet contracts. However, this must correspond to a cooling timescale comparable to the mass loss timescale, which we have shown to be unlikely to happen with these initial conditions (Section \ref{subsec:entropycooling}). In Figure 5 and 6 of \citet{Rogers24}, they still found a ``core-powered'' mass loss phase. However, this phase was shown to be very short-lived $\ll$ 1 Myr. 

According to these authors, the ``core-powered" phase results in a mass loss enhancement by a factor of two (as seen in the top panel of their Figure 5) due to the core luminosity. This effect is more significant than what our model predicts but less impactful compared to other studies on core-powered mass loss that used the GSS18 energy-limited wind model.

We argue, however, that the reported difference is not solely due to core luminosity. The reasons are as follows: (1) The core-enhanced mass loss effect should occur when the core luminosity maintains a larger planetary radius compared to the model without it. However, the planetary radii for the two model planets, with and without core luminosity, remain identical throughout the mass loss phase, as shown in their bottom panel. (2) The second panel from the bottom indicates that the cooling timescale is much longer than the mass loss timescale, suggesting that thermal evolution is inefficient, and thus the core luminosity is decoupled from the mass loss process. (3) The initial mass fractions between the two models differs by a few percent despite having the same initial radii. This suggests that the two planets have different initial envelope entropies, with the model including core luminosity being initially hotter. We suggest that the difference in final mass fraction is due to the initial entropy effect, as demonstrated in our work, rather than a core-enhanced effect. (4) Similarly, at the end of boil-off, these two planets should have the same entropy due to the decoupling effect if they start with the same initial entropy. Our model predicts that, with the same planetary parameters, their 2\% model (without core luminosity) should be at least 20\% larger in radius than the 1\% model (with core luminosity). However, they found identical radii for both planets, implying different entropies. With the initial entropy properly set to the same value, we would expect a much smaller mass loss enhancement in their model’s predictions.

Therefore, we consider this mass loss phase as part of boil-off, which we termed the transition phase, due to this insensitive behavior to the core luminosity. More importantly, the previous work argued that core-powered mass loss can be separated from boil-off by two other characteristics, i.e. the energy to lift H/He mass and the relation to the thermal evolution. From these aspects, as demonstrated and reevaluated in Section \ref{subsec:boil}, we do not think such a mass loss phase as found by \citet{Rogers24} should be interpreted as core-powered mass loss, given their non-energy-limited wind assumption.

In spite of differences in the treatment of physics (i.e., core cooling, disk dispersal, and accretion) used in \citet{Rogers24}, they found qualitatively similar results to our model that are modestly different in the quantitative details. First, they used a different core luminosity formulation than ours, with a constant core cooling timescale over time to parameterize the core luminosity. Following the discussion in Section \ref{disc:core}, a different core luminosity treatment did not qualitatively affect our evolution in boil-off, given the decoupling between the thermal evolution and mass loss.

Their model has a real disk dispersal phase and accretion phase, compared to our model, which we argue only slightly impacts our model results. In fact, a different disk dispersal time in their Figure 4 only changes the final mass fraction linearly, compared to the exponential behavior of the boil-off mass loss. The mass loss is controlled more by the initial entropy if the disk dispersal timescale is shorter than 1 Myr. Our initial entropy treatment yields a very comparable boil-off duration of 1 Myr and initial radius (entropy), to those from their work. In Section \ref{subsec:entropycooling}, we showed that our boil-off model is not sensitive to the initial mass fraction, and therefore the compositional (but not thermal) evolution history of the accretion phase is largely erased for a planet that has experienced a boil-off phase. Therefore, we find that the uncertainty from these different treatments is small.

\subsection{Internal energy advection in the deep radiative atmosphere}
\label{disc:adv}
Another physical process that needs additional attention is that of the internal energy advection. In Section \ref{subsec:envelope}, we suggested that the energy transport through the advective bulk motion is a consequence of the mass loss when the interior material is transported from a deep part of the envelope adiabat to a shallower region, which is eventually converted into the gravitational energy for the lifted material, instead of leading to an interior cooling. Even if such a cooling is assumed,  we have demonstrated that it does not impact boil-off in Section \ref{subsec:entropycooling}. Here we discuss its potential role as an energy source to power the outflow in the radiative atmosphere.

This advective energy transport is completely inefficient in the assumed isothermal outflow in the radiative atmosphere. However, in reality, the deep radiative atmosphere is not completely isothermal, which gradually transitions from the adiabatic envelope T-P profile, rather than the sharp transition between the adiabatic profile and the isothermal radiative atmospheric profile that is assumed in our model. The deepest part of the escaping radiative atmosphere below the critical optical depth to visible photons $\tau_h$, where it requires an additional energy source rather than the stellar heating, can be possibly located on a temperature gradient. In this case, the energy advection by the outflow can potentially play a role in draining the local internal energy to power the wind in this bottleneck region. Note that the energy advection in this case does not cool the interior as well. The efficiency of the advective energy transport depends on the temperature gradient between the RCB radius, where the H/He material is taken from, and the $\tau_h$ radius. Note that the discussion here is relevant only if $\tau_h$ occurs above the RCB in radius, which we find commonly happens to super-Earth mass planets $> 1 M_\oplus$.

This effect would assist in the atmospheric escape in the bottleneck region. Therefore, given the total amount of the internal energy advection and the intrinsic cooling energy is sufficient in the deep atmosphere to fuel the wind, the bottleneck of the boil-off mass loss then is the total stellar energy available to the radiative atmosphere (what we call bolometric-limited, Eq. \ref{mdot_bol}), instead of being intrinsic luminosity limited (Eq. \ref{mdot_int}). Understanding the relative importance of the internal energy advection and the radiative intrinsic cooling in the deep radiative atmosphere below $\tau_h$ requires a coupling between a radiative-transfer model and a hydrodynamic wind model.

In our model, our treatments of the temperature profile of the radiative atmosphere and RCB always correspond to an intrinsic-limited boil-off that converges to common final physical conditions that are found at the end of boil-off. The subsequent long-term evolution is therefore unaffected if we switch between the different decoupled wind assumptions as discussed in Section \ref{subsec:parker}. Consequently, our discussion above does not quantitatively affect our results.

\subsection{Simulating a population}
We have shown that the decoupled isothermal Parker wind starts to transition into the XUV-driven phase right after boil-off, when the planet starts to efficiently thermally contract.  At this same time the post-boil-off bolometric-driven escape comes to be inefficient and be completely quenched shortly later. As discussed above (Section \ref{subsec:comparison}), boil-off alone generates a planet occurrence distribution that is somewhat similar to, but cannot match the observed radius valley. A better match requires XUV-driven escape to further deepen the gap and shrink planetary sizes. 

Moreover, in the next few Myr after the transition time, there should exist a period when the post-boil-off Parker wind and XUV-driven escape are comparably efficient. In this phase the hydrodynamic wind is launched by the absorption of intrinsic cooling energy from the deep atmosphere and then stellar bolometric energy above $\tau_h$, and the wind is accelerated further and therefore the mass loss is enhanced by the heat deposited in the form of photoionizing high-energy flux at nbar pressure. Remarkably, such an enhanced isothermal Parker wind is coupled to the thermal evolution. We propose that these physical processes including boil-off, photoevaporation, and the enhanced escape should be further studied together, with a real hydrodynamic code and more complex energy transfer models in the core.  This should also be integrated into the planetary thermal evolution framework in future generations of sub-Neptune evolution models. 

With all of the improvements and physics discussed above included, one could generate a population of sub-Neptune models. The modeled demographics of the formed sub-Neptunes and super-Earths can be compared to the observed population and shed light on main physical processes sculpting their evolution.


\section{Conclusions} \label{sec:conclusion}
We developed a new one-dimensional python-based sub-Neptune interior and evolution model that incorporates both boil-off and core-powered escape, which allows us to assess core-powered escape with self-consistent physical conditions. We find that when considered outside of the boil-off phase, (the long-lived) core-powered escape is not able to drive significant mass loss more than 0.1\% of envelope mass fraction over 10 Gyr. The energy source for both boil-off and core-powered escape are reevaluated. The GSS18 analytical model is reproduced and compared to our numerical model, as a means to identify the physical assumptions that cause differences in results. Here are a number of key takeaways that we summarize from our study:
\begin{enumerate}
   \item{
   Boil-off and core-powered escape are primarily powered by stellar bolometric energy deposited in the radiative atmosphere above a certain optical depth to visible photons $\tau_h$. The energy and mass re-equilibration in the envelope is driven by convection, which does not require additional energy input. Instead, the energy source to overcome the gravitational force is the envelope internal energy that is released as a result of the mass loss. However, the energy supply of the deep radiative atmosphere where it is optically thick to bolometric radiation, can only be from a planet's envelope cooling energy (intrinsic luminosity). At a younger age, boil-off can be energy-limited by a scarcity of intrinsic luminosity before $10^4$ yr rather than stellar bolometric radiation that becomes freely available after only $10^2$ yr. However, such an intrinsic-limited boil-off always converges to a common evolution track from the non-energy-limited isothermal Parker wind model, due to the decoupling with the thermal evolution. Therefore, we argue that the non-energy-limited isothermal Parker is a good approximation for modeling boil-off. The post-boil-off mass loss happens late and therefore is not energy-limited.
   }

   \item{
   We prefer the term (post-boil-off) bolometric-driven escape over core-powered mass loss, as the Parker wind mass loss is not sensitive to core luminosity. It cannot be modeled alone without self-consistent initial conditions from a boil-off phase. With the better assessed physical conditions, we find that the post-boil-off mass loss is inefficient on a long timescale of 10 Gyr. As boil-off removes significant H/He mass from the planetary envelope within $\sim$ 1 Myr and therefore significantly reduces a planet's physical size, at the transition time when the thermal evolution timescale starts to be longer than the mass loss timescale, the mass loss rate is lowered by many orders of magnitude. Consequently, we find that it has no impact on exoplanet demographics.
   }
   \item{
   Boil-off is an efficient mass loss mechanism that can possibly remove a planet's entire convective H/He envelope if the core is not massive enough, or the atmosphere is highly irradiated, or both. The final mass fraction after boil-off depends primarily on the initial entropy and insensitively on the initial mass fraction if the initial physical conditions are properly assessed. We advocate that to quantitatively determine a planet's initial entropy and therefore radius, its initial Kelvin-Helmholtz contraction time should be comparable to the disk lifetime of $\sim$ Myrs. This generally corresponds to a RCB radius of $\leq 20\%$ of its sonic radius. 
   }
   \item{
   The wind-driven energy advection in the convective envelope is found to be a result of mass loss instead of driving envelope cooling. Although the advective cooling assumed in \citet{Owenwu16} can be many orders of magnitude greater than the radiative cooling at the early times of the boil-off it barely affects the thermal evolution during boil-off. Consequently, the cooling assumption is generally inefficient in impacting the post-boil-off mass fraction and entropy for an initial mass fraction smaller than $20\%$, unless the initial envelope entropy is unreasonably high.
   }
   \item{
   Once the photoionization energy deposition region penetrates below the sonic point, boil-off switches into the photoevaporation-dominated phase. We find for all planet models that have a boil-off phase, the transition happens very slightly before the end of boil-off. The subsequent atmospherics escape is therefore XUV-driven, if there is one. 
   }
   \item{
   The previous GSS16+GSS18 analytical model finds an efficient long-lived mass loss, because their ``core-powered escape'' from their coupled wind assumption comprises boil-off as part of the long-lived mass loss phase, namely bolometric-driven escape, as a result of simplified initial radii and H/He mass fractions. This should be disentangled by properly involving a boil-off phase that self-consistently provides the initial conditions for the later evolution. An appropriate criterion for the transition is crucial. We find that dividing the assumed coupled Parker wind into a boil-off phase and a long-lived ``core-powered'' phase with their criterion for the transition such that $R_{rcb} = 2R_c$ introduces a large uncertainty in evaluating the role of the long-lived wind. This is because the transition phase between the two stages is greatly prolonged, leading to slowly varying physical behaviors, e.g. the degree of coupling between the mass loss and thermal evolution and the mass loss enhancement from the core luminosity. Therefore, we propose that the transition should be set by the relative timescale between the mass loss and thermal evolution.
   }
   \item{
   We prefer a decoupled Parker wind (non-energy-limited, bolometric-limited or intrinsic-limited) to the coupled Parker wind assumed in GSS18, after reexamining the energy supply of the wind. The wind assumption largely impacts the behavior of the mass loss and thermal evolution. A number of other assumptions have led to an overestimation of the mass loss in the GSS18 analytical model, including a lack of energy flux that advects out of the interior through winds, no efficient shrinkage as a planet loses mass, and a pure H$_2$ composition (unreasonably low mean molecular weight) of the atmosphere.}
   \item{
   With self-consistent initial conditions and improved physics applied, our numerical model finds a flat demographic feature between $1.5-2.0 R_\oplus$ as a consequence of the vigorous boil-off mass loss, implying the potential role of XUV-driven escape in explaining the observed radius gap. A decrease in sub-Neptune occurrence above $5 R_\oplus$ can be explained by a boil-off based on the numerical model without the need for the high mass tail for the core-mass distribution. To match the observed radius cliff at $4 R_\oplus$, other subsequent mass loss mechanisms are required to further decrease the planetary radius. The radiative atmosphere is required in all planet models, as it increases the physical sizes of planets by 20-30\% above the RCB. Additionally, any version of what we believe are improved implementation of the GSS18 analytical model with different combinations of physical conditions with and without the radiative atmosphere does not show a radius gap similar to observations.
   }
   \item{
   Our results contrast with several more recent modeling studies that incorporate improved physical assumptions beyond GSS18 and similar models \citep{Gupta19,Gupta22} and report significant core-powered mass loss. However, we propose that this discrepancy arises from either other critical physical assumptions not accounted for in those models \citep{Misener21} or from what we believe are misinterpretations that overestimate the role of core luminosity \citep{Rogers24}.
   }
   
\end{enumerate}
With our numerical model, the thermal and mass loss evolution history of a sub-Neptune in the presence of boil-off and the long-lived Parker wind escape is determined under self-consistent initial conditions, with what we find is an appropriate decoupled Parker wind. In the future, other physical effects, e.g. photoevaporation and convective evolution of the rock/iron core, can be readily integrated into our model, enabling us to study these small planets on a population level in a more comprehensive way.

\section{Acknowledgments}
We thank James Owen, Hilke Schlichting, Eve Lee, Xi Zhang, Sivan Ginzburg, Akash Gupta, James Rogers, Kazumasa Ohno, Daniel Thorngren, Eric Lopez, Peter Gao, Madelyne Broome and William Misener for their comments and discussions on this work.

\bibliography{reference}{}
\bibliographystyle{aasjournal}

\appendix
\restartappendixnumbering
\section{Assessment of the energy needed to expand the envelope}
\label{app:envelope}
The most general form of Bernoulli's principle for irrotational and barotropic flow can be derived from the Euler equations of conservation of mass, momentum and energy, leading to the following expression:
\begin{equation}
\label{eqn:Bappendix}
\frac{\partial B}{\partial t} + v\cdot \nabla B = \frac{1}{\rho} \frac{\partial p}{\partial t} + \frac{\partial \phi}{\partial t} + \Gamma + \Lambda
\end{equation}
where $B=e + p/\rho + v^2/2 + \phi$ represents the Bernoulli constant (which is constant along flowlines in a steady-state flow). The external heating and cooling rates per unit mass, denoted by $\Gamma$ and $\Lambda$, are negligible in an adiabatic envelope at a single timestep. The gravitational potential $\phi$ is primarily determined by the core mass and changes slowly over a convection timescale as the envelope mass decreases, resulting in $\partial \phi/\partial t \approx 0$ compared to the other terms in Eq. \ref{eqn:Bappendix}. The barotropic condition is satisfied in isentropic envelopes. The above equation therefore simplifies to:
\begin{equation}
\rho\frac{DB}{Dt} = \frac{\partial p}{\partial t} 
\end{equation}
where $DB/Dt = \partial B/\partial t + v\cdot \nabla B$ is the Lagrangian time derivative of $B$.

Using the numerical model described in Sections \ref{subsec:thermal}--\ref{subsec:radatm}, we conducted differential calculations throughout the entire boil-off phase and across the envelope, confirming that $B$ varies radially by less than 5\% for all of the model planets we checked, so that $\nabla B$ small,  and ${\rho} \partial B/\partial t \approx \partial p / \partial t$. As discussed in Section \ref{subsec:envelope}, we ignored the kinetic energy term in the Bernoulli constant in these calculations because its contribution is small.

To further verify the assumption that $\rho \boldsymbol{v}\cdot \nabla B$ is smaller than the other terms in Eq. \ref{euler_energy} (or, equivalently, smaller than the other two terms in Eq. \ref{eqn:bern}), leading to Eq. \ref{euler_time}, we explicitly calculate and compare the individual terms in Eq. \ref{euler_energy}. We find that $\rho \boldsymbol{v} \cdot \nabla B$ is several orders of magnitude smaller than $\partial E_t/\partial t$ and $h_t\partial \rho/\partial t$, with the latter two terms being equal. For the purposes of this test, we calculate the fluid's velocity $v$ using the mass loss rate, accounting for the fact that the mass loss rate is not constant across the envelope. This suggests that in an adiabatic envelope with time-varying envelope mass, the Bernoulli constant $B$ readjusts itself radially on a fast enough timescale that it can be treated as radially constant.

\begin{figure}
\centering
\includegraphics[width=0.45\textwidth]{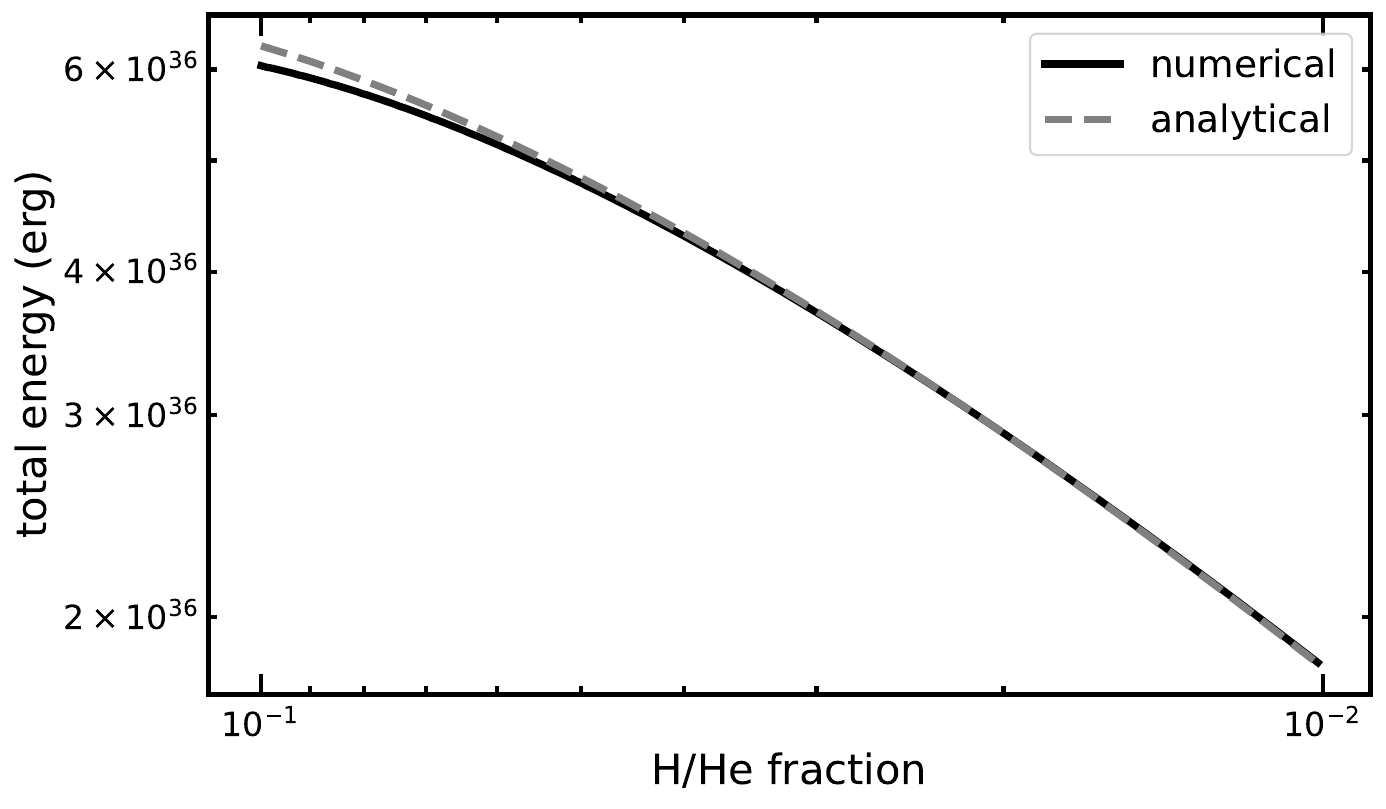} 
\caption{
By varying the envelope mass fraction, we show the total energy difference (black) of adiabatic envelopes between steps and compare it to the analytical estimate that corresponds to the energy needed to lift H/He mass from the surface of the planet (RCB). The agreement between the two quantities is consistent with our analytical assessment in Eq. \ref{denergy}.
}
\label{adaibaticstructure}
\end{figure}

We also showed analytically in Eq. \ref{denergy} that the total energy needed for an adiabatic envelope to lose H/He mass equals the Bernoulli constant $B$ times the mass loss $\Delta M$. We numerically calculated the structures and energy budgets of hydrostatic adiabatic (with constant adiabatic index) envelopes with the same RCB boundary conditions (pressure and temperature) but different envelope masses, assessed without the self-gravity. We decrement the envelope mass fraction from 10\% to 1\%, and record the difference between the total energy $E$ at each step:
\begin{equation}
\Delta E = \Delta\left[-\int{\left(\frac{G M_c m_i}{r_i}+\frac{k T_i m_i}{(\gamma-1)\mu}\right)}\right]
\end{equation}
where $m_i$, $T_i$ and $r_i$ are the mass, temperature and radius of each envelope mass shell, respectively. This is compared to the analytical estimate $(h+\phi)\Delta M$, where $\Delta M$ is evaluated as the envelope mass difference between adjacent steps. A good match is shown in Figure \ref{adaibaticstructure}, verifying that lifting mass from an adiabatic envelope to infinity (approximately the sonic point) does not require additional energy injection into the envelope other than the amount needed to expand the radiative atmosphere.

\section{Initial entropy effect and advective cooling effect for planets with an extreme initial radius}
\label{app:entropyadv}

\begin{figure}
\centering
\includegraphics[width=0.45\textwidth]{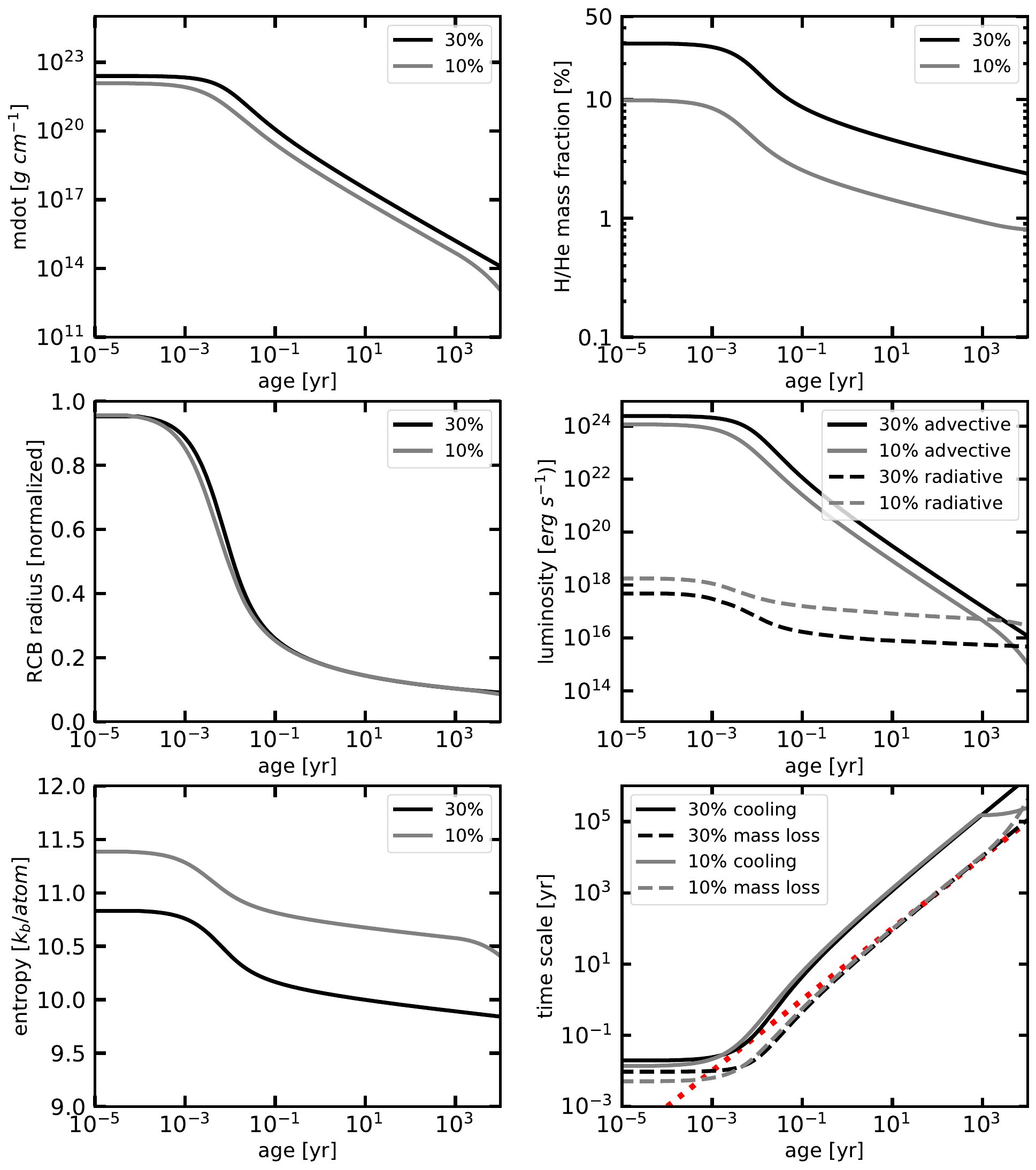} 
\caption{We show a comparison between two $3\ M_\oplus$ models ($30\%$ in $black$ and $10\%$ in $gray$) initially having substantially larger RCB radii (compared to Figure \ref{1v3entropy}) that are the same as the sonic point radii, both without core luminosity, similar to the setup shown in \citet{Owenwu16}. The advective cooling timescale is always comparable to the mass loss timescale. Consequently, advective cooling starts to efficiently remove the planetary interior energy at the time that the H/He mass fraction begins to decline after $10^{-3}$ yr, resulting in a more pronounced enhancement in the post-boil-off mass fraction between models with different initial mass fraction, compared to the models that initially have a smaller radius. However, we suggest in Appendix \ref{app:entropyadv} that the final mass fraction difference is also due to the initial entropy effect (Figure \ref{1v3rcbnca}), rather than only the advective cooling effect (Figure \ref{1v3rcbentropy}). We argue that those cases with the extreme initial radii discussed here do not occur given self-consistent initial entropies ($\sim 10 k_b/atom$) and initial mass fractions ($\leq 20\%$).
}
\label{1v3rcb}
\end{figure}

In a setup similar to \citet{Owenwu16} with extreme initial conditions, we model two $3M_\oplus$ planets initially starting with the RCB radii $R_{rcb,0}$ the same as their sonic point radii $R_s$ and both without any core luminosity. This leads to timescale ratio $t_{cool,ad}/t_{\dot{M}} \sim 1$. We examine the evolution of the mass loss rate, H/He mass fraction, radius, energy budget, entropy and relevant timescales, displayed in Figure \ref{1v3rcb}. Before $10^{-3}$ yr, the advective cooling is not efficient even though the cooling timescale $t_{cool}$ is comparable to the mass loss timescale $t_{\dot{M}}$, due to the lack of time for cooling. Since a significant mass loss is guaranteed during boil-off for this initial setup, the advective cooling that has an equally efficient timescale as the mass loss (for both the 10\% and 30\% models) always becomes effective at the same time when the H/He mass fraction starts to decline (after $10^{-3}$ yr in this case), which significantly enhances the post-boil-off fraction. Therefore, the advective cooling is always important with such an enormous initial radius $R_{rcb,0} \sim R_s$, as found in \citet{Owenwu16}. We warn that under these extreme initial conditions, our adiabatic and hydrostatic assumptions for the envelope discussed in Section \ref{subsec:boil} likely becomes invalid. 

\begin{figure}
\centering
\includegraphics[width=0.45\textwidth]{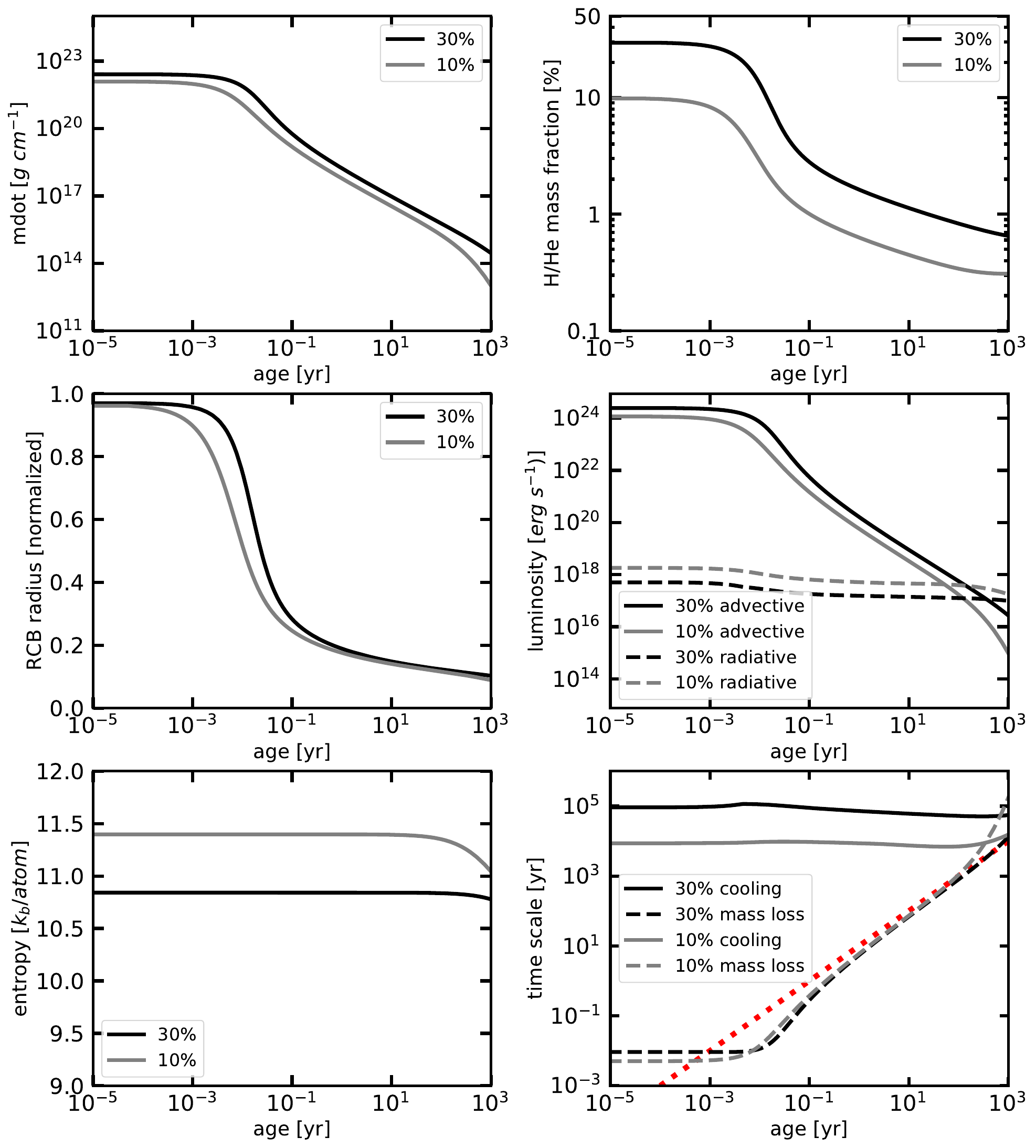} 
\caption{
We examine the importance of the initial entropy effect in affecting the final mass fraction at the end of boil-off by switching off advective cooling from the setup in Figure \ref{1v3rcb}, with other conditions the same. This eliminates the advective cooling effect. In the top right panel, we find the initial entropy effect accounts for a factor of 2 difference, implying that it is another important physical effect, in addition to advective cooling, in explaining the different final mass fractions found in \citet{Owenwu16}. The same physical quantities are plotted in each panel, as in Figure \ref{1v3entropy}.
}
\label{1v3rcbnca}
\end{figure}

\begin{figure}
\centering
\includegraphics[width=0.45\textwidth]{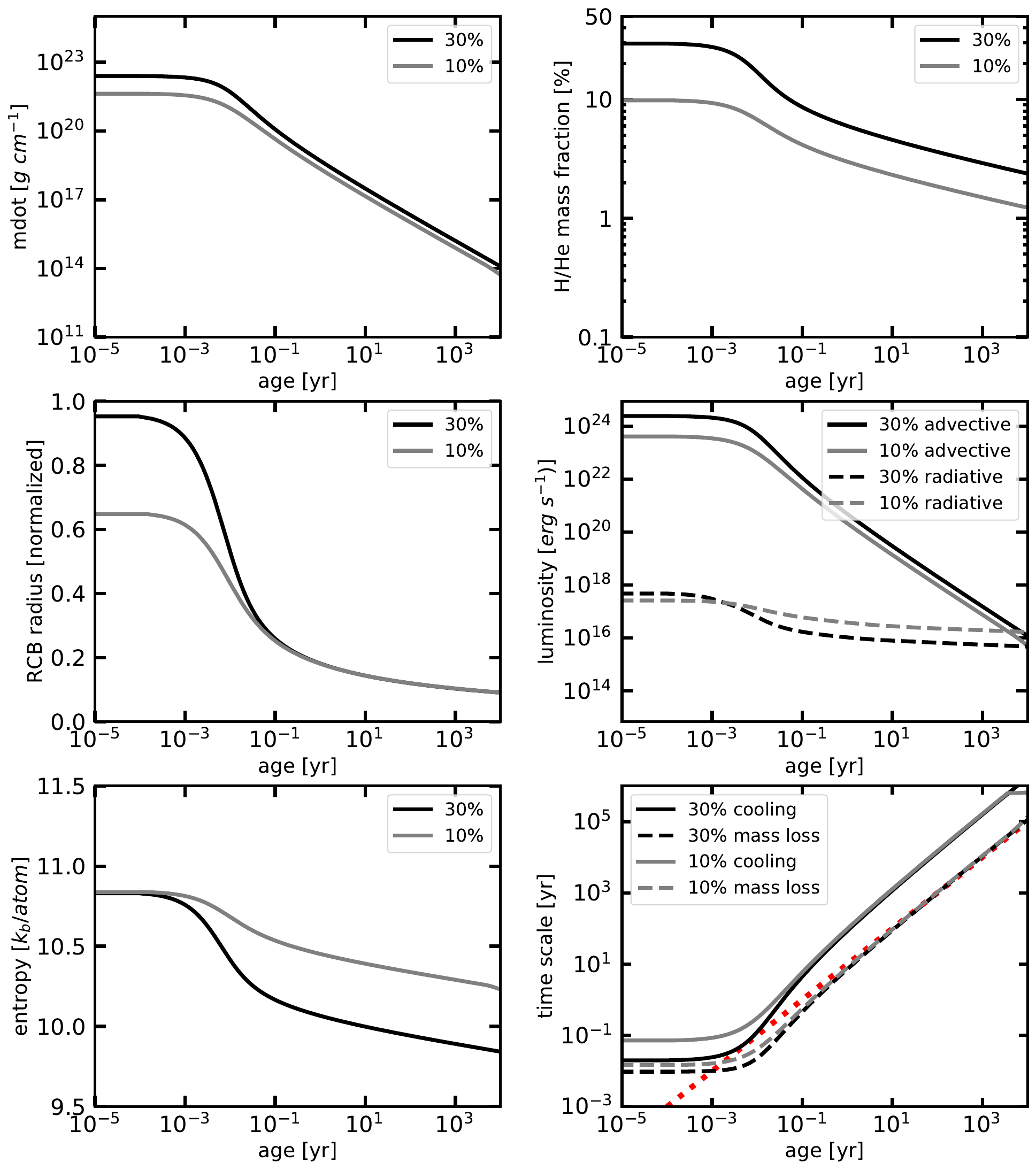} 
\caption{ 
We evaluate the importance of advective cooling by getting rid of the initial entropy effect from the setup in Figure \ref{1v3rcb}. We start both planets with the same initial entropy (bottom left), which gives different initial radii (middle left) with the 30\% model larger, same as the sonic point radius. In the top right panel, we find advective cooling generates an enhancement by another factor of 2, compared to the initial entropy effect, demonstrating that both effects are equally important with such an extreme initial radius. However, we argue that the advective cooling effect is typically minor if planets are born smaller, only 10\%-20\% of the sonic point radius $R_s$ with the self-consistent initial conditions. The same physical quantities are plotted in each panel, as in Figure \ref{1v3entropy}.
}
\label{1v3rcbentropy}
\end{figure}

We suggest that the elevated post-boil-off mass fraction from the setup above is not only from the advective cooling itself as described in \citet{Owenwu16} but also in part from the initial entropy effect, for the reason that the $30\%$ model always needs a smaller initial entropy than the $10\%$ model to inflate planets to the certain physical size $R_s$. To separate these two effects and evaluate their strength, we set up two model comparison experiments for the initially 10\% and 30\% models with the same $3M_\oplus$ core mass. In Figure \ref{1v3rcbnca}, we show the evolution tracks of the planets with the initial RCB radii the same as the sonic point radii $R_s$ but with the advective cooling switched off, which finds an enhancement of the final mass fraction at the end of boil-off by a factor of 2. In Figure \ref{1v3rcbentropy}, starting both models with the same initial entropy (but different initial radii, still comparable to $R_s$) to eliminate the initial entropy effect, we find advective cooling results in another factor of 2 difference in the final mass fraction. Therefore, the mass loss enhancements in Figure \ref{1v3rcb} and \citet{Owenwu16}, where both the 10\% and 30\% models start with the same initial RCB radii as their sonic radii, are from a combined effect due to both the advective cooling and the initial entropy. We find these two effects are equally important for planets with an extreme initial radius, which corresponds to a high entropy ($\geq 11 k_b/atom$) and high mass fraction ($\geq 20\%$). 

\end{document}